\documentclass[12pt]{article}

\usepackage{amsmath}

\usepackage[margin=1in]{geometry}     
\usepackage[font={small}]{caption}
\usepackage{graphicx}        

\begin{document}

\author{LUISA BONOLIS\\Max Planck Institute for the History of Science, Berlin, Germany\\Email: lbonolis@mpiwg-berlin.mpg.de, luisa.bonolis@roma1.infn.it}
\title{International  scientific cooperation during the 1930s. Bruno Rossi and the development of the status of cosmic rays into a branch of physics}

\maketitle

\begin{abstract}

\noindent
During the 1920s and 1930s, Italian physicists established strong relationships with scientists from other European countries and the United States. The career of Bruno Rossi, a leading personality in the study of cosmic rays and an Italian pioneer of this field of research, provides a prominent example  of this kind of international cooperation.  Physics underwent major changes during these turbulent years, and the traditional internationalism of physics  assumed a more institutionalised character. Against this backdrop, Rossi's early work was crucial in transforming the study of cosmic rays into a branch of modern physics. His friendly relationships with eminent scientists ---notably Enrico Fermi, Walther Bothe, Werner Heisenberg, Hans Bethe, and Homi Bhabha--- were instrumental both for the exchange of knowledge about experimental practises and theoretical discussions, and for attracting the attention of physicists such as Arthur Compton, Louis Leprince-Ringuet, Pierre Auger and Patrick Blackett to the problem of cosmic rays.  
Relying  on material from different archives in Europe and United States, this case study aims to provide a glimpse of the intersection between national and international dimensions during the 1930s, at a time when the study of cosmic rays was still very much in its infancy, strongly interlaced with   nuclear physics, and full of uncertain, contradictory, and puzzling results. Nevertheless, as a source of high-energy particles it became a proving ground for testing the validity of the laws of quantum electrodynamics, and made a fundamental contribution to the origins of particle physics.

\end{abstract}


\section{Introduction}

During the 1920s and 1930s, the  Italian physics community was formed by a small  number of competent and brilliant people.  The nation did not, however,  have  a consolidated tradition in modern physics, and the dicipline's academic weight was scarce and  was not enthusiastically supported by institutional centres \cite{Galdabini:1989lr}. Yet the new generation of  physicists born during the early years of the 20th century was able to overcome these difficulties  and succeeded in reaching many well-known achievements.  This probably derived partly from the fact that, during the end of the 19th and early 20th centuries, the international physics community had reached a supranational status which guaranteed effective forms of formal and informal knowledge transfer. During the 1910s and 1920s, the new physics developed around personalities like Ernest Rutherford, Niels Bohr, Arnold Sommerfeld, and Max Born, who played leading roles in providing  a map of reference centres across Europe.  These centres became a natural environment for the formation of research schools, exchange of knowledge, and scientific discussions. In 1919, when official  relations within the European scientific community were still tense in the aftermath of World War I, an International Research Council was created  to promote international co-operation in science, through the formation of a series of discipline-specific International Unions. During a meeting of the Council in Brussels in 1922 the International Union of Pure and Applied Physics (IUPAP) was founded to encourage and aid international cooperation in the field of physics and to help lower  barriers to the free circulation of scientists, information, and ideas.\footnote{Orso Mario Corbino, who would soon support Enrico Fermi's career, convinced of the importance of relaunching Italian physics, was among the ten distinguished physicists forming the executive committee who prepared rules, regulations, and activities of the new organization.}  

At that time, however,  policies of ostracism were still advocated by the International Research Council against the Central Powers. The League of Nations' International Commission on Intellectual Cooperation also excluded German and Austrian scientists when twenty-five chemists convened in Brussel for the first Solvay Chemistry Congress in 1922.\footnote{The only German chemist invited to any Solvay council during the 1920s was Hermann Staudinger (in 1925), who was a pacifist and  in different situations had taken position against chemical warfare.}
 In the postwar Solvay Physics councils, Schr\"odinger was the only physicist from the former Central Powers who attended a meeting before 1927, and Einstein, who had been invited in 1921 and 1924, declared that in any case he would not  participate as long as his German colleagues were excluded. The consequences of the war affected the traditional international conferences of mathematics, too. Only in 1928, in Bologna, a German delegation of 67 mathematicians headed by David Hilbert attended, for the first time since the war, an international meeting of mathematicians. Hilbert himself expressed his satisfaction `that after a  long, hard time all the mathematicians of the world are represented here [\dots] all limits, especially national ones, are contrary to the nature of mathematics. It is a complete misunderstanding of our science to construct differences according to peoples and races, and the reasons for which this has been done are very shabby ones.
Mathematics knows no races [\dots] For mathematics, the whole cultural world is a single country.'\footnote{C. Reid.  {\it Hilbert}. Berlin: Springer-Verlag,  1970, p. 188.}

In recognizing the necessity of a broad communication with the international community, IUPAP provided in these difficult years a more institutional base to a field that already functioned through the  spread practise of using journals, conferences, international symposia, visiting, and post-doctoral fellowships. The gradual restoring of international ties during the 1920s, and the establishment of new facilities at  research centres creating productive environments throughout the world, both favoured close ties among physicists and their common efforts to probe the new frontiers of the discipline.  Physics was  more and more becoming a field with no international boundaries. 

It is  well known how Fermi, after getting his laurea degree in 1922, had the opportunity of having personal contacts with prominent members of the international scientific community since the time of his first sojourns in G\"ottingen, with Max Born, and in Leyden, with Paul Ehrenfest. That same year Einstein was invited by Paul Langevin to lecture at the Coll\`ege de France, which marked the first public appearance by a German scientist since the war.

Official international relations  continued to be difficult until the second half of the 1920s.\footnote{Relationships between Mussolini and Germany were in particular tense mainly because of the South Tyrolean question (S\"udtirolfrage) which became and continued to be an international issue.} 
 It is thus remarkable ---a sign of the changing situation--- that at the international physics conference organized in Como in September 1927, on the occasion of the centenary of Alessandro Volta's death, several prominent German scientists were invited: Max Born, James Franck, Max von Laue,  Arnold Sommerfeld, Wolfgang Pauli,  Werner Heisenberg, Lise Meitner, Friedrich Paschen, Max Planck, and Otto Stern. Other great figures in the world of physics like Hendrik A. Lorentz and Ernest Rutherford were also there, while Niels Bohr presented for the first time his principle of complementarity, and Sommerfeld described a series of results showing the importance of Fermi's new quantum statistics for interpreting the behaviour of electrons in metals.\footnote{The following year, Heisenberg invited Fermi to the first {\it Leipziger Universit\"atswoche} on `Quantum theory and Chemistry' which he organized together with Peter Debye in June 1928, to which Fermi contributed with a talk on the `Application of Statistical Method of Problems of Atomic Constitution.'} As recalled by Emilio Segr\`e,  `it was easy to see that Fermi was the only Italian who counted in the eyes of the foreign participants.' \footnote{\cite[p. 46]{Segre:1993qy}. Segr\`e immediately added `What I had seen at the conference tipped the scales in my decision to switch from engineering to physics,' and indeed during the following months he became Fermi's first student at the Royal Institute of Physics at Via Panisperna in Rome.} 
    \begin{figure}[h]
\centering
\includegraphics[width=0.6\linewidth]{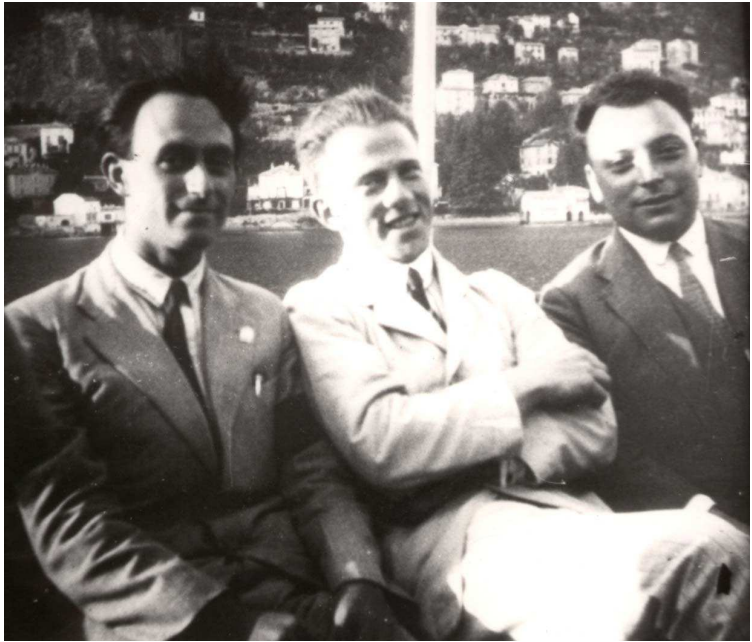}

\caption{Enrico Fermi, Wolfgang Pauli and Werner Heisenberg on the Como Lake during the Volta Conference of September 1927 (Amaldi Archive,  Sapienza University, Rome).}
\end{figure}
\par\noindent
 
 The Como conference officially opened a new era in  international physics conferences,  and launched  Enrico Fermi onto the international stage.  Fermi had recently been appointed to the Chair of Theoretical Physics in Rome, a position which was created in 1926. Next October, the 1927 Solvay conference dedicated to `Electrons and photons' had not only Schr\"odinger, Born, Planck, Einstein among its participants, but also the young Pauli and Heisenberg,  two of the future leading figures of European physics. Fermi was invited in 1930, standing near Heisenberg in the official photograph of the conference. 

It is not by chance that the Italian concept of the solitary physicist, aided only by a laboratory assistant, gradually came to an end with the new generation of physicists born in the early 20th century. Strong  relationships developed around this time between physicists in Italy and other European countries such as Germany, Great Britain and France, as well as with some physicists of the U.S. scientific community.  These connections formed particularly in connection with the birth of Enrico Fermi's school in Rome during the late 1920s, and the establishment of a research tradition in cosmic-ray physics originating from Bruno Rossi's early studies in Florence.    Some general information is known about Italian physicists visiting the most important centres abroad, especially during the 1920s and 1930s. On the other hand, after Fermi was appointed to a professorship in Rome, many European physicists began travelling there. It was not only young physicists who began to visit Italy: visitors included Samuel Goudsmit, George E. Uhlenbeck, Hans Bethe, Felix Bloch, Christian M\o ller, Georg Placzeck, Edward Teller, Fritz London, Rudolf Peierls, Homi Bhabha, and Eugene Feenberg. On these occasions new bonds were also created  with Fermi's young collaborators, notably with Edoardo Amaldi, who would become a leading figure of Italian physics when Fermi left Italy at the end of the 1930s.

 However, a systematic and specific study of how these issues affected the Italian physics community during the above-mentioned years has yet to be performed. The only notable exception is that of the scientific collaboration between Giuseppe Occhialini and  Patrick Blackett at the beginning of the 1930s,\footnote{For a detailed account see \cite{Bustamante:2007kx}, \cite{Leone:2008fk} and \cite{Leone:2011kx}.} 
which, however, was not put in a more general international context.\footnote{Another specific study has recently taken into consideration the relationship between Viktor Hess and Domenico Pacini within the context of the discovery of cosmic rays in the 1910s, from the point of view of nationalism and internationalism in science \cite{Carlson:2011uq}, especially related to the failed recognition of Pacini's contribution at an international level.} 

 The study of the details of these displacements, of their motivations, and of the impacts they had on the scientific careers of the protagonist, could help to paint a picture of the changes in physics from the late 1920s to the end of the 1930s. At that time, the interest shared by so many physicists in quantum electrodynamics, nuclear physics, and cosmic ray studies provided a common ground for interaction.  This common ground also materialized in regular scientific meetings in different European centres, which helped to shape the informal network connecting these poles of attraction.

This kind of enquiry can benefit from the use of a biographical key as an organising principle.  Such an approach can provide a better understanding of the role of the individual within the international research network which was taking shape in connection with the explosive development of physics. It appears that personal relationships, nurtured by frequent contact at the physics conferences which intensified in those years, played an important role especially in the development of an international network at a European level.  This network would prove especially important after World War II.
 
  The case of Bruno Rossi, a leading personality in the study of cosmic rays and a pioneer in the emergence of this research field as a branch of physics since the early 1930s, is a prominent example  of this spirit of international cooperation.\footnote{The main biographical information on Bruno Rossi can be found in his autobiography: \cite{Rossi:1990aa}\label{Rossi:1990aa} and in \cite{Clark:2000fk}. Rossi's early work, up to the end of 1930s--beginning of 1940s, has already been discussed  in: \cite{Bonolis:2011fk}; and especially in: \cite{Bonolis:2011ak}, where the birth and development of coincidence methods in cosmic ray physics have been analysed in greater detail. For this reason,  the present work will explore some more specific aspects of Rossi's activity especially related to his personal interaction with physicists of his time.} Rossi was born in Venice, in 1905.  Like many physicists of his generation he was only a boy during World War I, and so was not involved in the international problems typical of the previous generation. His university studies ended in November 1927, when he got his laurea degree in physics at the University of Bologna.  Rossi then moved to the Physics Institute of Florence located on the hill of Arcetri, not far from the house where Galileo Galilei lived part of his life and died in 1642.  Giorgio Abetti, the director of the Astrophysical Observatory, organised seminar presentations by prominent international scientists in Arcetri, and these seminars helped to broaden Rossi's horizons. Enrico Persico's lectures on the new quantum mechanics were also a main event in Florence.\footnote{Together with Fermi, Persico had won the first Italian competition for a chair of Theoretical Physics.} Rossi and Giulio Racah compiled the first Italian university manual on this topic. All these novelties stimulated the young physicists' fantasy:  fascinated by Persico's lectures, Bernardini and Rossi  decided to verify experimentally the predictions of wave mechanics by repeating, with slow electrons, one of the classical experiments of optical interference.  Failure of this project, as well as of an attempt to photograph the spectra of comet tails for the purpose of discovering their chemical composition, generated a sense of frustration in the young Rossi: `I was beginning to wonder whether my ambition had not led me into a blind alley and whether the time had not come to lower the aim.'\footnote{\cite{Rossi:1990aa} (note \ref{Rossi:1990aa}), 6.}
  
While Enrico Fermi was building up a research group in Rome that would soon redirect its interest from atomic to nuclear physics, Bruno Rossi in Florence was eager to  to start working at  some experimental project addressing itself  `to the fundamental problems of contemporary physics.'
Then, in the autumn of 1929, the  paper by Walther Bothe  and Werner Kolh\"orster  on the nature of the extraterrestrial penetrating radiation appeared,\footnote{\cite{Bothe:1929aa}.\label{Bothe:1929aa}} 
which was `like a flash of light revealing the existence of an unsuspected world, full of mysteries, which no one had yet begun to explore'  \cite[p. 43]{Rossi:1966aa}. This article had a lasting impact on Bruno Rossi's life and science.

\section{Cosmic ray research becomes a branch of physics}

Cosmic rays were characterized by a penetrating power far exceeding that of any other known rays. Of these, the most penetrating were the $\gamma$-rays, whose penetrating power  was actually expected to increase steadily with increasing energy according to current theories. The problem of the nature of cosmic rays did not attract general attention, probably because of the widespread belief that the answer was already known: the astonishingly penetrating cosmic rays could not be anything else but $\gamma$-rays of very high energy.\footnote{During the previous years Robert Millikan had provided theoretical justifications  for his ideas on their `cosmic' origin. According to his theory,  cosmic rays were the `birth cry' of atoms in space, being born in the form of   $\gamma$-rays, from the energy set free in the synthesis of heavier atoms through fusion of primeval hydrogen atoms; see for example: \cite{Millikan:1928aa,Millikan:1928vn}. For an accurate reconstruction of Millikan's theories on the origin of cosmic rays see \cite[Ch. 3]{Galison:1987kx}\label{Galison:1987kx}, and \cite[Ch. 3]{Russo:2000ab}.\label{Russo:2000ab}}  
$\gamma$-rays were known to ionize through the intermediary of secondary charged particles  which they generate in matter; therefore, it was expected that cosmic $\gamma$-rays  traveling through matter would  be accompanied by a flow of  secondary electrons resulting from the Compton effect, which were presumed to be the  ionizing agent recorded by the measuring instruments. 
 A direct  study of this corpuscular radiation could thus clarify the nature of cosmic rays.

Up to that time, absorption measurement  carried out with highly sensitive ionization chambers had been a major tool in the study of the penetrating radiation.\footnote{The ionization chamber used during the first era of cosmic-ray experiments, even being a reliable and stable instrument, was not suited for detailed investigations of the properties of cosmic rays, only permitting the determination of the total intensity of radiation averaged over all directions of incidence.} But at the end of the 1920s, brand-new instruments were being developed that opened the door to the investigation of the physical properties of the local radiation, thus starting a revolution in cosmic ray experimental research and transforming it into a completely new  research field. A  novel counting tube  had just been developed in Kiel by Geiger and his pupil William M\"uller \cite{Geiger:1928kx} who had announced their invention  on 7 July 1928, during a meeting of the German Physical Society.\footnote{For the origin of the GM counter see \cite{Trenn:1986lr}.}
 Actually,  this  device  very soon appeared to respond to an external  radiation, which Geiger and M\"uller strongly suspected to be cosmic radiation. At the time, however, they left open the question of whether all the spontaneous counts should be attributed to cosmic rays. 

Hints of that same `local' reality had already been captured by  the Russian physicist Dimitri Skobeltzyn \cite{Skobeltzyn:1927fk},  the son of a professor of physics, working at the {\it Leningrad  Physicotechnical Institute}.
 During an informal conference organized in Cambridge U.K. at  the end of July of 1928, he presented a talk on `The intensities of $\gamma$-rays' and showed some photographs of what he thought to be cosmic-ray tracks in his Wilson chamber.
 Following his remarks, Geiger announced that Bothe and Kolh\"orster `were working on a 
method to register cosmic rays by the coincidence of pulses in two  GM counters, and that  they hoped to be able to study the penetrating power of the rays by this method' \cite{Skobeltzyn:1985vn}.  Physicists attending the conference probably learned for the first time about the existence of   this new kind of counter. 

Bothe had been a pioneer in the coincidence methods which  he had already used  in his contributions to the understanding of the particle-wave dualism of light.\footnote{For a discussion on Bothe's development of the coincidence method,  see \cite{Fick:2009fk}.} 

His investigations  had in the meantime stimulated Werner Kolh\"orster, who was working on $\gamma$-ray experiments in Geiger's laboratory at the  {\it Reichsanstalt} as a permanent guest,  to  place two of the  GM  counters  side by side in a beam of $\gamma$-rays.  
As it turned out, that was the first step toward submitting the $\gamma$-ray hypothesis to a crucial experimental test. In  a  preliminary note dated 6 November  1928,\cite{Bothe:1928vn} Bothe and Kolh\"orster  reported their innovative attempts to  measure the absorption of those secondary electrons by recording the coincidences between two superimposed GM counters interleaved with lead plates of increasing thickness.\footnote{They sent a new short note to {\it Nature} appearing on the issue of  April 27, 1929: `Up to the present time the view that the penetrating radiation consists of short gamma rays has been prevalent chiefly because the large penetrating power which these rays possess is associated with radiation of gamma ray type. Our recent experiments, however, indicate that this radiation is of corpuscular nature'. \cite{Bothe:1929fk}.} 
In a new experiment they placed one GM counter above another one, with a 4.1-centimeter layer of gold between them.  On 28 June  1929, a final and detailed paper was submitted.\footnote{\cite{Bothe:1929aa} (note \ref{Bothe:1929aa}).} In the opening lines of the article Bothe and Kolh\"orster clearly highlighted the problem of cosmic rays: `Research into the high-altitude radiation has so far taken a strange course, for the  most diverse features of the radiation, such as intensity, distribution, absorption and scattering, and even its origin, are investigated and debated, {\it whilst the really essential question regarding the nature of the high-altitude radiation has hitherto found no experimental answer}  [emphasis added]'. From  data recorded with their mixed arrangement of counters interleaved with a thick gold shield, they argued that coincidences could be produced  {\it only by individual ionizing particles crossing both counters}, which they thought to be high energy electrons. The high penetrating power of such particles excluded the possibility that they could be Compton electrons generated by the `primary' $\gamma$-radiation. 

Bothe and Kolh\"orster's experiment, in providing evidence for the enormous potential of the coincidence method, actually represented the very first attempt to determine the nature of cosmic rays, and helped to focus the physicists' interest on  the radiation found at the place where measurements were made.  Both the novelty of the research topic and the low cost of the necessary tools were the key ingredients of Bruno Rossi's excitement in the fall 1929. 
He immediately set to work with the help of his students ---
particularly Giuseppe Occhialini and Daria Bocciarelli--- and of his colleague Gilberto Bernardini. Within a few weeks the first counter was in operation. Being an inexpensive detector, within the reach of every small laboratory, the GM  counter required only good scientific intuition and experimental skill to get useful results in the new fields of nuclear physics and cosmic rays. 
The technique of building counters on which the Arcetri group became quite skilled was later `exported' from Florence to Rome, and indeed the Geiger-M\"uller technique which Rossi introduced in Italian physics would play a crucial role in the well-known discovery of the radioactivity induced by neutrons and the related discovery of nuclear reactions brought about by slow neutrons made by Enrico Fermi and his group in Rome in 1934 \cite{Leone:2005aa}. 
  
Immediately after building the first functioning counters, Rossi tackled the coincidence technique, 
which was at the core of the Bothe--Kolh\"orster  experiment.  With incredible insight and skill, he
succeeded in fully developing the capabilities of the method. His first publication in the field was  submitted on February 7,  1930, and appeared in the April issue of {\it Nature}.\footnote{\cite{Rossi:1930mz}.\label{Rossi:1930mz}} He had the simple but  ingenious idea of using an electronic amplification device as an \textit{automatic switch}. In his circuit for triple coinciding impulses with three counters, the positive electrodes were coupled to  three valves in such a way that the current flow was interrupted in the valves only when all the counters were working at the same time. This meant that in his ``mixed arrangement'', consisting of GM-tubes and valves, a pulse would only be emitted when two or more other pulses were delivered to the circuit {\it at the same time}.

On 12 November 1929 Bothe had submitted an article which appeared in the January issue of  the  \textit{Zeitschrift f\"ur Physik}, describing a method for registering simultaneous pulses of two Geiger counters \cite{Bothe:1930lr}. However, Rossi's improved version of the coincidence circuit, which offered a tenfold improvement in time resolution, was conceptually different from Bothe's scheme, which employed a single tetrode vacuum tube and could register only {\it twofold} coincidences. The possibility of arranging three counters in coincidence, or more in the $n$-fold version of the circuit, greatly reduced the rate of chance coincidences, thus allowing  observations with increased statistical weight.  Moreover, the time correlation among associated particles crossing different counters could be established.\footnote{For details on the use of coincidence circuits in cosmic-ray research see \cite{Bonolis:2011ak}.\label{Bonolis:2011ak}}

 Now, the coincidence concept applied by Bothe in his first researches on the Compton effect was  assuming a much stronger significance in conjunction with vacuum tube circuits. If the Geiger-M\"uller counter was an instrument of precision, being a tool more discriminating than the ionization chamber, Rossi's $n$-fold electronic coincidence circuit radically changed  the view on the problem of cosmic rays, and opened a ``new technological window'' to explore the universe.\footnote{On this aspect, which is a fundamental key to understand Rossi's scientific activity, see L. Bonolis, ``Bruno Rossi and the opening of new windows on the universe'', to be published on the Astroparticle Physics, http://www.sciencedirect.com/science/article/pii/S0927650513000832.\label{astroparticlejournal}} The possibility of arranging more counters, in whatever geometrical configuration, eventually opened new possibilities of investigation. Two or three aligned counters allowed the   study of any kind of directional effects, while the non-aligned configuration would  soon prove fundamental for  studying  secondary effects, such as the production of new particles  from interactions between cosmic rays and matter.   These arrangements were the precursors of the AND logic circuit later used in electronic computers. 

    Realized in electronic circuits, the coincidence counting method came to be of vital importance for all experiments with several electronic detectors.\footnote{For a wide and detailed discussion on the electronic methods see \cite[ch. 6]{Galison:1997rm}.}  During the 1930s this technique was adopted around the world, in the study of cosmic rays, in `nuclear'  physics, and in the nascent particle physics.\footnote{Actually, until the end of the 1930s, nuclear physics  was a wide field including high energy processes which later would give rise to  particle physics.}

\section{In Berlin with Walther Bothe}

In the meantime Rossi had written to Bothe, telling him that he would like to spend some months in his laboratory at the {\it Physikalisch-Technische-Reichsanstalt} in Berlin-Charlottenburg.\footnote{Rossi's work is often mentioned and analysed in Bothe's manuscripts; however, letters written by Rossi cannot be unfortunately found in Bothe's papers at Max-Planck-Gesellschaft Archives in Berlin, nor in Bothe's small fund preserved at Deutsches Museum Archives in Munich, even if Rossi himself thanked Bothe in some of his articles for `frequent exchanges of correspondence'.}
 Bothe's answer was a positive one, so Bruno Rossi left Arcetri in the late spring-beginning of summer 1930 supported by a grant from the Italian National Council of Research. At the end of 1920s,  Germany ---the cradle of quantum mechanics--- was nearly a paradise for Italian physicists. Berlin was a town  where arts and sciences met and flourished in the melting pot of a  sophisticated innovative culture.  This would certainly have been impressive for the young Italian arriving from a town as small as Florence, even given its rich artistic and historical tradition.
Rossi  was only 24 years old, and the memory of that summer was still very vivid in  his mind many years later:\footnote{\cite{Rossi:1990aa} (note \ref{Rossi:1990aa}), 15.}

\begin{quotation}
\small

Berlin was then the heart of modern physics. For the weekly seminars the lecture room was crowded with physicists of all ages. The first row of benches looked like a hall of fame. Sitting there were scientists whose names were known to me as the creators of the new science. Unconsciously, I had felt that they hardly would look like ordinary human beings. And yet there they were, attentive and unassuming, Albert Einstein   Max Planck  Otto Hahn  Lise Meitner  Max von Laue  Walther Nernst and Werner Heisenberg.

\end{quotation}

\normalsize

Patrick Blackett  came, too, visiting from England. He had spent the academic year 1924--1925 at G\"ottingen, being among the first to reopen contacts with Germany after World War I. Their friendship, begun on that occasion, would  later turn out a very important  relationship in Rossi's future life.\footnote{Since Blackett's wife was Italian, they `easily became quite friendly'.  Notes on P.M.S. Blackett, interview by John L. Heilbron, December 7, 1962, Niels Bohr Library \& Archives, American Institute of Physics, College Park, MD USA (from now on referred to as NBL\&A).\label{blackettinterview}}  Blackett was a recognized expert on cloud chambers, so that Rossi asked him about the possibility of sending his collaborator Gilberto Bernardini to Cambridge in order to learn this important technique. 

In that same period, during the winter of 1929--1930, the young Giovanni Gentile Jr., after a short stay in Rome with Fermi, had visited Berlin where he worked on molecular physics with Erwin Schr\"odinger's assistant Fritz London. In May 1930 Gentile Jr. had moved to Leipzig, where he had collaborated with Felix Bloch,  one of Heisenberg's assistants. Gian Carlo Wick,   who had graduated in Turin, was also in Leipzig. Heisenberg was still a very young professor; his fascinating personality also attracted Ettore Majorana, who spent some time  in Leipzig slightly later, in 1933. That was the time when the new generation of Italian physicists systematically visited the most important European centers. They  felt the need to do research in foreign institutes and laboratories, and experience different atmospheres and new work styles, as well as to learn new techniques. Edoardo Amaldi went to Leipzig, too, but being an experimental physicist, he chose to work with Peter Debye  on the X-diffraction of liquids. Like Emilio Segr\`e, Amaldi had always worked in spectroscopy, and had been trained in the art of experimental physics by Franco Rasetti, who was professor of Spectroscopy at the University of Rome. Rasetti himself had been the first to take off in 1928, spending about nine months at the California Institute of Technology at Pasadena, under Robert Millikan.\footnote{As Rasetti recalled, when interviewed by Judith R. Goodstein,  he had the chance to meet Heisenberg, who had been invited to give lectures on quantum mechanics. Rasetti remarked  how they became very good friends (they were both born in 1900). They  liked mountain climbing ---a very common sport among physicists--- and made an excursion together.  February 4, 1982, Archives of the California Institute of Technology,  available at http://oralhistories.library.caltech.edu/70/1/OH\_Rasetti.pdf.\label{RasettiGoodstein}} 
At Caltech, Rasetti  had quickly accomplished work on the recently discovered Raman effect, which provided some fundamental clues about the quantum behaviour of diatomic molecules. Later, in 1931—1932, Rasetti was in Berlin-Dahlem with Lise Meitner and Otto Hahn, where he studied the penetrating radiation emitted by beryllium under $\alpha$-particle bombardment, confirming that it consisted of a mixture of neutrons and gamma radiation. Gilberto Bernardini and Lorenzo Emo Capodilista from Florence also visited their laboratory in the early 1930s. They  were all interested in learning the techniques  of the cloud chamber and the manipulation of radioactive substances.  In 1931 Segr\`e attracted the interest of Peter Zeeman, a Nobel Prize winner and the discoverer of the celebrated Zeeman effect, and worked for some time in his laboratory to study forbidden spectral lines. Later he got a Rockefeller fellowship and spent it in Hamburg, with Otto Stern. Many of these sojourns were indeed financed by the Rockefeller Foundation,  whose guiding concept from the beginning of the 1930s  became `the advancement of knowledge.' Extending its longstanding attachment to science, the Foundation committed itself to scientific research as a means of enabling human progress, thereby favouring the flourishing of international exchanges. In Italy people involved in the selection were renowned mathematicians like Vito Volterra and Tullio Levi-Civita, both of whom, like many members of the Italian mathematics community, had a longstanding tradition in international bonds.\footnote{Here it will suffice to recall the scientific correspondence on the problem of gravitational field exchanged during the 1910s  between Einstein and Levi-Civita.  Together with the mathematician Gregorio Ricci-Curbastro, Levi-Civita was the founder of tensor calculus, known to mathematicians as Absolute Differential Calculus, a basic tool for the formulation of Einstein's general theory of relativity.}

Bruno Rossi was thus one of the first Italians of the new generation, bred by Orso Mario Corbino in Rome and by Antonio Garbasso in Florence, to visit European laboratories.  His choice was actually quite natural. 
After Bothe and Kolh\"orster's work, the astrophysical and the physical aspects of the cosmic-ray problem had become well defined: on one side there was an interest in establishing the nature of the primary cosmic rays, as this knowledge could throw some light about their place of origin and their production mechanism; on the other side there was the local radiation, found at the place where measurements were made. Investigations of the local radiation became the object of Rossi's research programme, on the ground of the same reason stated by Bothe and Kolh\"orster: {\it the real problem of the day was the nature of the H\"ohenstrahlung}. And Rossi was determined to seek an answer to this question. 

Rossi spent the whole summer of 1930 in Berlin-Charlottenburg. From a conceptual point of view he extended their work,  but now he made a step forward \textit{based} on the working hypothesis that the penetrating radiation had a {\it corpuscular nature}. His results provided a more direct proof of Bothe and Kolh\"orster's conclusion that a corpuscular radiation could be observed at sea level which was not a secondary effect generated from ultra-$\gamma$-radiation. However, the principal novelty of Rossi's experiments in Berlin was that he made a direct comparison between the coincidence rates recorded when a 9.7 cm lead absorber was placed \textit{above} the two counters of the coincidence circuit with those recorded when the same absorber was placed \textit{between} the counters. The  first configuration actually revealed  an excess of about 4\% in the coincidence rate, which Rossi duly interpreted as the probable production of `secondary corpuscular rays' in the metal absorber.  This was an early hint of the complexity of the interaction of the penetrating radiation with matter. It was his first encounter with the evidence of the unexpected existence of secondary showers generated in the shields, which he would explore in depth during the two following years. 

Last but not least, is to be remarked that Bothe must have been very confident in the young Italian physicist, in order to reveal to him the secrets to preparing reliable Geiger-M\"uller counters.  Tt the time this was still  a very difficult art,   considered by many to be a kind of `witchcraft'.  In fact, as it will be explained in the following section, Bothe  also told Rossi about new possibilities of testing the corpuscular nature of cosmic rays, of which both would soon become the main proponents.

\section{The geomagnetic effects and the beginning of the Rossi--Heisenberg correspondence}

There was one serious objection to the conclusions reached by Bothe and Kolh\"orster. The interpretation of their experiments was based on an arbitrary extrapolation of the known properties of photons and electrons at low energies. It was conceivable, for example, that the energies of cosmic-ray photons might be much greater than those computed from their mean free path according to the Compton scattering formula  of Oskar Klein and Oshio Nishina  \cite{Klein:1929fj}, based upon Dirac's theory, and which was known to be valid for energies of the order of 1 MeV, characteristic of some particles emitted by radioactive sources. If this were the case, the secondary electrons would have had a greater range  and more of them would have penetrated the gold block between the counters, and they might have produced much the same coincidence effects as a primary corpuscular radiation.

However, if primary cosmic rays were charged particles, they would also be affected by the geomagnetic field before entering the terrestrial atmosphere. When Rossi was in Berlin, Bothe made him aware of the problem related to the existence of the latitude effect, a geomagnetic effect connected with the charged nature of cosmic ray particles indicating a lower intensity of cosmic rays near the equator, where the horizontal component of the geomagnetic field is stronger. Since 1927 Jacob Clay had found that the cosmic-ray intensity dropped by about 15 percent as he neared the equator, as the result of experiments  made carrying  ionization detectors onboard ships that traversed an extensive latitude range, on three different voyages between Java and Holland \cite{Clay:1927fk,Clay:1928fk}. That same summer Bothe and Kolh\"orster went for an expedition to the North Sea and the northern Atlantic Ocean to study the dependence from magnetic latitude between Hamburg and Spitzbergen, although it produced no positive results \cite{Bothe:1930uq}. Negative results had been found by Millikan  and G. H. Cameron, too, \cite{Millikan:1928uq} during recent experiments carried out between Bolivia and Canada.

The possibility of using the Earth as a magnet to analyse cosmic rays was important for the followers of the corpuscular nature of cosmic rays. In this regard, Bothe called Rossi's attention to the foundational work that the Norwegian geophysicist  Carl St\o rmer and his pupils had carried on since 1904 on the very complicated mathematical problem of determining the motions of charged particles in the field of a magnetic dipole, a good approximation to the Earth's magnetic field. In their paper reporting measures effected during their expedition to the North, Bothe had remarked how `St\o rmer's theory appeared  to be so complex as to rule out the possibility of applying it to cosmic-ray problems'.  However,  `cosmic-ray physicists were interested in a simpler problem; they wanted to know whether particles of a given energy could or could not reach a given point of the Earth in a given direction'.\footnote{\cite{Rossi:1990aa} (note \ref{Rossi:1990aa}), 27--28. There is a small notebook among Rossi's papers regarding the motion of electrons in the Earth's magnetic field (the second page bears the title  `Elettrone in c.m. terrestre') which, according to the content, he most probably bought  and used in Berlin.
Rossi Papers (MC166), MIT Institute Archives and Special Collections, Cambridge MA (from now on referred as MIT Archives), Box 1, Folder 13.}   
  St\o rmer had shown the existence of a special class of trajectories: in the \textit{bounded} trajectories the particles remain forever in the vicinity of the magnetic dipole.   Rossi found out that the answer he was searching was contained in a formula derived by St\o rmer, from which it was possible to determine whether or not a cosmic ray particle was moving along this kind of trajectory.

  On July 3rd, during his stay in Berlin, Rossi sent a short note to the {\it Physical Review} where he  arrived at the following conclusion:  besides the already-expected latitude effect, a second phenomenon must exist, the  East-West effect,  which would be revealed by an asymmetry in relation to the variation of the Earth's magnetic field with the geomagnetic meridian, with more particles coming from East or West, depending on the negative/positive charge of the particles \cite{Rossi:1930ve}. 

The article appeared in the issue of 30th August, and did not contain any detail on the derivation of the final formula, providing information on the angle related to the azimuthal asymmetry.
Werner Heisenberg frequently came to Berlin from Leipzig, particularly to attendweekly colloquia. While Rossi was still in Berlin, they must have discussed  this most interesting topic,  and Rossi probably even sent Heisenberg an advance draft of the paper, because on August 19 Heisenberg wrote him asking  to summarize his calculations on the motion of an high-energy electron in the Earth's magnetic field.\footnote{Trace of the Rossi-Heisenberg dialogue on cosmic rays is contained in a series of letters (8 written by Heisenberg and copies of three letters written by Rossi) now preserved in the Archives of Padua University (unfortunately not accessible at the moment) which have been published in 1993 on a rather specialized journal: \cite{Gembillo:1993lr}. This most interesting correspondence has been completely ignored by historians. A first attempt is made here to put this exchange of ideas in the proper context.}
 As a theoretical physicist at the frontiers of quantum mechanics, now turning to the physics of atomic nuclei and of elementary particles, Heisenberg had immediately become aware of the interesting new perspectives opened by the latest experiments on the nature of cosmic rays.  On September 13th, during the last days of his German sojourn, Rossi answered reporting a series of detailed passages, arriving at last at the same formula contained in his article on the {\it Physical Review}.

As announced in his letter to Heisenberg, immediately after his arrival at Arcetri Rossi began to search for proof of the existence of the East-West effect.  If found, this proof might provide  evidence for the corpuscular nature of the cosmic radiation, as well as a precious  indication of the sign of its charge. However, within error bounds, the experiment gave a negative result \cite{Rossi:1931pd}. He was not too surprised, being aware that the asymmetry would become pronounced only at low geomagnetic latitudes and at sufficiently high elevation to afford the observation of cosmic rays of comparatively small energy. For this reason he planned an expedition to Asmara, the capital of the Italian colony in Eritrea, a town rising at an altitude of 2370 m, and at a geomagnetic latitude of 11$^{\circ}$ 30' N. 

On November 19th Rossi wrote to Heisenberg in reply to a letter sent on October 6. Rossi apologized for the great delay, and said that only in the previous days he had been able to look at Heisenberg's notes `on the nature of the Ultrastrahlung'. He compared his calculations on the borders of the `forbidden regions' with Heisenberg's ones observing that their respective results were nearly coinciding, even if they had been obtained by different methods.
Rossi announced, too, that he was going to publish on {\it Die Naturwissenschaften} a preliminary article on the work carried out in Germany, which was in fact dated October 25 \cite{Rossi:1930qf} and commented that his observations implied the possibility of  a penetrating $\gamma$-radiation; nevertheless, he remarked, this interpretation was difficult to accept ``basing on the hypothesis that the whole observed corpuscular radiation is produced in the Earth's atmosphere.'' At the beginning of 1931 a most complete article appeared on the {\it Zeitschrift f\"ur Physik} reporting a detailed description of work performed at Bothe's laboratory in Berlin.\cite{Rossi:1931ul} 

 By the summer of 1931, Rossi had performed a series of experiments (deviation with an electromagnet \cite{Rossi:1931lq} and study of intensity of cosmic rays with a counter-telescope at different inclinations to the vertical line\footnote{\cite{Rossi:1931dq}.\label{Rossi:1931dq}}) aiming with different means at studying the nature and behaviour of cosmic rays.  All these studies attracted the attention to the new vision of the problem of cosmic rays. In assuming that the Bothe-Kolh\"orster experiment had ``reasonably'' proved that the coincidence effect was caused by an ionizing material particle, L.  M. Mott-Smith  had independently attempted  to produce a magnetic deviation making use of triple coincidence counting.\footnote{\cite{Mott-Smith:1931fk}. Mott-Smith mentioned Rossi's work in an article appearing on February 1, in which he described a further attempt at magnetic deviation  for which `it was essential to have an entirely automatic device': \cite{Mott-Smith:1932uq}.}  Mott-Smith decided in a later work with G. L. Locher  to determine whether the cloud chamber and the counters were detecting the same phenomenon, in what may be considered a first attempt to create a device using both visual and electronic means.\footnote{\cite{Mott-Smith:1931kx}.   They interposed a  cyclic-expansion cloud chamber  between two counters ``so that every particle which operates the counters by passing through them must also pass through the chamber.'' The signal lamp lighted only when a coincidence in the two counters occurred. \label{Mott-SmithLocher}} 
 They concluded that ``the coincidence effects are directly due to ionizing material particles, to the definite exclusion of photon,'' but left open the question of the nature of such particles and the related problem of the general significance of this result for cosmic-ray theory.

Bothe and Kolh\"orster's pioneering article, Rossi's early researches,  and other attempts employing the new electronic techniques were instrumental in transforming the field into a branch of modern physics.  For the first time, the physical nature of cosmic rays had become accessible to experimentation, and cosmic rays themselves became the very object of research.  This research developed along two main lines: one concerned with the cosmic rays themselves  --- what are they, where do they come from, how do they reach the space surrounding the Earth --- and the other focussed on the interaction of this windfall of high-energy particles with matter.

 \section{International conferences and early collaborations on cosmic rays}
\label{intconferences}
In the meantime, from May 20 to May 24, 1931, an international conference   ---even if  European in character---  mainly focused on nuclear physics was organized by Paul Scherrer and Wolfgang Pauli  at the Eidgen\"ossische Technische Hochschule (ETH) in Z\"urich. 
On that occasion Rossi met Marie Curie, Fr\'ed\'eric Joliot, George Gamow (who delivered the opening lecture), Patrick Blackett, Lise Meitner, Bothe, and Pauli's pupil, the young Rudolf Peierls.\footnote{See Rossi's notes on the conference, Notebook  `Quaderno Zurigo', MIT Archives, Box 1, Folder 7.}    He  also had the opportunity to become acquainted with  Maurice De Broglie and in particular with Louis Leprince-Ringuet, who since 1929 had joined  de Broglie's private laboratory where research on X and $\gamma$-rays were performed.

 Both the French physicists, and James Chadwick, who had worked with Geiger  at the Technical University of Berlin during the academic year 1913--1914, were making use of the valve counter invented by the Swiss physicist Heinrich Greinacher, as a mean of detecting the passage of protons generated during the disintegration process.\footnote{In 1924 Greinacher had the idea of combining valve circuits with the point counters and  used an amplifying valve of specially high grid-insulation (electrometer valve) combined with a small ionization chamber, to register the primary radiation without the `natural disturbance' of the secondary ionization by collision: \cite{Greinacher:1924lr,Greinacher:1926lr}.} 
Greinacher's method of `rein elektronische Verst\"arkerung' (`purely electronic amplification') spread all over Europe. The method was taken up in laboratories from Wien \cite{Ortner:1929uq}   to England \cite{Ward:1929lr}, where it was particularly appreciated by Rutherford and his collaborators. In a short time, Greinacher's method was adopted in Maurice de Broglie's laboratory, in Paris, thanks to the young Louis Leprince-Ringuet \cite{Leprince-Ringuet:1931fk,Leprince-Ringuet:1931lr}.

 From his acquaintance with Leprince-Ringuet, and encouraged by the interest in his new valve counter technique, known only in two or three laboratories in Europe, Rossi began to consider a visit to Paris. 

 Rossi's belief in the importance of the problem of the nature of the secondary radiation generated by the interaction of cosmic rays with matter became particularly strong at the Conference on Nuclear Physics. This conference was organized in Rome in the Autumn of 1931 by Enrico Fermi with the support of Orso Mario Corbino. The first international meeting in the field, at which U.S. physicists such as Millikan and Compton participated, was held from October 11 to 17, under the aegis of the Italian Royal Academy, with Guglielmo Marconi as president. The presence of leading researchers in the field of nuclear physics and cosmic rays gave the event enormous importance, and put the seal on the new disciplinary identity, helping  to familiarize young Italian physicists with current problems. A well-known photograph taken during the Conference shows the group of physicists gathering on the stairs in front of the Physics Institute of Via Panisperna.\footnote{See Proceedings of the Conference: {\it Convegno di Fisica Nucleare}. Rome. Reale Accademia d'Italia, 1932. Among the others there were in particular Marie Curie, Louis Brillouin and Jean Perrin from France, Arnold Sommerfeld, Paul Ehrenfest, Werner Heisenberg, Walther Bothe, Hans Geiger, Otto Stern, Lise Meitner, Peter Debye from Germany and Holland, Francis W. Aston, Patrick Blackett C. D. Ellis R. H. Fowler and Nevill Mott from Great Britain, Niels Bohr from Copenhagen, Wolfgang Pauli from Z\"urich, Arthur Compton Samuel Goudsmit and Robert Millikan from US. Ernest Rutherford was absent, so a telegram was sent by the group.}

The fourth day of the Conference  was dedicated to the topic of cosmic rays, and Fermi invited Rossi to give an introductory speech on problems in the field.. This choice makes clear that in having his name in the list of prominent speakers like Arnold Sommerfeld, Niels Bohr, Walther Bothe, Ralph H. Fowler, Fermi fully recognized the importance of Rossi's pioneering work.  
 
 The opening lines convey the `mystery still surrounding this phenomenon',  especially related to the high energies involved, and  which worried theoreticians:\footnote{\cite[p. 51]{Rossi:1931rr}.\label{Rossi:1931rr}} 

\begin{quotation}
\small

The most recent experiments have produced evidence of such strange events that we are led to ask ourselves whether the cosmic radiation is not something fundamentally different from all other known radiations; or, at least, whether in the transition from the energies which come into play in radioactive phenomena to the energies which come into play in cosmic-ray phenomena the behaviour of particles and photons does not change much more drastically than until now it was possible to believe.
\end{quotation}

\normalsize
 
 After presenting the general aspects of the phenomenon, and the experimental results which proved that  cosmic rays were `most likely of extraterrestrial origin', Rossi tackled the hottest issue of the day: the problem of the origin and nature of the penetrating radiation. He presented a detailed and cogent discussion.  He also explained the reasons why he thought that Millikan's assumption, according to which cosmic rays are born from the synthesis of elements in the Universe, could not be correct, a view that Bothe strongly supported.
  Millikan did not like Rossi's discourse; he considered him only an arrogant young man and for a number of years thereafter chose to ignore his work altogether. According to what Bethe  later told Judith Goodstein: `Rossi was perhaps the first person who proved Millikan wrong [\dots] Rossi was quite remarkable [\dots] And I don't think Millikan would forgive him that'.\footnote{H. Bethe interviewed by J.  Goodstein, February 17, 1982, January 28, 1993. Archives California Institute of Technology, available at http://resolver.caltech.edu/CaltechOH:OH\_Bethe\_H\label{bethegoodstein}.} 
   On the other hand, Arthur Compton, who was developing a great interest in the nascent nuclear physics, was quite impressed by Rossi's talk in Rome, which he studied in detail together with other works on cosmic rays.\footnote{See Compton's research notebooks, ``cosmic ray abstracts'', December 9, 1931, quoted in \cite{Russo:2000ab} (note \ref{Russo:2000ab}), 144.}

  \begin{figure}[ht]
\centering
\includegraphics[width=\linewidth]{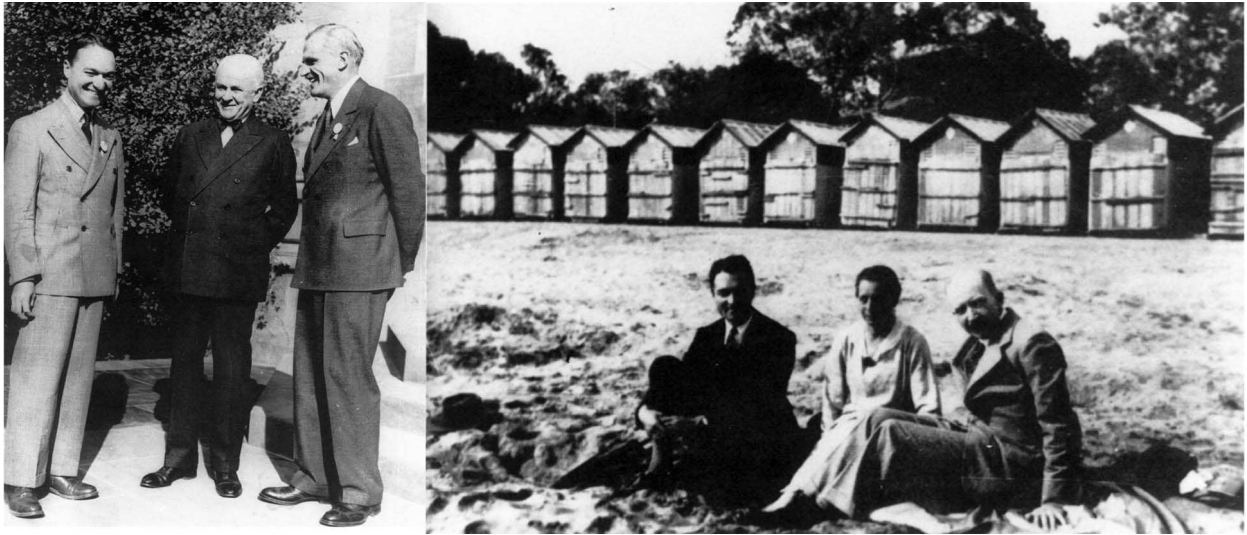}

\caption{On the left, Bruno Rossi, Robert Millikan, and Arthur Compton during the International Conference of Nuclear Physics held in Rome in October 1931. On the right, Rossi with Lise Meitner and Walther Bothe on the Lido beach of Venice, during their visit after the Conference (courtesy of Bruno Rossi family).}
\end{figure}
\par\noindent
 
 As Rossi would later recall, the conference `provided the first occasion for the proponents of the new corpuscular theory to present this theory to the scientific community, still strongly attached to the old wave theory. So this conference marked the beginning of the historical debate about the nature of cosmic rays, which was to continue for several years'.\footnote{For a reconstruction of  some aspects of this debate,  which in particular saw   Millikan  and Compton on opposite sides, see \cite{De-Maria:1985vn}  and  \cite{De-Maria:1989aa}.} The  conference was mainly an important occasion for discussing new techniques and new results in the study of cosmic rays. The new way of looking at related phenomena, which was emerging from Rossi's and a few others' research, began to make clear how problems arising within this new field were relevant both to nuclear physics and to the nascent quantum electrodynamics (QED),  a field in which both Pauli and Heisenberg had already made pioneering contributions.
 
  A report on the conference appeared in \textit{Nature} \cite{nature:1931fk}. After mentioning the opening addresses, the unknown author remarked that  Rossi's talk on  `penetrating radiation' `led to a very interesting discussion': `It appears that we are yet a long way from understanding this phenomenon, but a variety of new methods of investigation are now being applied, which at least promise to yield important information'. The writer concluded in recalling the `admirable arrangements which were made for the scientific discussions, but also for the magnificent hospitality', and added that `the success of this conference was largely due to the untiring effort of Prof. O. Corbino and the secretary to the conference, Dr. E. Fermi,   who managed to combine both the necessary firmness in directing the conference with the freedom which is so essential for fruitful discussion'.  
  
  We have also Marie Curie's impression through a letter she sent to her daughter Ir\`ene:  
  \begin{quotation}
\small

  There are a lot of people at the conference which is not lacking of interest but is however tiring. I do not know [them] all. You saw many of them in Zurich [\dots]  I try  to follow the talks as far as I can, which is not always easy, owing to the many technicalities and especially because of the lack of clarity in the elocution of some. I think I shall have  to say some words on the occasion of the discussion on the talks about radioactivity phenomena [\dots] I have up to now very few things to tell you, except that Bohr insists on the impossibility at present to apply quantum mechanics within the nucleus. 
  \footnote{Marie to Ir\`ene Curie, Rome, October 13, 1931. Personal Archives Pierre and Marie Curie, Biblioth\`eque Nationale de France, Manuscripts D\'epartement, Paris, NAF 28128. Actually, shortly after the Roman conference  the French invited Fermi  to summarize the latest developments in nuclear physics at the 5th International Congress on Electricity held in Paris on July 1932.\label{curie}} 
  \end{quotation}
    \normalsize

  During the conference Pauli discussed with Fermi and Bohr his daring idea of a new neutral particle of spin 1/2 whose existnce he had postulated   the previous year. Pauli postulated this particle in order to remove fundamental theoretical difficulties in  both nuclear structure and the explanation of the continuous $\beta$-ray spectrum, which seemed not to obey the law of conservation of energy.\footnote{In his talk on the hyperfine structure, Samuel Goudsmit mentioned Pauli having expressed at a meeting at Pasadena in June 1931 `the idea that there might exist a third type of elementary particles besides protons and electrons, namely `neutrons' [\dots] The mass of these neutrons has to be very much smaller than that of the proton, otherwise one would have detected the change in atomic weight after $\beta$-emission'. Goudsmit also recalled Pauli's belief that his neutral particle might `throw some light on the nature of cosmic rays'. \cite[p. 41]{Goudsmit:1932fk}.}
   Bohr did not like Pauli's solution to the apparent violation of conservation laws, and preferred to think that such laws would break down within nuclear distances. Fermi on the contrary was very much in favor of the idea. His reflections on this important topic and on the related issues discussed during the conference led him in the autumn-winter of 1933 to formulate his theory of $\beta$-decay.\footnote{ \cite{Fermi:1934qy}.\label{Fermi:1934qy}}  Fermi introduced the so-called weak interaction, one of the four  fundamental forces, responsible for some nuclear phenomena including $\beta$-decay.

 In the concluding remarks of his contribution to the proceedings of the Roman conference, Rossi had again stressed that, `Until we have proof to the contrary, we must thus consider it as a corpuscular radiation [\ldots] Moreover, several different experimental facts let us think that, in going through matter, it is generating a secondary corpuscular softer radiation. The more urgent research aim  is now the problem of clarifying the nature of the corpuscular penetrating radiation, of measuring its energy, of establishing its origin and {\it of investigating which role is playing its secondary radiation in the observed phenomena} [emphasis added]'.\footnote{\cite{Rossi:1931rr} (note \ref{Rossi:1931rr}), 66.} 
 
The possibility that the secondary radiation could, at least in part,originate in the atomic nucleus, contributed to arouse  interest in this phenomenon among nuclear physicists. Rossi's interest  on nuclear research  is also confirmed by the grant he received at the beginning of 1932 from the Volta Foundation attached to the Italian Academy. According to his written application, he intended to use the grant `to go abroad and specialize in researches regarding radioactivity and penetrating radiation'.\footnote{MIT Archives, Box 28, Folder `Bruno Rossi (Biographical information)'.}

During the winter of 1931--1932 Rossi visited Paris, where he became close friends with Leprince-Ringuet and Pierre Auger. It was an important occasion for the exchange of knowledge about Geiger-M\"uller counters,  unknown in Maurice De Broglie's  laboratory.  Simultaneously, Rossi learned how to build and use the valve counter set up by  Heinrich Greinacher, a device with which which Leprince-Ringuet had become quite familiar. 

Around this time, the study of elements bombarded by $\alpha$ particles would soon turn out novel results. Valve amplifiers, which started to become widespread around 1930, were instrumental in increasing the speed of counting $\alpha$- and other heavy particles in scattering experiments.  In particular, valve amplifiers worked much more quickly than scintillation counting, with which most of the historic discoveries had been made at the Cavendish. 
And in fact, in a few months Chadwick's  work  showed the possibility of detecting neutrons, which do not themselves produce ionization. The valve counter became an indispensable tool for nuclear research.\footnote{Actually Joliot was probably also interested, because he asked Leprince-Ringuet to move to the Curie Laboratory.  However, the latter did not want to renounce to the great `liberty' he had in De Broglie's private laboratory.}

Rossi's students Occhialini and  Bocciarelli had just employed Geiger-M\"uller counters to detect radioactive processes. It is thus clear how interesting  the new detecting device might be, which at the time could be found only in very few laboratories. It was a tool which  might shed light on the interaction of particles and radiation with matter, as well as on the possible connection of nuclear physics with cosmic ray phenomena. On the other side, Rossi's arrival in Paris was instrumental in piquing Leprince-Ringuet's interest in cosmic rays: `Rossi's stay was of the most fruitful: by a very natural exchange, the laboratory of the duke de Broglie was introduced to the electron counter and to the coincidences which allow to distinguish the very penetrating radiation in the middle of other, softer ones, and to define their direction. It was the door opened towards the wonderful world of cosmic  rays'. \cite[p. 300]{Leprince-Ringuet:1991nx} Leprince-Ringuet was `seduced' by cosmic-ray research and soon began to build counters all alone (`J'ai fabriqu\'e 100 \`a 200 compteurs qui marchaient plus ou moins bien') \cite[p. 64]{Puyo:1976uq}. He was the first to build a coincidence circuit in France, which he immediately presented  at the Paris Exhibition of instruments and experimental techniques organized by the French Society of Physics.\footnote{\cite{Bustamante:2010zr}. On the collaboration Leprince-Ringuet-Auger see also \cite{Bustamante:1994aa}.}   

The seed Rossi had sown flourished immediately. Leprince-Ringuet proposed to Pierre Auger to combine their efforts, and they decided to work together on a project about the latitude variation of cosmic rays.
Auger had collaborated since 1929 with the Soviet physicist Dimitri Skobeltzyn, working with a cloud chamber at Marie Curie's laboratory.\footnote{According to the correspondence, Skobeltzyn had asked Marie Curie as early as 1927 for permission to spend some time  in Paris  in her Laboratory. He arrived  on 7th April 1929 with a Rockefeller grant and left in 1931. `Liste du personnel du Laboratoire Curie 1904--1934', Curie Archive, Paris. The strong international tradition of Marie Curie's laboratory is outlined in a study study presenting a global overview of all the researchers of the Curie laboratory, then focusing in particular upon its female researchers: \cite{Pigeard:2012fk}.} 
In the Soviet Union, Skobeltzyn had studied the paths of $\beta$-rays in a magnetic field of 1,700 gauss, and had taken photographs of `very fast corpuscular rays', which he considered to be `Compton  electrons produced by penetrating radiation'.\footnote{\cite{Skobeltzyn:1929uq}. From about 600 pictures obtained with a Wilson chamber in the uniform magnetic field, Skobeltzyn found 32 pictures with tracks originating outside of the Wilson chamber and not affected noticeably by the magnetic field. He assigned these tracks energies greater than 15,000 eV, and speculated whether  `{\it One should assign these $\beta$ rays to the secondary electrons created by Hess ultra-$\gamma$-rays} [emphasis added]'.\label{Skobeltzyn:1929uq}}  
On the 1st July 1929, during a meeting of the Academy of Sciences, Auger and Skobeltzin presented a paper on the nature of cosmic rays \cite{Auger:1929fr}.  They discussed Bothe and Kolh\"orster's preliminary note on {\it Die Naturwissenschaften}, where the corpuscular hypothesis had been put forward for the first time. They did not agree with this hypothesis, and again confirmed that the  tracks observed with their cloud chamber must be due to Compton `ultra-$\beta$' electrons  produced by an ultra-$\gamma$ radiation. 
 
But now, after  Rossi's  visit to Paris, Auger and Leprince-Ringuet found the coincidence circuit,  merging as it did electronic methods with counters, to be an appealing device for the study of cosmic rays. As they later clearly explained in their main article reporting the obtained results, they used the coincidence method during their  round-trip voyage from Hamburg to Buenos Aires,  in fall 1933, which provided quantitative proofs for  the existence of a  variation of intensity with latitude \cite{Auger:1934kx,Auger:1933ys}. Together with Compton's campaign of the same period, it confirmed the corpuscular nature of cosmic rays at sea level.\footnote{See also Pierre Auger's diary written during the voyage. Auger Papers, Acad\'emie des Sciences Archives, Paris. As an important acknowledgement of  their work on the problem of the latitude effect, they were asked to present their results during the joint International Conference organized by IUPAP and the Physical Society in London in the autumn of 1934.} 
At the mountain international laboratory of Jungfraujoch, they studied the soft component found by Rossi, and the hard one, `which was able to traverse 10--15 cm of lead' \cite{Auger:1936uq}, and widely cited him in all their articles in connection with the experimental techniques used during their research. 
From then on Leprince-Ringuet dedicated himself to cosmic ray research \cite{Leprince-Ringuet:1935zr,Leprince-Ringuet:1936kx}. In fact, his doctoral dissertation was entitled  `Recherches sur l'interaction avec la mati\`ere des particules de tr\`es grande \'energie: \'electrons d'origines diverses et particules du rayonnement cosmique' [Research on the interaction of very high energy particles with matter: electrons of different origin and cosmic-ray particles].\cite{Leprince-Ringuet:1935vn} The experimental work was carried out using a cloud chamber of the `Blackett-Occhialini' type.  This was a new ingenious system of controlling the time of expansion of a cloud chamber  by a Rossi electronic circuit, which had been developed by Blackett and Occhialini in 1932 (see Section  \ref{blackettocchialini}). In a short time Leprince-Ringuet was asked to  became professor of physics at the \'Ecole Polytechnique, where he established a laboratory dedicated to the study of nuclear and cosmic-ray physics attached to his chair.  This chair was to become one of the most important in the field. 

Pierre Auger, on the other hand, continued hisresearch along the line inaugurated by Rossi.  He applied the coincidence method to study  the secondary effects of cosmic rays, and the absorption of $\gamma$-rays.\footnote{ \cite{Auger:1935uq,Auger:1935kx}. Owing to the great interest of Auger and his group's later achievements, it is worthwhile going in some details regarding his scientific path up to the discovery of Extensive Air Showers, which actually would deserve a much more extended discussion.}
 In 1935 Auger also constructed a transportable Blackett-Occhialini cloud chamber to be used at high altitudes \cite{Auger:1935fk}. He was especially interested in the existence of  the two different components Rossi discovered during his early research.  This study led him and his collaborators  to analyse the secondary showers produced in metal shields by  cosmic rays at high altitude  \cite{Auger:1936qf,Auger:1936bh}, the properties of the penetrating component \cite{Auger:1936dq}, their angular distribution \cite{Auger:1937fp,Auger:1937cr}, and the mechanism of production of electrons and photons showers  according to Bhabha and Heitler's theory.\footnote{\cite{Bhabha:1937yq}\label{Bhabha:1937yq}.} During the course of these investigations, Auger and his collaborators were led to `explore a very large surface [\dots] progressively arranging counters more and more wide apart in order to reveal more and more dispersed atmospheric showers'. This procedure led then to suggest that `A considerable part of the group of the electronic corpuscles, at sea level, is constituted by the branches of showers created in the atmosphere by the high-energy rays which cross it'\cite{Auger:1938fr,Auger:1938rt}. After about 5 years they had re-discovered the extensive cosmic ray showers already observed by Rossi and his assistant Sergio De Benedetti during their Eritrea campaign in 1933.\footnote{Rossi's commitment to the problem of geomagnetic effects will be outlined in one of the following sections. For a discussion about  E. R. Regener, Bothe and Kolh\"orster reaching  an independent conclusion about the existence of  the phenomenon of air showers see \cite[Section 2]{Kampert:2012fk}.}

 The large impact of Rossi's  short stay in Paris  is thus well established.  He did not use the whole grant, and returned to Italy in order to participate in a competition for a chair of experimental physics sponsored by the University of Ferrara.  The competition was officially announced on April 13, 1932.\footnote{This proved a lucky circumstance: when he was  under pressure in September 1938 after being dismissed from his position at the University of Padua,  he was able to have some money and the visa on his passport in a short time, after requesting to complete his stay abroad for research reasons.}  
 When Fermi asked him to give a talk on cosmic-rays during the Roman Conference of 1931, he  was well aware that Rossi was working at the frontier of the `new physics', and was conscious of the importance of his results.\footnote{Emilio Segr\'e participated to the same competition, but 
Fermi, who was the only person of the commission to understand the value of research on modern physics, struggled for two days to convince his colleagues that scientific results were more important than age. At last he succeeded in imposing only the name of Bruno Rossi. E. Fermi to B. Rossi, October 30, 1932. Archives of the University of Padua.} 
 Rossi presented a detailed and well-explained description of his research and 27 published articles.\footnote{Archivio Centrale dello Stato, Ministero della Pubblica Istruzione, Direzione Generale, Busta 60, Fascicolo 359.} The impressive amount of work carried on during only two years and a half makes clear why Fermi was so determined in his choice.  From that time onward they had a very close relationship, which ended only with Fermi's premature death in 1954.\footnote{According to Martin Annis, a former student of Bruno Rossi at MIT, it appears that Fermi's  attitude towards Rossi could be especially perceived when both were among other people, and Fermi changed his expression in turning towards his old friend. Martin Annis, interview by the author, Cambridge M.A., September 30, 2006. Their special friendship is particularly testified by many episodes told to the author by the late Mrs. Nora Rossi Lombroso, to whom I am still very grateful.}

\section{The two components of cosmic rays and Heisenberg's theoretical speculations}

On 4th November 1931, immediately after the Roman conference, Millikan had been invited by Heisenberg in Leipzig to hold a talk  on the  `current state of the research in the field of the penetrating cosmic radiation'.\footnote{\cite[p. 881]{Rechenberg:2010kx}.\label{Rechenberg:2010kx}} It was an opportunity for Heisenberg to reflect on the problem of  high energy processes. His work carried out with Wolfgang Pauli in the late 1920s on the foundations of quantum electrodynamics had a natural extension in the analysis of elementary processes in high energy cosmic-ray interactions. 
Like other theorists, he wanted to check the validity of  quantum electrodynamics as formulated up to that time, in order to understand how  special difficulties, which appeared  at high speeds and impulses, could be removed by suitable changes of the relativistic quantum theory. 

The divergence difficulties in the newly developing  quantum theory of fields  appeared to be connected to the nature of high energy cosmic ray processes, so theoreticians studying these topics in cosmic-ray physics considered Rossi's experiments especially interesting.\footnote{For a discussion of Heisenberg's works on cosmic-ray phenomena see \cite[p. 241--246]{Heisenberg:1993qy} and especially \cite{Rechenberg:2010kx}(note \ref{Rechenberg:2010kx}) 882--892.}  On November 18, immediately after Millikan's conference in Leipzig, and after about one whole year since the exchange of letters of autumn 1930, Heisenberg wrote again to Bruno Rossi:  `I have lately made some calculations on cosmic radiation and I collected the formulas in a small manuscript. I would be very grateful if you could be so kind as to read it and write me how much of its content is trivial, well known, and  wrong, and then send it back [\dots]  You probably will laugh at the theoretician's list of wishes [\dots] I still remember the Roman days with joy, the congress was really one of the best and instructive I ever participated until now'.  It is thus clear that Heisenberg considered Rossi's opinion on his work on cosmic rays literature important enough to send him his  notes, which on December 8 Heisenberg  asked to get back because he needed them for a discussion to be held during the following week-end.\footnote{Heisenberg outlined the first results of his survey in two manuscripts; it appears that the one with formulas mentioned in the letter to Rossi might be related to a series of handwritten notes listing 39 not numbered formulas entitled `Formelsammlung f\"ur H\"ohenstrahlung' and clearly prepared before  the Zurich talk. Archiv der Max Planck Gesellschaft, Va. Abt., Rep. 57, Werner Heisenberg.}  
 In a short survey on experiments  he discussed  `Counters or ionization chamber', and the Millikan-Cameron curve.
 
The discussion following his talk at the October Conference in Rome had reinforced Rossi's opinion that  further measurements of the penetrating power of cosmic-ray particles would produce important information about the nature, properties, and origin of these particles.  Up to that moment the absorption of the `corpuscular radiation' had been measured only by Bothe and Kolh\"orster with 4.1 cm of gold between two counters, and by himself using about 10 cm of lead during his recent experiments.  However, by indirect arguments, he had become convinced that many particles must have much greater ranges. His aim was now to extend knowledge about the absorption curve of corpuscular radiation up to thicknesses comparable with its penetrating power.  From his notebooks we learn that in those very days Rossi was  completing a new experiment which consisted in counting coincidences between counters aligned vertically one above the other, and separated by suitable absorbers of variable thickness. However,  `with two counters placed sufficiently far apart to allow these absorbers to be placed between them, the number of `true' coincidences would have been smaller than the number of `chance' coincidences, i.e., of coincidences due to the almost simultaneous passage through the counters of two unrelated particles'.  He overcame this difficulty `by inserting a third counter between the other two and recording threefold instead of twofold coincidences, thereby cutting down the frequency of chance coincidences to a negligible value'.\footnote{\cite{Rossi:1990aa} (note \ref{Rossi:1990aa}), 19.} 

\begin{figure}[h]
\centering
\includegraphics[width=0.5\linewidth]{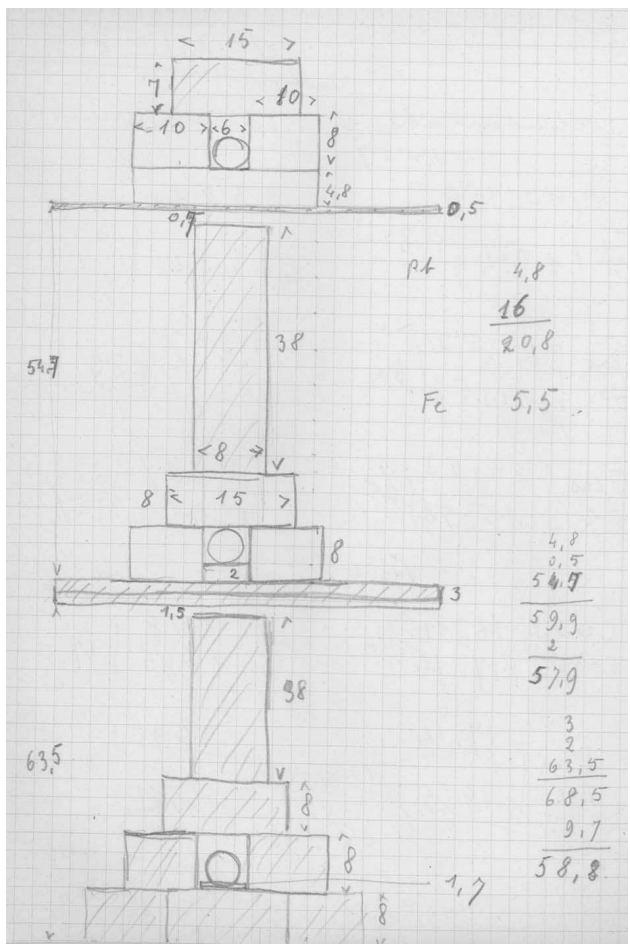}
\caption{Original drawing showing Rossi's experimental arrangement which proved that the cosmic rays  contained particles capable of traversing one meter of lead. Coincidences were recorded by the three counters $C_{1}$, $C_{2}$ and $C_{3}$ separated by a thick lead layer. Threefold coincidences instead of twofold coincidences were used, in order to reduce the chance coincidence rate (Courtesy of MIT Archives).}
\label{Figure3}
\end{figure}

On December 1, a drawing of the experimental arrangement  appeared on his notebook (Fig. \ref{Figure3}).\footnote{B. Rossi, Notebook,  MIT Archives, Box 1, Folder 8.}
Rossi's data regarding the absorption trend measured between 0 and 10 cm, between 10 and 25 cm, and between 25 and 101 cm of lead showed that the fairly rapid decrease in the rate of coincidences in the first 10 cm of lead became so slow that about 50\% of the particles emerging from 10 cm of lead had still the capability  of traversing one meter of lead. 

 On December 16 he sent a short letter to \textit{Die Naturwissenschaften}.\footnote{\cite{Rossi:1932cr}.\label{Rossi:1932cr}}
 His conclusions again supported, with much stronger evidence, the opinion already expressed after the research carried out in Berlin. The first announcement meant a definite blow to the hypothesis of $\gamma$-ray nature for cosmic radiation: `A fairly large proportion of the corpuscular radiation found at sea level has still the capability of traversing more than 1 m lead, that is about more than the thickness of the whole atmosphere'.  He stressed again that such corpuscular radiation had a too large absorption coefficient (if compared to the absorption coefficient measured by Millikan and Cameron) to be a secondary radiation generated by the primary ultra-$\gamma$-rays.

On December 12, Heisenberg wrote Rossi again.  He thanked him for sending back his manuscripts, and mentioned  the final draft of the article on the absorption of cosmic rays in 1 meter lead: `I am really burning with interest in the manuscript you announced, please  send it as soon as  possible. Since I am a little bit unhappy about the fact that the experiments which I know always seem to agree with  theory and  that there are, however, evidently other ones from which one can draw definitive conclusions against the $\gamma$-ray hypothesis'.

 Later Rossi explained  how hard it was for the majority of the scientific community to accept his results on the existence of particles capable of such penetrating power, given  that the most penetrating particles known at that time ---$\beta$-rays from radioactive substances--- could go only through a fraction of a millimeter of lead: `Doubts were expressed as to the legitimacy of the coincidence method, and I had to perform further experiments to dispel these doubts'.\footnote{\cite[p. 40]{Rossi:1981aa}.\label{Rossi:1981aa} Millikan in particular had explicitly attacked the `so-called Geiger counter coincidence measurements': `I have been pointing out for two years in Pasadena seminars, in the Rome Congress on nuclear physics in October, 1931, in New Orleans last Christmas at the A.A.A.S. meeting, and in the report for the Paris Electrical Congress, that these counter experiments never in my judgment actually measure the absorption coefficients of anything. I shall presently show that no appreciable number of these observed ionizing particles ever go through more than 30 cm or at most 60 cm of lead  [\dots]  \textit{These figures cannot both be correct without carrying with them the conclusion that the primary rays at sea level and below are not charged particles} [emphasis added]' \cite[p. 663]{Millikan:1933uq}. Millikan did not even mention Rossi's experiments, which were of course the main target of his criticism.}

On December 20 Heisenberg thanked Rossi for the  final draft: `from which I learned very much. Now I really well understand which difficulties represent a major obstacle to the $\gamma$-ray hypothesis and I still want to reflect about all the related connections'.  He also mentioned the so-called `transition effect,' which had been discovered by G. Hoffmann.  This effect manifested itself as a discontinuous change in the slope of the absorption curve measured with the ionization chamber, when the cosmic radiation passed  from one absorber to another of a different atomic number.\footnote{In 1927 Hoffmann found in one of his ionization chambers a sudden jump of the line marking the position of the electrometer, as if a large number of ions were produced at once in the gas of the chamber: \cite{Hoffmann:1927fj}.}
In sending his Christmas greetings, Heisenberg concluded: `Actually I do not know whether the transition effects may prove a serious problem, as you say. I have got the impression that these problems may be more related to the geometrical arrangement of the apparatus. However I want to think further about. So that for the moment  thank you for your manuscript, I will write more exactly about it in some time.' 

Around this time, on 15th December 1931 Heisenberg wrote to Niels Bohr mentioning his new commitment to `H\"ohenstrahlung,' and in particular commenting the `transition effect.' He   had reached the following conclusion: `The transition effects are  all deriving from slow electrons which are released by fast electrons. If one calculates classically the distribution of these slow secondary electrons, everything comes out in order, the experiments are appearing quite quantitatively correct'.\footnote{\cite{Rechenberg:2010kx} (note \ref{Rechenberg:2010kx}) 883.}

After Christmas holidays, on January 9, 1932, Heisenberg wrote again to Rossi: 

\begin{quotation}
\small
I reflected more in detail on your work and I also widely discussed it with Bohr. Now --- supposing that experiments are all right--- we are both of your same opinion: either the radiation is a `corpuscular' one (i.e. electrons), and thus St\o  rmer's theory is failing for some reason, or the primary radiation is absorbed much more rapidly than must be expected basing on some reasonable theory. The Klein-Nishina formula should be wrong about a factor of 10. At the moment it is difficult to say which of the two hypotheses is the worst. 

I was not able to get a clear idea about the transition effects. Here it would in any case be urgently necessary to do further experiments. May I keep the manuscript for some more days?
\end{quotation}
\normalsize

In the following days\footnote{The draft of Rossi's letter has no date, but it is clearly an answer to Heisenberg's previous letter.} 
Rossi answered saying that he was quite happy  that both Bohr and Heisenberg agreed on his conclusions about coincidence experiments. As far as the transition effects were concerned, he too was  convinced that the experimental data were not sufficient to draw any clear conclusion on their origin. He mentioned further experiments he panned to carry out, about which he would tell him in the near future.

 As a matter of fact, the real hot issue was the production of  secondary radiation generated in the metal shield above the counters. Rossi had already observed their first hints during his experiments in Berlin. At the Rome Conference both Rossi and Bothe, who were at the moment the main proponents of the corpuscular hypothesis, had emphasized the growing evidence for the production of a softer secondary radiation by penetrating cosmic-ray particles in matter,\footnote{\cite{Bothe:1932fk} and \cite{Rossi:1931rr} (note \ref{Rossi:1931rr}).}  also keeping in mind the first cloud chamber photographs published by Skobeltzyn  and showing multiple tracks of apparently the same age.\footnote{Skobeltzyn had found four photographs among 27 showing multiple tracks: three with two and one with three `branches'. Since the probability of two or three independent particles traversing the chamber during he expansion was very small, Skobeltzyn interpreted the observed groups as primary cosmic rays particles accompanied by one or two secondary electrons knocked out of the atoms by elastic Compton collisions. \cite{Skobeltzyn:1929uq} (note \ref{Skobeltzyn:1929uq}).}  The following year, Skobeltzyn published a new article \cite{Skobeltzyn:1930fj}.

  On the other hand, no double track had then been observed by Mott-Smith and Locher with their cloud-chamber whose lightning mechanism was triggered by counters.\footnote{\cite{Mott-Smith:1931kx} (note \ref{Mott-SmithLocher}).} Rossi became convinced  that {\it the coincidence between  three non aligned counters} could provide a much more effective  method to obtain evidence of  `branches' generated along the trajectories of corpuscular radiation.  This would provide definite proof of the existence of a softer secondary radiation, evidence for which up to that moment was still of a rather indirect nature. From the middle of December 1931 until the end of September 1932, Rossi performed a series of remarkable experiments to study both the soft secondary radiation generated by the `penetrating radiation' and the absorption curve of the latter.

In the first experiment   three GM counters  were placed out of line, in such a way that a single particle traveling on a straight line could not possibly discharge all of them. A threefold coincidence, therefore, could only be produced by a group of two or more particles arriving simultaneously and going through at least two of the counters. In order to avoid any effects caused by external radioactivity, the counters were placed  inside a thick lead shield. Actually, 35.5$\pm$1.3 coincidences per hour were recorded when the counters were completely surrounded by a lead shield. By removing the upper part of the shield the counting rate was reduced to 10.0$\pm$ 0.5 per hour. The large increase of coincidences due to its presence thus afforded direct evidence for the production of one or more secondary particles by a cosmic ray traversing the matter.\footnote{The coincidences observed without the top shield were still in excess of the expected number of chance coincidences, about 5 per hour, but their number also decreased when the lead {\it under}  the counters was removed. {\it It was thus clear that associated groups of particles emerging simultaneously from the shield itself gave rise to new radiation generated  by the interaction of the cosmic rays with the shield itself}.}

Rossi felt sure that the existence of such secondary radiation was clearly established, and a method for its investigation was made available.  However, no known process at the time could explain the abundant production of secondary particles revealed by his experiments.\footnote{A single recoil electron is produced during the Compton process, at the time the only known interaction of high-energy photons. Electrons knocked out of atoms during the ionization process generally have a very small energy, and the double Compton  collision involving the same photon, is a very rare phenomenon, as Bothe and Kolh\"orster  had already remarked.}
 Such was the `novelty' of the results, so contrary to common sense, that the experiment appeared too incredible to the editors of the scientific journal to which Rossi submitted his short note, most probably \textit{Die Naturwissenschaften}   where he had already published preliminary results of his experiments. They refused to publish it. As Rossi would later recount, his note, received on February 10, 1932 by {\it Physikalische Zeitschrift},\footnote{\cite{Rossi:1932fk}. \label{Rossi:1932fk}} was accepted only after Heisenberg had vouched for its credibility. The correspondence here discussed at last provides a concrete explanation and a much wider background for what Rossi outlined very concisely in later accounts of his life.\footnote{\cite{Rossi:1981aa} (note \ref{Rossi:1981aa}), 41.}

Notwithstanding the difficulty he encountered in getting his results accepted, Rossi's experiment showing the unexpected abundance of particles generated in metal shields marked the beginning of extensive experimental work on the secondary effects of cosmic rays. The `Rossi coincidence method', in parallel with the suitable use of  shields arranged in  most different configurations,  soon became part of the experimental machinery used by physicists following the same research philosophy.\footnote{In April his article on the \textit{Physikalische Zeitschrift}(note  \ref{Rossi:1932fk}) was quoted by Thomas H. Johnson and Jabez C. Street, who claimed to have observed a similar phenomenon. They used only two counters, but they placed  two blocks of lead above, arranging them on either side of the counters in positions such that straight line paths through both counters were impossible for secondary rays originating in the lead: \cite{Johnson:1932uq}.}

 On  February 10 Heisenberg thanked Rossi for his `new work'  which he had `studied with great interest'.  Heisenberg also mentioned how curious he was about what he would obtain regarding the `angular distribution of secondary particles'. He also announced that he had `put together a rather long article on all what theory has to say about the problem of cosmic radiation. It is not so much but it is worthwhile collecting all together. I will have the draft sent to you. Something has changed regarding the ``grouping of formulas''.'

 The issues discussed in this correspondence with Rossi and presented by Heisenberg during his Zurich talk of January 25, became the basis of his  first  long article on cosmic rays received by {\it Annalen der Physik} on 13th February 1932, where Heisenberg discussed in detail `the most important experiments on cosmic radiation from the point of view of the existing theories,  to determine at which points the experiments roughly agree with  theoretical expectation, and where such large deviations occur that one has to be prepared for important surprises' \cite[p. 430]{Heisenberg:1932ys}.
 Based on the current theory, he discussed the passage  of very fast electrons through matter, the absorption and scattering of hard $\gamma$-rays, and several typical cosmic-ray phenomena, such as those observed in the absorption curves. 
  Heisenberg began with Skobeltzyn's experiments, and after examining the transition effects he dedicated a long discussion to coincidence and absorption experiments, notably to  Rossi's recent works.\footnote{Heisenberg cited \cite{Rossi:1931dq} (note \ref{Rossi:1931dq}), \cite{Rossi:1932cr} (note \ref{Rossi:1932cr}), and \cite{Rossi:1931rr} (note \ref{Rossi:1931rr}).}
   He emphasized how Rossi observed \textit{threefold} coincidences, thus confirming the novelty of his experimental approach.
From his experiments it was clear, remarked Heisenberg, that at sea level $\gamma$-radiation did not ultimately play any role.  A dark cloud still hovered over the corpuscular hypothesis, according to which the intensity of cosmic rays should depend on the geomagnetic latitude.  This hypothesis hadyet to be definitely proved.  Heisenberg concluded that, comparing the data with the Klein-Nishina formula, discrepancies existed between theory and experiments. From the results obtained it followed, if no new physical hypothesis was introduced to explain the effects, either that the Klein-Nishina formula for the absorption of highly energy light quanta gave a value about 25 times too small, or that the formula obtained on the basis of the classical and quantum theories for the  slowing down of fast electrons gave too small a value. Heisenberg was also convinced that `Dirac's radiation theory or the equivalent quantum electrodynamics'  were  `failing in principle', so that `a satisfactory estimate of the frequency of the secondary radiation processes basing on the past quantum theory' seemed barely possible.  On the whole, Heisenberg's article reflected the general confusion at a time when theory was not able to cope with experiment.  However,  his detailed discussion probably helped to focus general attention on the problem of cosmic rays and on Rossi's work, which he had widely quoted in his article.

 In those very days Guido Beck was writing from Leipzig to Fr\'ed\'eric Joliot: `I think you'll be interested to know that according to recent research by Mr. Rossi in Florence, in treating the theoretical problem Mr. Heisenberg is rather sure that  the primary `cosmic rays' are very fast electrons, thus meaning that your $\gamma$-rays are  the hardest known electromagnetic radiation up to now.' This sentence shows that Heisenberg did not share Millikan's views about cosmic rays being a ultra-$\gamma$ radiation.\footnote{Guido Beck to Fr\'ed\'eric Joliot, Leipzig 23th February 1932. Joliot-Curie Archives, Paris, box F144.}  However, it would not be until the following year that the corpuscular view would be definitely confirmed by several experiments investigating the geomagnetic influence on cosmic radiation.
 
At the moment, the {\it annus mirabilis} of nuclear and particle physics had been inaugurated when Harold Urey announced the discovery of deuterium. On February 27, Chadwick's article on the `possible existence of a neutron' appeared on  {\it Nature} \cite{Chadwick:1932fj}. 
The subsequent discovery of the positron, and the discussions on its role in Dirac's theory of the electron as well as at nuclear and cosmic-ray level, would deeply change  the general view. Before the end of 1933, Fermi's theory of $\beta$-decay would provide a new perspective to theoreticians, and Heisenberg's attention would later be attracted by new phenomena like nuclear disintegration caused by cosmic rays,  and by the brand new problem of mesotron decay. But now, the new perspectives which had  been opened by the neutron intrigued Heisenberg, and in the following months  led  to his neutron-proton nuclear model published in the well-known series of three articles.

On March 21, Heisenberg  wrote a new letter ---he was in Munich at the moment--- in which he put a series of questions to Rossi regarding coincidences in Rossi's 1 meter lead experiment. He also made a  sketch of the experimental set up showing the three counters interleaved with lead shields.\footnote{`1.  With which exactness can one measure coincidences? i.e. what is the shortest interval which is still separable?; 2. How big is the number of coincidences of A and C alone, compared with the triple coincidences from A, B and C?; 3. Is the difference  between both numbers to be explained by accidental coincidences from A and C and how many  accidental coinc.[idences] are there? Or do the secondary electrons play here an essential role?'.} 
His last message to Rossi contained in the mentioned group of letters is dated 9th May 1932. He thanked him for `the interesting letter,' which from the discussed issues certainly regarded Rossi's recent important work analysing the secondary radiation presented on May  1 at the Accademia dei Lincei,\footnote{\cite{Rossi:1932uq}.\label{Rossi:1932uq}} on which Rossi had been working in parallel with the 1 meter lead experiment. 

After the discovery of the abundant production of secondary radiations in metal shields, Rossi had investigated more deeply into the origin of this unexpected phenomenon. From his notebooks we learn that he used a great variety of configurations of the `triangular' counter arrangement, changing the position and the thickness of the layers of matter placed above them, and also inserting absorbing shields in different positions. In making this `logical analysis' he also compared the behaviour of different materials such as lead and iron. 

 His most significant results were summarized in a curve, later known as the `Rossi transition curve', representing the variation in the number of coincidences recorded by three counters in a triangular array  as a function of the thickness (in mass per unit area) of layers of lead and iron placed above them and emitting the secondary particles. The initial rise of the curve was  readily accounted for by the increasing number of secondary particles generated in increasing thicknesses of lead. Beforehand one would expect the transition curve to reach a maximum for a thickness about equal to the average range of the secondary particles and then to decrease slowly with a slope corresponding to the rate of absorption of the `primary' radiation. Actually, the curve drops much more rapidly than the absorption curve of cosmic rays at sea level. According to Rossi, this  was  due to the fact that what they called `primary' radiation is in part composed of  softer corpuscles of {\it secondary origin} generated by the first ones in the atmosphere or in the ceiling of the room and totally absorbed in a few centimeters of lead, thus contributing to a steeper slope than expected.\footnote{Actually, since spring-summer 1931 Rossi had observed that, by inclining more and more to the vertical line his `counter telescope', the slant rays were `softer and not harder than the vertical ones,'  contrary to what  one would expect. The filtering effect of the increasing thickness of the atmosphere should in fact result in a hardening of the revealed radiation. He had already stressed that: `This result may be accounted for by assuming that the corpuscular rays generate in the atmosphere a softer secondary corpuscular radiation [\dots]', \cite{Rossi:1931dq} (note \ref{Rossi:1931dq}).} He was thus able to conclude that :  `{\it [\dots]  the soft components, and not the hard components of the corpuscular incoming radiation, more actively contribute to the production of secondary radiation.} To put it simply (considering that,  as for the soft component, it is mainly of secondary origin) {\it the probability that a  secondary corpuscle will generate a tertiary one} should be much greater than the probability that a primary corpuscle is generating a secondary one' [emphasis added]. \cite[p. 256]{Rossi:1932kx} 
 
 Rossi remarked that  what he had previously assumed to be \textit{secondary} radiation generated in metal shields by `primary' cosmic rays at sea level was in fact a `\textit{tertiary} radiation  producing most of the observed threefold coincidences observed with 1 cm lead,'  and that the soft radiation he had observed `along strongly slanting directions' which he considered  `practically all of secondary origin,' was at the root of  the phenomena. He  stressed in fact that most Wilson chamber `bifurcate tracks'  were due to  radiation generated in matter,\footnote{\cite{Rossi:1932uq} (note \ref{Rossi:1932uq}), 739.} because only rays of this type could produce tracks whose curvature in magnetic fields and ionization could be estimated.  This clarifyied for the first time the nature of the tracks observed in Skobeltzyn's cloud-chamber photographs. It is to be remarked that  other absorption measurements of cosmic ray particles  had shown the existence of a hard and a soft component of the cosmic radiation. However, Rossi's experiments were the first to prove that the hard rays on one side, and and the soft rays ---the shower-producing component--- on the other, are {\it fundamentally different in character}   and do not differ merely on account of their energy.\footnote{For a more detailed discussion on Rossi's remarkable studies carried out during the period 1930--1932 see \cite{Bonolis:2011ak} (note \ref{Bonolis:2011ak}).}
 
\begin{figure}[htb]
\centering
\includegraphics[width=0.5\linewidth]{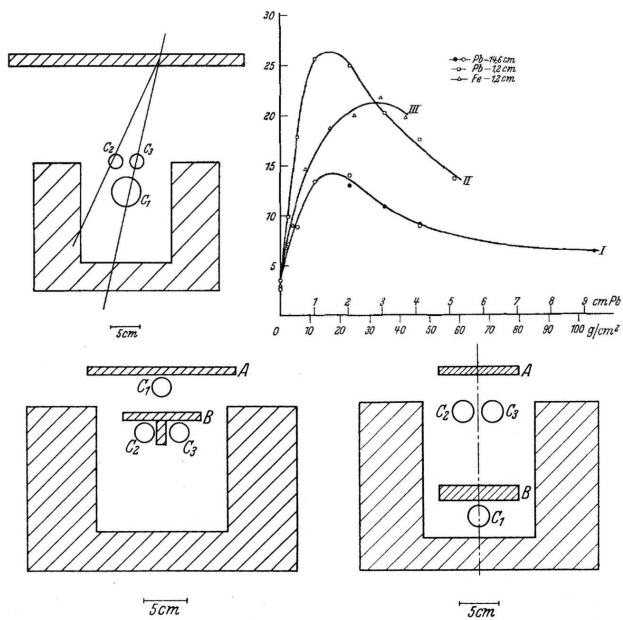}
\caption{Triangular array of G-M counters enclosed in a lead box screening from background radiation for the study of the soft and hard components of cosmic rays. The coincidence rate is displayed as a function of the thickness of the lead or iron screen placed above the counters. Each curve results from the superposition of two terms, with the soft component mainly contributing  to the coincidence rate at small thickness of the  screen, and quickly dying off (rapid rise and initial decrease in the shower production) while the hard component contribution continues up to a large thickness, and decreases very slowly (long tail of the lead curve). The soft component  contribution increases with the screen atomic number. Below, left: Arrangements for absorption measurements of shower particles obtained placing shower producing layers of constant thickness {\it above} the counters. Coincidences between the counter $C_{1}$ and $C_{2}$--$C_{3}$ are recorded with layers of different thickness {\it between} $C_{2}$ and $C_{3}$. Without the absorbing wall, a shower produced by a single ray in the absorber  B can discharge both of the counters underneath. These coincident discharges are produced by low energy shower particles which are easily stopped by some absorber between the counters. Below, right: Measurements of coincidences produced in $C_{2}$--$C_{3}$ and $C_{1}$ by penetrating particles generating a shower in  the screen A and traversing absorber B.}
\label{Figure4}
\end{figure}

 These results were summarized and discussed in detail in a long paper published in March 1933.\footnote{\cite{Rossi:1933vn}.\label{Rossi:1933vn}}
  There, Rossi showed the different configurations he had used for his analysis, and which had led, by way of direct evidence, to his remarkable conclusion: a soft secondary radiation was produced during the interaction of cosmic rays with matter (see Fig. \ref{Figure4}). This implied, in particular, that the soft component he had been able to reveal in the atmosphere since 1931, increasing in inclined directions from the vertical line, was actually  all of  `secondary' origin. At the end of the article, virtually closing the earliest stage of a new era for cosmic ray physics, Rossi warmly thanked Bothe, Heisenberg and Fermi `for several inspiring discussions.'   
  In a note to {\it Nature} of July 3, entitled `Interaction between Cosmic Rays and Matter,' he further specified that `these rays are therefore to be regarded as a {\it secondary radiation of the primary cosmic rays}, the equilibrium value of which is roughly three to four times greater {\it in air} than in lead.'\footnote{He immediately remarked that `the shower-producing rays are more readily absorbed by elements of higher atomic number [\dots]'. From this and from the observation that showers more frequently occur in elements of high atomic number, he concluded that `the production of showers must be the main reason for their absorption.'  \cite{Rossi:1933qy}.}

In April 1932, during the annual gathering of the world's leading quantum physicists organized at Niels Bohr's Institute for Theoretical Physics in Copenhagen, a parody of Goethe's {\it Faust}  was staged to commemorate the 10th anniversary of Bohr's Institute and the hundredth anniversary of Goethe's death. The satirical play, entitled{\it The Blegdamsvej Faust}, was written primarily by Max Delbr\"uck, and satirized the dramatic changes and  discoveries that were revolutionizing physics. Bohr  was identified as the Lord, and the plot centred around Pauli-Mephistopheles' attempts to sell the unbelieving Ehrenfest-Faust (Pauli's most stubborn opponent)  the newly invented neutrino. The  meeting was held in the spring, a short time after Chadwick's demonstration of the existence of the neutron,  in the atmosphere of friendly collaboration  thatcharacterized Bohr's  Institute. Chadwick's discovery clarified the difference between `Pauli's neutron' ---the `neutrino', as it had  just been renamed by Fermi and his group in Rome--- and the new constituent of the nucleus. In the second part of {\it The Blegdamsvej Faust}, Rossi's investigations  played a role:  the arrival of Millikan-Ariel  was announced by roaring Cosmic Rays. Wilson Chambers, Counting Tubes, Cosmic Rays, Protons and Electrons were mentioned in connection with Heisenberg  (`definitely in a bad mood'), as well as with Rossi and Hoffmann (both very `nervous')\dots  

On the whole, the Copenhagen Faust testifies to the state of the field at the time: the confusion, the conflicting views, the turmoil within the discipline, the individual efforts, and even the uneasiness and despair. Cosmic-ray research epitomizes  the general situation during the early 1930s: `This field is still in its infancy and full of uncertain, contradictory, and puzzling results [\dots],' as  Otto Frisch wrote  to an unnamed colleague in 1935.\footnote{ \cite[p. 86]{Roque:1997qy}.\label{Roque:1997qy}}

At the same time, the parody  provides clear evidence for  the bonds that existed in a community characterised by a constructive and cooperative attitude.  This community had a sense of humour about itself, and was even capable of criticizing  its most beloved deities ---like Bohr and Einstein--- and of reflecting on the state of the new physics, and joking about its inner tensions and difficulties. It was not by chance that the play took as its epigraph one Bohr's most famous sayings: `Nicht um zu kritisieren', `Not to criticize \dots'.

	\section{The Blackett-Occhialini cloud chamber: cross-\\fertilization  during the infancy of particle physics\label{blackettocchialini}}
	
In the summer of 1930, during his sojourn at Bothe's laboratory in Berlin, Rossi had met Patrick Blackett, a great expert of the cloud chamber technique and a  skilled experimentalist. Blackett's wife was also Italian, which provided another reason to establish a friendship with his Italian colleague. They met again in 1931, in Zurich and in Rome. Blackett's interest in nuclear physics and cosmic rays was triggered by the arrival in Cambridge of Rossi's young collaborator Giuseppe Occhialini in July 1931, substituting his colleague Gilberto Bernardini. 
The Cavendish Laboratory was a very modern one, but Occhialini had something to offer.  He was the bearer of the GM counter technique, not  known in Cambridge, as Blackett  himself recalled: `It was a curious fact that the Geiger counter was not in use earlier in the Cavendish Laboratory. Partly this may have been because Rutherford was fixated on the alpha-particles [\ldots] Geiger counters were not necessary for studying alpha-particles' \cite[p. XXXV]{Blackett:1969fk}. He also brought the coincidence counting circuit developed by Rossi. It is indeed remarkable that Chadwick mentioned Occhialini at the very end of his second article announcing the existence of the neutron, in connection with a coincidence experiment performed by the young Italian researcher showing that `the neutrons very rarely produce coincidences in tube counters under the usual conditions of experiment \cite[p. 708]{Chadwick:1932kx}. 

In his Nobel Lecture, Blackett recalled that in that same autumn of 1931, in collaboration with Occhialini, he `started to study the energetic particles found in cosmic rays by means of the cloud method.' They thought of using a Rossi circuit with two GM counters in coincidence placed above and below the chamber. They would detect cosmic rays traversing the chamber and triggering its expansion, essentially making cosmic rays take their own photographs. This system allowed the detection of many more significant events than could be obtained by the random exposure method.\footnote{ Jabez C. Street, who had already worked with W. F. G. Swann using a cloud-chamber,  later  joined Johnson, already working on coincidence experiments. After developing together  a circuit for recording coincident discharges they  made another  early attempt of an instrument employing both visual and electronic devices: \cite{Johnson:1932qv}.}  They succeeded in getting the first photographs by this new method in the early summer of 1932.\cite{Blackett:1932mz} Occhialini had come for three weeks: he stayed for three years.
  Their device can  be considered the perfect fusion of {\it image and logic}. It marked  a watershed in the history of physics and a turning point in Blackett's scientific path. From then on cosmic rays became the focus of his interests until the end of the 1940s.

Occhialini's sojourn in England is a typical example of a strategy implemented by Italian physicists to learn new experimental practises and theoretical tools not available in their  home nation, and to experience a richer scientific context. During the 1930s, Germany (Berlin, Hamburg, Leipzig, G\"ottingen, Heidelberg, Frankfurt), Great Britain (Cambridge, London), France (Paris), Denmark (Copenhagen), Sweden (Stockholm, Uppsala), and Switzerland (Zurich) attracted Italian physicists slightly younger than Fermi.

The Arcetri Group's decision to send a physicist to the Cavendish Laboratory went hand in hand with the contemporary failed attempts to build a cloud chamber in Rome. Since 1930 Fermi,  Rasetti, and Amaldi  had started to make counters. Rasetti would later recall, when interviewed by Thomas Kuhn about 30 years later, that these had been `poor counters, poor cloud chambers'.\footnote{F. Rasetti and E. Persico interviewed by T. S. Kuhn on April 8 1963, NBL\&A.} 
According to Rasetti,  all their knowledge of Geiger counters `came from Rossi, from occasional contacts with Rossi'.   But after their decision to abandon atomic spectroscopy, by that time an exhausted research field in which Rasetti was a real master, they realized that their nuclear techniques  were `very primitive.'  The strong interest in the possibility of building a reliable cloud chamber in Italy was again one of the reasons for Rasetti's choice to spend a second fellowship of the Rockefeller foundation at the Kaiser Wilhelm Institut f\"ur Chemie in Berlin-Dahlem, where he had the opportunity to learn the techniques of preparing radioactive sources, of building counters, and ionization and cloud chambers under Lise Meitner.\footnote{See F. Rasetti, Biographical notes and scientific work of Franco Rasetti,   Edoardo Amaldi Archive, Physics Department of Rome University Sapienza, Box E8, Folder 2, and \cite{Goodstein:2001fk}.} 
Only in 1933, after Rasetti had returned from Germany, did the researchers in Rome begin their intensive investigation of nuclear subjects. In particular, they built some GM counters, designed and built a cloud chamber, and began to prepare Po+Be neutron sources. 

In the meantime, the cross-fertilization between Rossi's methods and Blackett's extraordinary competence in cloud-chamber techniques marked the beginning of a new era in the physics of detectors and experimental methods. The  extension of the basic concept  to any kind of detector controlled by the most sophisticated electronics implied the possibility of selecting  `special'  events and using them to trigger such devices. A key experimental tool was thus born for the development of both nuclear and elementary particle physics.  It became a longstanding tradition for these fields to evolve when new kind of  technologies became available. At the moment the Blackett-Occhialini counter-controlled chamber provided a clear demonstration of the existence of the secondary showers observed by Rossi, and could also place the `positively charged particle  having a mass comparable with that of an electron' just observed by Anderson \cite{Anderson:1932fk} in the perspective of the annihilation process implied by Dirac's theory of the relativistic electron.\footnote{As most of the shower particles were recognized to be positive and negative electrons, Blackett and Occhialini suggested that they were created in pairs  by photons with energy exceeding 2m$_{e}c^{2}\simeq$1MeV  absorbed in the neighborhood of a nucleus, and associated the mechanism of their creation with `Dirac's theory of electrons': \cite[p. 713]{Blackett:1933fj}.}  
This suggestion was later confirmed experimentally by Chadwick, Blackett, and Occhialini \cite{Blackett:1933rt}.

In the meantime Carl Anderson read Blackett and Occhialini's article `In which for the first time, as far as I know, the idea of pair production was clearly brought out. I very well remember reading that paper and being wholly convinced on the first reading that this was the proper explanation.'
He had learned of their apparatus from a publication, either from their first article describing the mechanism controlling the cloud chamber, or from the second one where they provided photographs obtained with the method: 

\begin{quotation}
\small

We knew that they were building an apparatus to study cosmic ray particles, cloud chamber apparatus in a magnetic field. We did not know that they were using the Geiger counter triggering mechanism, which was a big step forward technically [\dots] We did not know they were doing that until publication in the spring [\dots] So then the minute we learned that Blackett and Occhialini had this counter control, {\it we immediately set about building one}. In fact, Bill Pickering built the electronics for the first Geiger counter control of our cloud chamber [emphasis added].'\footnote{Carl Anderson interviewed by Charles Weiner, June 30, 1966, NBL\&A.}
\end{quotation}
\normalsize

According to Blackett, `Bohr was at first unconvinced by Anderson's evidence, but was persuaded by the extensive evidence offered by the photographs of Blackett and Occhialini.' Dirac worked very closely with them; in fact, he was often at the laboratory. When asked how long they had known about Dirac's theory, Blackett replied he wasn't quite certain, but that it didn't matter anyway `because nobody took Dirac's theory seriously.'\footnote{P.M.S. Blackett interviewed by John L. Heilbron, note \ref{blackettinterview}. About the skepticism on Dirac's theory see also J. Cockcroft interviewed by Thomas S. Kuhn, May 2, 1963. NBL\&A.} 
At the beginning of 1933, Blackett and Occhialini's observation of showers of positive and negative electrons provided both new  evidence in favor of Rossi's corpuscular programme  and its emerging connection with quantum electrodynamics, and a sound experimental base for Dirac's relativistic theory of the electron and its connection with antimatter in a period when the integration of theory and experiments was still far from complete. 

 But probably the most interesting impact from Blackett and Occhialini's clear statements about the {\it creation} of new particles (`both the negative and positive electrons in the showers must be said to have been created during the process [\dots]  one can imagine that negative and positive electrons may be born in pairs during the disintegration of light nuclei')  was on Enrico Fermi. It became a fundamental ingredient in Fermi's thought leading to his theory of $\beta$-decay. Fermi grasped the idea of creation and transferred it to  the electron and neutrino pair created in the act of decay, according to an analogy with photons in QED. He had already showed a great interest in Pauli's hypothesis of the neutrino since the Roman conference of 1931. Now, at the Solvay conference, in which both took part, Pauli's remaining doubts were dispelled by new experimental evidence about energy conservation in the process. His idea of the neutrino as the solution  to the problem of continuous-spectrum $\beta$-decay was published in the proceedings. This definitely provided Fermi with the main ingredient for his theory, which took form immediately after the conference.\footnote{\cite{Fermi:1934qy} (note \ref{Fermi:1934qy}).}

 \section{Dialoguing  with theoreticians: Enrico Fermi, Hans Bethe,  and Homi Bhabha}
 
The results Rossi derived from St\o rmer's theory had provided an answer to the question of whether particles of a given energy \textit{could} or {\it could not} reach a given point of the Earth in a given direction. 
However, after the prediction of the East-West effect, Rossi's experiments performed at Florence had failed to reveal the expected asymmetry. He was aware that this negative result might be ascribed to atmospheric absorption, if the energy losses of cosmic-ray particles in air were much greater than those due to ionization alone. However, before reaching any definite conclusion, it was necessary to improve the theory of geomagnetic effects. 

After Rossi's departure from Florence, his contacts with Rome were less frequent, mainly because of the distance. But this was the right occasion, as recalled by Gilberto  Bernardini: `when we had something that we considered both interesting and disquieting. Only then we went to speak with Fermi.'\footnote{G. Bernardini, `First round-table discussion' in \cite[p. 273]{Brown-Hoddeson:1983uq}.\label{Brown-Hoddeson:1983uq}} Rossi visited Fermi during a weekend before the beginning of February 1933 and both had a discussion  concerning the influence of the Earth's magnetic field on the intensity distribution of cosmic rays. As recalled by Rossi himself:\footnote{B. Rossi, introduction to their joint paper  reprinted in  E. Fermi, {\it Collected Works. Note e Memorie, Italy 1921--1938}, Vol. 1, edited by E. Amaldi et al., Accademia dei Lincei and University of Chicago Press, 1962, p. 509:  \cite{Rossi:1933rt}.\label{FermiRossi}}

\begin{quotation}
\small

I was very much interested in this effect whose study offered the possibility of determining the sign of the charge of primary cosmic ray particles [\dots] When I presented this problem to Fermi he pointed out that Liouville's theorem was applicable to the case in question, and provided a very simple solution showing, in fact, that the intensity was the same in all allowed directions.
On the basis of this result, we re-examined the data obtained in the Florence experiment, as well as the results of the experiments of Clay  and others on the latitude effect, and came to the conclusion that an abnormally large atmospheric absorption provided indeed the most likely interpretation of all available data [\dots] Assuming the numerical value of the atmospheric losses indicated by our analysis, we then concluded that in the vicinity of the equator the east-west effect should have been clearly observable.'

\end{quotation}
\normalsize

On the strength of this prediction, Rossi decided to speed up the organization of the already-planned expedition to Africa.

 Like Heisenberg and Fermi, other theorists were highly interested in the interpretation of cosmic-ray experiments, and particularly in the energy loss of high-energy particles through matter.  At that time, the young but very gifted theorist Hans Bethe, an expert on the interaction
 of radiation and matter, a topic on which he had just published a landmark paper  \cite{Bethe:1930aa},  spent several months with Fermi  in
 1931--1932 with a Rockefeller scholarship.  From Rome, Bethe  wrote enthusiastically about Fermi to Arnold Sommerfeld.\footnote{H. Bethe to A. Sommerfeld: 9 April 1931(HS 1977-28/A,19),  20 April 1932 (HS 1977-28/A,19),  25 April 1931 (HS 1977-28/A,19),  29 Juli 1931 (HS 1977-28/A,19),  1 Mai 1932  (HS 1977-28/A,19), 11 April 1933  (HS 1977-28/A,19), Deutsches Museum Archives, Munich.}
 He was going to spend the rest of his Rockefeller fellowship in Cambridge, and was very sorry  to leave Italy: `The stimulus I have here by Fermi, is larger by orders of magnitude [\dots] Dirac is well known for speaking only one word per light year, and the other people in Cambridge are far from having the general view of the quantum theory that Fermi has.'\footnote{H. Bethe to A. Sommerfeld, Rome, April 9, 1931 in {\it Arnold Sommerfeld Wissenschaftlicher Briefwechsel}, Band 2: 1919-1951, edited by M. Eckert and K. M\"arker (Deutsches Museum - GNT Verlag 2004), pp. 322-322. Bethe was already a skilled researcher, but he would later recall the important influence Fermi had upon him: 
 `Fermi gave me the wonderful method of doing things quickly and easily, any problem can be solved by sitting down for twenty minutes and thinking about it starting from first principles. Sommerfeld never did that. Sommerfeld said `Well here is the title of your problem, now you do it' and then you had to put in differential equations and if possible Bessel functions. For Fermi that didn't matter. You just did the mathematics the best way that came to your mind, and the physics was clear by the time you started [\dots] Fermi was a most important influence, in fact I consider Sommerfeld and Fermi as equal in ---as my teachers.' Video interview with Hans Bethe, Session 29, available at http://www.webofstories.com/play/4466.}  
 
 By 1932 Bethe had also calculated the stopping power of charged particles of relativistic velocity in matter using M\o ller's theory.
On the other hand Fermi had just written his  famous landmark paper on quantum electrodynamics  \cite{Fermi:1932lr}, and was interested in the various ways the interaction of electric particles could be formulated in relativity theory:\footnote{The problem of high-energy relativistic collisions and of the stopping of charged particles in matter revived by cosmic-ray experiments was a main issue in the agenda of theoretical physicists during the early days of quantum electrodynamics. For an analysis of the development of M\o ller's theory see \cite{Kragh:1992uq}. For a study of the relationship between the M\o ller formula and quantum electrodynamics see \cite{Roque:1992fk}.} 

\begin{quotation}
\small

`[\dots] his main interest was in the M\o ller interaction, which was just the first order in $e^{2}$ of the result of QED.
So he proposed that we write a joint paper on the various expressions for the interaction of relativistic charged particles: the full result of QED, the M\o ller approximation, and Breit's interaction, which was valid to order $v^{2}/c^{2}$. We soon had done the algebra, which left only the writing of the paper. Fermi had no secretary, so he did the typing. He would speak every sentence in German ---which he knew well from a year as a postdoc in Germany. I then had a chance to suggest corrections. I made very few, some in language, fewer in content; then he would type it. My job was to write the few formulas, which would then be inserted by hand into the typed manuscript.' \cite{Bethe:2002fk}

\end{quotation}
\normalsize

It was June 1932. In a few hours the article was done  and sent for publication to the {\it Zeitschrift f\"ur Physik} \cite{Bethe:1932aa}. These problems continued to hold a special place in Bethe's mind. Not long thereaftre, he co-authored a fundamental paper with Walther Heitler on the problem of electron-positron scattering, and wrote his well-known works on the passage of fast electrons through matter.\footnote{\cite{Bhabha:1936uq}; \cite{Bhabha:1937yq} (note \ref{Bhabha:1937yq}).}
 
During his stay in Italy, Bethe visited Rossi in Florence and learned `to appreciate Italian cuisine' at Rossi's mother's home.\footnote{\cite{Rossi:1990aa} (note \ref{Rossi:1990aa}), 45.}  
Since that time Bethe began to build his  extensive knowledge of cosmic-ray data, becoming an expert on the quantum treatment of the physics of cosmic ray absorption. He completely disagreed with Millikan's  `birth cry'  theory which `quite evidently did not make any sense,' and  regarded Rossi's `promising' researches with great favor.\footnote{H. A. Bethe interviewed by P. Galison, (note \ref{Galison:1987kx}), 102.} 
They were nearly the same age, and it was easy for the two men to establish a warm relationship, which later proved crucial in Rossi's life.\footnote{In that same summer 1932 Bethe was offered an assistant professorship at T\"ubingen, but after Hitler's ascension to power in January 1933 he was dismissed  because his mother was of Jewish origin. After finding a temporary position in Manchester during the period 1933--1934, he  left Europe and emigrated to the United States to accept a position in Cornell during the summer of 1934. In 1940, when Rossi was looking for a stable position in the U.S., Bethe succeeded in calling him to Cornell University, where Rossi remained until he became involved in the Manhattan project and moved to the newly built top-secret laboratory at Los Alamos in the summer of 1943.}

  Theoreticians like Heisenberg, Fermi, Bethe, Heitler, and Bhabha held a deep commitment to the nascent quantum electrodynamics, the archetype of quantum field theories. They read articles  on cosmic-ray experiments with great attention  and compared  data with their own theoretical analysis in trying to explain the high energy behaviour of charged particles involved in cosmic-ray researches. According to Rossi's experiments on the rate at which showers occurred in different substances, a given mass of lead had been found  much more effective than the same mass of lighter elements.\footnote{\cite{Rossi:1933vn} (note \ref{Rossi:1933vn}), Fig.~ 6 on p. 164, and Table 13 on p. 172.} Other experiments showed that lead, rather than the lighter elements, was a more effective absorber of the radiation generating the showers, when absorbers of the same mass per cm$^{2}$ were compared. These results were presented in particular at a conference held in Zurich in 1933 \cite{Rossi:1933uq}, where the  young Homi Bhabha was also a participant. 
  
  Bahba opened his very first  scientific paper  (`Zur Absorption der H\"ohenstrahlung'), concerned with the essential role that showers play in the absorption mechanism, with a comment  on Rossi's talk \cite[p. 120]{Bhabha:1933fj}: `In the beautiful Rossi's experiments on the absorption of cosmic rays in lead presented during the Zurich Physics Conference, at least three particles were necessary in order to generate triple coincidences, and this is very close to identifying these processes with the `showers' observed by Blackett and Occhialini.'\footnote{Bhabha, who was writing in German, used the English word `Showers', just invented by Blackett and Occhialini. He thanked Rossi for providing specific experimental data which he was able   to compare with his theoretical calculations.} 
   And indeed Bhabha, who was visiting Pauli in Zurich coming from Cambridge ---and had just spent some time in Rome with Fermi during the period 1932--1933--- was quite aware of the connection that Blackett and Occhialini had established between the showers visualized by their cloud chamber triggered by an electronic circuit,  with the abundant production of secondary particles revealed by Rossi's experiments of the previous two years.\footnote{Also thanks to this important work, in 1934 Bhabha was awarded the Isaac Newton Studentship at Cambridge which enabled him to complete his PhD under Ralph H. Fowler.}

    In his presentation at the Zurich conference, Rossi had explicitly talked of `nuclear processes' being responsible for the absorption of this shower-producing radiation, and his concluding words were that the `primary Ultrastrahlung' might contain positive electrons which, in encountering negative electrons, would generate $\gamma$-rays. These, in turn, `in colliding with nucleus, would create the observed groups of particles, which probably  will again fragment in positive and negative electrons.' In his detailed theoretical analysis, based mainly on Rossi's recent work, Bhabha started from the idea that `a primary ray can create a shower in going through matter.' But he made a distinction between the different particles  generated within a group, which he called `shower-particles' and `secondary particles'.  Only the latter, he argued,  still retained enough energy to create a new shower within the traversed matter. The production of showers was thus the reason for the absorption of such secondary particles in interactions with matter.
    
     At that time Walther Heitler was turning away from his previous topics of research in quantum chemistry and moving into the field of quantum electrodynamics, which then `represented the fundamental unsolved problem.'
He too thought that `high energy phenomena would give some key to the further development of quantum electrodynamics [\dots] The interest in high energy of course led me into cosmic radiation [\dots] Later on Blackett discovered cosmic ray showers. Now these showers were the next major step in my work.'\footnote{W. Heitler, interviewed by John L. Heilbron, March 19, 1963, NBL\&A.\label{HeitlerInterview}} 
    Having   put forward the theoretical premises for the cascade idea,  Bhabha continued to keep an eye on the whole phenomenology deriving from Rossi's investigations.  Within a few years, these investigations would become the testing ground for the Bhabha--Heitler theory of electromagnetic showers based on QED.

 \section{Arthur Compton and the international  campaign for the observation of the geomagnetic effects of cosmic rays}
 \label{geomagnetic}

In the early 1930s  the practise of using Geiger-M\"uller counters was only beginning. The literature on the subject consisted of only Geiger and M\"uller's original 1928 article, a few German articles, and some U.S. articles that did not discuss construction details. Blackett, for example, recalled that  the Geiger counter was a very delicate instrument: `In order to make it work you had to spit on the wire on some Friday evening in Lent. One had to be initiated into all the mysteries in order to get any results at all.'\footnote{P.M.S. Blackett interviewed by John L. Heilbron, NBL\&A), note \ref{blackettinterview}.} According to Luis Alvarez,  who  graduated at the beginning of 1932, `No one in Chicago had seen one. I built the first counter that anybody had seen in Chicago.' It became his first research project following J. Barton Hoag's suggestion. Without any application in mind he tackled all the problems connected with the witchcraft of building a reliable Geiger counter: `[\dots] it was pretty rudimentary  [\dots] people who could build good Geiger counters with low backgrounds were thought to be wizards or sorcerers.'\footnote{L. Alvarez interviewed by Charles Weiner and Barry Richman, Berkley, February 14, 1967, NBL\&A.\label{alvarez}}  
After he had finished his first counter, given his first seminar on the counter, and demonstrated it, Arthur Compton  proposed him to upgrade his counters to make cosmic-ray measurements.

 Since his participation in the conference on nuclear physics in Rome, Compton had been highly impressed by Rossi's arguments and research. The beginning of their relationship represented a real turning point in his scientific life. Compton became convinced of the importance of demonstrating `once and for all'  the nature of cosmic rays, and he later  acknowledged Rossi's role in providing the strong motivation for his worldwide research programme aiming at studying cosmic rays at different altitudes and geomagnetic latitudes.\footnote{This was recalled by Rossi in his autobiography (note \ref{Rossi:1990aa}, 18).} Alvarez learned from him about the coincidence technique, and accepted his suggestion  to build a Geiger counter telescope. When Compton  became his graduate adviser, he had just decided to change his main research interest moving on from the X-ray work ---for which he had been awarded the Nobel Prize--- to the ambitious programme of organizing a world-wide cooperative project involving tens of physicists. The project aimed at measuring the intensity of cosmic rays at widely spaced locations all over the world in order to demonstrate its correlation with {\it magnetic} rather than {\it geographic} latitude.\footnote{Compton  convinced the Carnegie  Institution to fund a world survey between 1931 and 1934.} 

As recalled by Alvarez:\footnote{See note \ref{alvarez}.}

\begin{quotation}

\small

Clay had shown that the cosmic rays dropped off as you went towards the equator, but nobody believed him because Millikan's experiment showed that they were constant over the whole surface of the globe. The only persons apparently to believe Clay's work were Clay himself and Compton, who had apparently known Clay and knew he was a reliable person. So Compton designed a series of cosmic-ray meters where there was a calibration possible by means of a standard radium source that was placed a meter away. So he had a good standard in his cosmic-ray measurements that the other people didn't have. Everybody else measured the ionization current, and that can depend on changes in a number of variables that you don't have any control over. As Compton said, `If I put this standard source a meter away and see how the ionization rises, that change in ionization will always be the same regardless of the background'.\footnote{Each set of apparatus designed by Compton consisted of the same 10 cm spherical steel ionization chamber, filled with argon at a pressure of 30 atmospheres. Ionization current generated in the main chamber by cosmic rays was mainly compensated by a constant current from a small subsidiary chamber with an uranium radioactive source. The value of the resulting current was recorded continuously on a moving tape by Lindemann's electrometer.} 

\end{quotation}
\normalsize

To obtain the  data needed to determine the nature of the primary cosmic rays, Compton organized and led expeditions to all parts of the world to measure the cosmic ray intensity over a wide range of geomagnetic latitudes and longitudes and at many elevations above sea level. The globe was divided into nine regions, and roughly 100 physicists divided into smaller groups and sailed oceans, traversed continents, and scaled mountains, carrying identical detectors to measure cosmic-ray intensities. Such a detailed survey was one of the first examples involving many physicists all over the world working on the same research project.
 His wife Betty recalled the spirit of this programme of heroic proportions: `I think the brightest idea Arthur ever got was to get himself out of the basement room laboratory watching a spot of light going across a scale, and get himself traveling around the world.'\footnote{Betty Compton interviewed by Charles Weiner, April 11, 1968, NBL\&A.}

Within a year, Compton's world-wide campaign  confirmed and greatly extended Jacob Clay's earlier observations, showing  that the intensity of cosmic rays is systematically correlated with geomagnetic latitude and  altitude \cite{Compton:1932kx,Compton:1933vn}. It proved that at least a significant fraction of the primary cosmic rays are charged particles, and thus are subject to the influence of the Earth's magnetic field.
In proving the existence of a strong latitude effect, his results excluded the possibility that all the charged particles detected at sea level were secondaries generated within the Earth's atmosphere by cosmic $\gamma$-radiation. This dispelled all  doubt, and definitely contradicted  Millikan's views.
 
 Compton's interest in cosmic-ray research, was the beginning of a warm relationship between Bruno Rossi and the more mature U.S. Nobel Prize winner. This relationship, too, would prove important after a few years, when Rossi was obliged to leave Italy in 1938 after the enactment of fascist laws targeting Italian Jews.
  In September 1933 Rossi and his collaborator De Benedetti, who  had graduated  in Florence during the academic year 1933--1934,\footnote{The title of De Benedetti's thesis   was `Measurement of cosmic rays in different zenital directions'; his adviser was  Gilberto Bernardini, a former collaborator of Rossi in Florence. I am grateful to the family of Sergio De Benedetti for a list of his works and for biographical material.}
   went to Eritrea to carry out  experimental tests on the existence of the East-West effect Rossi himself had previewed during his stay in Berlin in 1930. 

During the  journey from Spalato to Massaua, which took about a week, they measured the dependence of cosmic-ray intensity on magnetic latitude \cite{Rossi:1933fk}. They performed these experiments with a greatly improved ionization chamber provided by Compton, thus participating in his world-wide research programme.\footnote{Tables on results from intensity measurements show dates starting from September 1, and continuing through September 7. B. Rossi, notebook, MIT Archives, Box 1, Folder 13. See also letter from Rossi to Compton from Asmara, November 25, 1933 (Compton papers, Personal Correspondence, Washington University Libraries), where he described his measurements and thanked Compton for helping with language and publication of their short articles  on the {\it Physical Review}.} 

However, by the time they left Italy, experiments proving the existence of a difference in the intensity of cosmic rays between East and West had already been  announced in two letters  sent to the Editor of the {\it Physical Review} by Luis Alvarez and Compton,\cite{Alvarez:1933yq} as well as by Thomas Johnson \cite{Johnson:1933fj}. Both had used a GM counter `telescope' for studying the East-West asymmetry.  However, they did not mention Rossi's prediction of the effect, nor his joint paper with Fermi.\footnote{See \cite{Rossi:1933rt} (note \ref{FermiRossi}).} Instead, theycredited Lamaitre and Vallarta, whose paper had been published three years later  \cite{Lemaitre:1933qy}, and which from then on was generally mentioned in connection with these results. Soon after, the effect ---actually a more pronounced one--- was also observed by Rossi and Sergio De Benedetti in Eritrea, who had unfortunately lost the priority of this important discovery  by a few months \cite{Rossi:1934aj}. The study of the asymmetry predicted by Rossi definitely confirmed the corpuscular hypothesis of the nature of the  cosmic radiation, and in addition provided a most important clue: cosmic rays at sea level are largely positively charged.

Rossi mentioned a `further result' of their observations in Eritrea, which later turned out to be of the greatest importance, and which it is here reported in his own translation from the original paper:\footnote{\cite{Rossi:1990aa} (note \ref{Rossi:1990aa}), 36--37.}
\begin{quotation}
\small

	The frequency of the coincidences recorded with the counters at a distance from one another, shown in the tables as `chance' coincidences appears to be greater than would have been predicted on the basis of the resolving power of the coincidence circuit measured in Padua [\dots]  Those observations made us question whether all of these coincidences were actually chance coincidences. This hypothesis appears to be supported by the following observations [\dots]\footnote{In about 21 hours they registered 14 coincidences between three counters at a distance from one another disposed in such a way that the same particle could not go across all of them. They  established that, according to the resolving power they should expect much more casual coincidences (about 200), while they observed only 6. Moreover, they often observed coincidences contemporaneously occurring in both two-counter circuits (used to measure intensity in both azimuthal and the zenithal directions) when casual coincidences were measured in one of them.} Since the interference of possible disturbances was ruled out by suitable tests, it seems that once in a while the recording equipment is struck by \textit{very extensive showers} of particles, which cause coincidences between counters, even placed at large distances from one another. Unfortunately, I did not have the time to study this phenomenon more closely.\footnote{In the original article Rossi used the Italian word `sciami', which is a term corresponding to `swarms', generally used for a great number of insects like bees. However, the word `shower' had just been coined by Blackett and Occhialini, the first to visualize the phenomenon of a spray in fine streams of particle tracks in their cloud chamber triggered by an electronic  circuit. The term `sciami', however, continued to be used by Italian cosmic-ray physicists.}

\end{quotation}
\normalsize
	
	This appears to be the first observation of Extensive Air Showers (EAS). A few years later these were `rediscovered' and studied in detail by Pierre Auger and his collaborators (Section \ref{intconferences}),  and after World War II  they became the object of fundamental major research projects by the MIT cosmic-ray group founded by Rossi.
	At the moment Rossi's remark  went completely unnoticed, mainly because it was included in an article published in an Italian journal. He himself did not make any explicit connection with his previous experiments clarifying the existence of a soft radiation of secondary origin in the atmosphere, increasing in inclined directions from the vertical line, probably fearing to propose such new phenomenon in a period when coincidence experiments were still under attack.

	\section{Positrons, Geiger-M\"uller counters and cloud chambers in Paris: collaborating with the Joliot-Curie}

	 Phenomena  like the formation of electron-positron pairs observed by Blackett and Occhialini were also on Fr\'ed\'eric Joliot's mind. In 1932 Joliot and Curie had missed  the discovery of the positron,\footnote{The early history of the positron is discussed in \cite{Leone:2010qf}.  For an in-depth analysis of the discovery and its acceptance within the scientific community see  \cite{Roque:1997qy} (note \ref{Roque:1997qy}).}  
	 but in 1933  they published the first photograph of  the materialization of an electron-positron pair \cite{Curie:1933yt},  and Joliot continued to work  on the problem of pair annihilation \cite{Joliot:1933fk,Joliot:1934uq}. 
	
	Rossi and Joliot met in London  in 1934.\footnote{Rossi had probably met Joliot for the first time in Zurich, because he and Curie  had participated to the conference of May 1931. Only Marie Curie was present at the  nuclear conference organized in Rome in October of the same year (see note \ref{curie}). On that occasion Rossi had contacts with her,  both as an organizer and as lecturer. He  probably met Joliot and Curie again in Paris during the winter of 1931--1932, during his stay at Maurice De Broglie's private laboratory.
 He might even have participated in the  5th Congr\`es international d'\'electricite  of July 1932, where Fermi had been invited to present a report on `The present status of the physics of the atomic nucleus.' In any case,
 in December of that same year, Rossi sent  them an illustrated postcard  showing the `Ponte Vecchio' in Florence. B. Rossi to I. Curie and F. Joliot, Padua, 12 December 1932, Joliot-Curie Archives, Paris, Box F144.} 
	They were both invited  to the international conference organized by the IUPAP and the Physical Society which had taken place from 1st to 6th October.  The conference was the first of this type to be held in Great Britain. An important part of the conference was dedicated to a discussion of nuclear physics, and a session was concerned with `new types of radio-activity', a most recent discovery made in Paris by Joliot and Curie using $\alpha$-particles, and later in Rome by Fermi using neutrons as projectiles for bombarding light nuclei \cite{Joliot:1934fk,Fermi:1934uq}.
	 Both Fermi and the couple Ir\`ene Curie and Fr\'ed\'eric Joliot were invited to give talks on their exciting discoveries,
	  while Rossi  discussed recent results arising from the study of cosmic rays.\footnote{On that occasion Millikan, Bowen, and Neher, as well as Anderson and Neddermeyer, still postulated that the primary radiation consisted mostly of photons.}   
	 The programme of the conference, which was inaugurated by Ernst Rutherford,  captures the extent to which quantum electrodynamics, nuclear physics and cosmic ray studies were  still enough in their infancy and strongly interlaced.   From this `turbulent confluence,'   particle physics was beginning to take its first solitary steps.\footnote{\cite{Brown-Hoddeson:1983uq} (note \ref{Brown-Hoddeson:1983uq})  4.}

	During the 1930s cosmic rays represented the most important source of data on the high-energy behaviour of both quantum electrodynamics and nuclear forces.\footnote{The period 1934--1938  is discussed in \cite{Brown:1991fk}.\label{Brown:1991fk}}  
	The recent discovery of the positron marked the advent of cosmic rays as a tool in the exploration of the particle world.  It furthermore opened the door on a new area of investigation, beyond nuclear physics, which  was itself entering reaching maturity after the detection of the neutron. People like Fermi, Joliot, and Rossi were inevitably involved in both fields. Fermi had just published an article with Rossi concerning the influence of the Earth's magnetic field on the intensity distribution of cosmic rays, and his theory of $\beta-$decay.
Joliot and Curie, by that time acknowledged stars of `nuclear physics,'   following Chadwick's announcement concerning the existence of the neutron,  carried out research at the Jungfraujoch mountain research station in Switzerland in April 1932.  Their goal was to highlight the possibility that cosmic rays might be formed mostly by neutrons.\footnote{ \cite{Curie:1933kx}. These experiments did not support the idea and they went back to their work on nuclear physics.}

	 Even if the connection with Dirac's hole theory was considered well established by the end of 1933,  the problem of the annihilation and pair-formation process and its related radiation, and more in general the necessity of putting the  positron in a much wider perspective was, still a topic of active research in the autumn of 1934, at the time of the London conference.  Heitler, who was studying  high-energy phenomena within quantum electrodynamics and within the theory of the positive electrons, had one special problem in mind: `Is there any breakdown of the theoretical predictions at high energies?'\footnote{Heitler interview, footnote \ref{HeitlerInterview}.} At first Bethe and Heitler believed that to be the case.  Both German emigrants to Great Britain on account of the Nazi's racial legistlation, they had already published their landmark paper `On the Stopping of Fast Particles and on the Creation of Positive Electrons'.\footnote{\cite{Bethe:1934fj}.\label{Bethe:1934fj}} The theoretical results contradicted the high energy experiments on cosmic radiation. The Conference was thus a good opportunity to compare their deductions from the standard quantum electrodynamical theory with the latest high-energy data from cosmic radiation.
Fermi himself, in late summer 1933, had published with Uhlenbeck an article where for the first time the one-photon annihilation cross sections were derived in the non-relativistic limit \cite{Fermi:1933fk}. After participating in the Solvay conference held in October, he had formulated his $\beta-$decay theory, which had a great influence on his experimental discovery of neutron-induced artificial radioactivity of March 1934.\footnote{In this regard, Gian Carlo Wick, at the time a young theoretician working with Fermi in Rome, wrote Joliot on 17th February 1934 about his attempt to apply Fermi's theory  to the Joliot-Curie's discovery of artificial radioactivity with emission of positrons. Joliot-Curie Archives, Paris, Box F144.  On the following March 4, Wick's extension  of Fermi's theory to positron emission and to electron K-capture by the nucleus was presented by Fermi himself at the Accademia dei Lincei: \cite{Wick:1934fk}. The influence of his $\beta-$decay theory on Fermi's subsequent discovery of artificial radioactivity is discussed in \cite{Guerra:2009fk}.} 
	 
	 Up to the winter of 1933--1934, Joliot  had performed experiments with the assistance of his wife Ir\`ene using a cloud chamber to reveal positron tracks  \cite{Curie:1933ys,Curie:1933zr,Curie:1933vn}. But now,  in  occasions related to already mentioned experiments on the materialization of electron-positron pairs appearing immediately before  the momentous   discovery of the artificial $\beta$ radioactivity with positron emission, Joliot used  a Geiger-M\"uller counter built   by his postdoctoral student Wolfgang Gentner.  Gentner, who already had some experience with GM counters, had arrived from Frankfurt as a postdoctoral researcher in January 1933.\footnote{See report on Gentner's activity at Curie laboratory dated 4th July 1935 and signed by the director Andr\'e Debierne. Archive Curie, Paris, Box 20.} At this time, when the  Joliot-Curie couple was becoming fully  involved in the `positron affair': `In Paris nobody knew really how to build Geiger counters,'  recalled Gentner. Moreover, Marie Curie did not know this technique, so she did not agree with using counters to reveal $\gamma$- and X-rays.\footnote{`She told me ---continued Gentner--- that I had to do this with ionization chamber [\dots] and with Hoffman electrometer and things like this. I knew these things, but I thought that it was much easier to do it with a Geiger counter. And she said, `No, with a Geiger counter you don't know really what you are measuring, and I would prefer you to work with ionization chambers.' I had a long discussion with her about this, finally compromising, and she decided that I could perhaps do both, compare Geiger counter measurements with ionization chamber [\dots]'. Wolfgang Gentner interviewed by Charles Weiner, 15th November 1971, NBL\&A.\label{gentnerinterview}}   
	
	 Joliot, on the other hand, was very interested in GM counters, so after a short time Gentner became the specialist  at the Institute and built counters and amplifiers for Joliot. Actually, Gentner's GM counters were at the core of the discovery of artificial radioactivity with emission of positrons in January 1934. After irradiating aluminum with $\alpha$-particles from polonium, Joliot  observed that the aluminum foil continued to emit positrons even after removing the $\alpha$-particle polonium source. For this reason he initially thought that something was wrong with the counter.As Gentner would later recall: 
`If something didn't work, he went always to call me, I should look what was happening  [\dots] when he discovered artificial radioactivity, I worked with him in the same room [\dots] he called me, `The counters are not working all right' [\dots]'.
After Gentner had assured that the counter was  all right, Joliot said `Oh, I think I know what it is, it's perhaps some kind of radioactive isotope [\dots]'. So after a week, the JoliotCurie couple sent the publication out.\footnote{For an historical reconstruction of the discovery see \cite{Guerra:2011uq}.}
	
	 But in the autumn of 1934, Gentner was planning to leave Paris in a few months.  Thismay explain Joliot's interest in having a new young researcher acquainted with the Geiger-M\"uller technique, and in particular with coincidence methods, such as Rossi's assistant Sergio De Benedetti.  At the same time, Joliot had a great knowledge of how to use the cloud chamber, a tool which Rossi was planning to have in Padua.  An exchange in this sense must have appeared interesting for both. Joliot had just set up a new Wilson chamber with variable pressure,  which he had used to study individual disintegrations  and the trajectories of the recoil particles \cite{Joliot:1934qf}. Moreover, the successful collaboration between Blackett and Occhialini had just opened the way to a fruitful marriage between cloud chambers and counters.  Something similar might be introduced in Paris, too. 

	Active discussions about the above issues were probably held during the London conference because of the strong relationship between their research.  Rossi and Joliot must have decided to have  Sergio De Benedetti in Paris.
	
	 Roman physicists, too, wanted contact with the French. Actually, at the beginning of 1934 Franco Rasetti had asked Ir\`ene Curie to accept the chemist Oscar D'Agostino in her laboratory `pour se perfectionner dans la chimie des \'el\'ements radioactifs' and to  `apprendre les manipulations n\'ecessaires pour obtenir les sources radioactives les plus fr\'equemment employ\'ees par les physiciens [\dots]'.\footnote{Rasetti's letter was originally dated 20th December 1933; he changed the date in 26th January, but forgot to change the year. This circumstance is confirmed by the fact that he mentioned having met Fr\'ed\'eric Joliot-Curie in Leningrad, where actually a conference on nuclear physics had taken place from 20th to 25th September 1933, and where Joliot had been invited (A. Joff\'e to F. Joliot from Leningrad, July 1933, Joliot-Curie Archives, Paris, Box F144). See also the letter written by Fermi to Marie Curie: E. Fermi to M. Curie, 26th January 1934 (Laboratoire Curie, 2118), Joliot-Curie Archives, Paris).} 
	Around that time Fermi and Rasetti, who were interested in experimental techniques in nuclear physics, had built a gamma-ray spectrograph using a bismuth crystal. On February 9 Fermi sent a  single bismuth crystal  to Joliot, together with a letter to D'Agostino and copies of their article published on `La Ricerca Scientifica'.\footnote{\cite{Fermi:1933kx}. E. Fermi and F. Rasetti to O. D'Agostino from Rome, February 9, 1934. Fondazione Oscar D'Agostino, Avellino.}
	D'Agostino had some experience with the chemistry of radioactive elements, having worked with Rasetti at the separation of RaD. As Rasetti told the Joliot-Curies in Leningrad, they had obtained  `\`a peu pr\`es 110 millicurie de Ra D tr\'es pur'. In the meantime Rasetti was working at a cloud chamber and hoped to visit them in Paris during the long Easter holidays.\footnote{Franco Rasetti to Fr\'ed\'eric Joliot, 26th January 1933, Joliot-Curie Archives, Paris, Laboratoire Curie 2022.} 
	Some days before, on January 19th, Fermi had as well written to Maurice de Broglie announcing D'Agostino's arrival: `Je vous serai tres reconnaissant si vous pourrez lui montrer quelque chose des beaux travaux que on fait dans votre laboratoire [\ldots]'.\footnote{E. Fermi to M. de Broglie from Rome, January 19, 1934, Fondazione Oscar D'Agostino, Avellino. I am grateful to G. Acocella for a copy of this letter.}
But D'Agostino stayed in Paris for only  a short time.  ,He was asked to remain in Italy after the Easter holidays, when  research on the radioactivity induced by neutrons had just begun at the Institute of Physics in via Panisperna.

 On 15th November 1934 Rossi announced De Benedetti's arrival at Joliot-Curie's laboratory on November 20: `Please present my respects to M.me Joliot and accept the expression of  my gratitude and my deepest friendship.'\footnote{De Benedetti's sojourn in Paris during the period 1934--1936 (mentioned by Lucia Orlando  in her paper on the contribution of Italian Jews to the new physics, \cite{Orlando:1998jt}), has never been discussed,  the attention being always   attracted by his move to Paris after being forced to leave Italy in October 1938.}  Actually Rossi had really a debt of gratitude to the Joliot-Curie couple, as they  had succeeded in getting a CNS (Caisse National des Sciences) grant for De Benedetti's sojourn in his laboratory.\footnote{See document contained in the folder `De Benedetti', as well as the list of collaborators of Curie Radium Laboratory and the letter written by De Benedetti from Padua on 27th April 1938, Box I20. Curie and Joliot-Curie Archives, Paris.}   

On 29th December Rossi again thanked Joliot for the `warm welcome' that his collaborator had received (`I am sure that the months he will spend with you will be of great utility for his career').\footnote{Bruno Rossi to Fr\'ed\'eric Joliot, Padua 15th November and 29th December 1934.   Joliot--Curie Archives, Paris, Box F144.} 
	It is no wonder that, during the period November 1934--January 1936, De Benedetti published three articles related to the creation of electron-positron pairs in different chemical elements, the first f which  was presented by Paul Langevin at the Acad\'emie des Sciences already on April 15th, 1935.\footnote{\cite{De-Benedetti:1935ve}. His work was mentioned by Dimitri Skobeltzyn in a letter to Joliot of 24th November 1935 discussing the problem of absorption of $\beta$-rays, where he compared results obtained by Geiger-M\"uller counter and cloud chamber  techniques. Joliot-Curie Archives, Box F144.} 
	The following two papers were related   to the absorption of $\gamma$-rays from radioactive sources and the simultaneous emission of electrons and  the materialization of electron-positron pairs.\footnote{\cite{De-Benedetti:1936ys,De-Benedetti:1936vn}. 
	A later article on the same issues, regarding work performed during his second stay in Paris, appeared  in 1941 in the {\it Physical Review}: \cite{De-Benedetti:1941bh}. 
	 By the time this latter article was published he had left France, immediately before the German occupation of Paris, for the United States. After working in several universities, he joined the faculty of Carnegie Institute of Technology of Pittsburgh in 1949, becoming  a pioneer in the use of positrons as probes to study metals, gases and liquids.}
	
	In December 1935 the Joliot-Curie couple was awarded  the Nobel Prize for chemistry for their  work on the synthesis of radio-elements.\footnote{They received hundreds of congratulation messages for this achievement. A card was sent by Rossi on November 16: `Cher Monsieur Joliot, veuiller agreer, pour vous et pour M.me Joliot, l'expression de mes f\'elicitations les plus vives pour le prix Nobel, qui couronne les admirables r\'esultats de vos recherches. Bien amicalement \`a vous, Bruno Rossi.' The Florence group sent a telegram on November 15: Saluons joyeusement troisi\`eme prix Nobel famille, Bernardini, Bocciarelli, Franchetti, Occhialini, Racah. Joliot-Curie Archives, Paris, Box JC16.} Soon after, in January 1936, De Benedetti  left the `Laboratoire Curie' to return to Padua. Within a few months another Italian physicist arrived in Paris  to work with Fr\'ed\'eric Joliot: Bruno Pontecorvo, the youngest member of Fermi's group, who had received his laurea degree in 1933 and was a promising researcher. He had participated in  activities leading to the October 1934 discovery of nuclear reactions brought about by  slow neutrons, and had learned  the technique of building Geiger-M\"uller counters initially exported by Bruno Rossi from Florence to Rome. His custom-built small counter with very thin walls for detecting $\beta$-rays, later known as the `counter of the Coll\`ege de France', was still used as a model for building such devices in the 1940s.\footnote{Pierre Radvanyi and Jeanne Laberrigue, personal communications to the author.} 
	The four years  Pontecorvo spent with Joliot working in his brand new laboratories of the Coll\`ege de France and of Synth\`ese Atomique at Ivry, from spring 1936 to early summer 1940, deeply changed the course of his personal and scientific life. After Fermi, Joliot became his second `maestro'. On his side, Joliot always considered Pontecorvo `his best pupil.'
	
	When De Benedetti arrived in October 1938, following the racist politics of Mussolini's government,\footnote{For the same reason Salvatore Luria, who had a laurea degree in medicine, was accepted with a CNS grant at Curie laboratory in December 1938. Joliot-Curie Archives,  Paris, Box I20.}  
	Pontecorvo was still in Paris, working at his  pioneering research on nuclear isomerism \cite{Pontecorvo:1938kx,Pontecorvo:1938uq,Pontecorvo:1939fk}.They left Paris together by bicycle, immediately before the arrival of the German troops on June 14, 1940, and both emigrated to the United States.

\section{Cascade theory and  mesotron decay:  the triumphs of the logic experimental tradition}
\label{cascade}

During the 1930s, when quantum electrodynamics appeared to break down at the high energies involved in cosmic rays, experiments using counters became one of the fundamental tools for testing the new physics.  In early 1937,  simultaneous papers by J.\ Franklin Carlson and J.\ Robert Oppenheimer \cite{Carlson:1937kx}, and by Homi Bhabha and Walter Heitler\footnote{\cite{Bhabha:1937yq} (note \ref{Bhabha:1937yq}).}  explained shower formation on the basis of the quantum electrodynamics cross-sections calculated by Bethe and Heitler.\footnote{\cite{Bethe:1934fj} (note \ref{Bethe:1934fj}).} The process may begin with high-energy electrons and positrons or photons.  When they interact with matter, they will respectively emit high-energy photons through a “bremsstrahlung” process or materialize into  electron-positron pairs.  The further production of pairs and photons  will repeat this cascade process until the energy of the produced particles falls below a critical value.  The electromagnetic cascade theory resurrected relativistic quantum field theory toward the end of the 1930s.\footnote{\cite{Galison:1983fk}; \cite{Brown:1991fk} (note \ref{Brown:1991fk}).\label{Galison:1983fk}}
 Investigations showed that all the phenomena connected with the so-called soft component of cosmic radiation could be understood more or less quantitatively on this basis,  and that the laws of quantum theory were valid for electrons up to the highest energies known at the time.  The logic tradition was instrumental in dealing with large samples of events connected to a single kind of phenomenon.  And indeed, Bahbha and Heitler used the `Rossi transition curve'  to test the correctness of their theoretical results, and were able to make  quantitative estimates based on the statistical analysis of a great number of events. They concluded that `The comparison with experiments shows that Rossi's transition curve and Regener's absorption curve in the atmosphere can be understood on this theory'.

The agreement between the theoretical results and the experimental data was so good that at least a large fraction of the  cascade production of showers by high-energy electrons or photons was  considered to be one of the best-established facts about cosmic-ray phenomena. The core of the theory was the explanation of the activity regarding the soft part of cosmic rays, an unequivocal success of quantum electrodynamics. However, the hard component of the radiation was left unexplained.\footnote{Only after the war, in 1947,  would Cecil F. Powell's group, including Giuseppe Occhialini, make it clear that the mesotron of cosmic rays was actually the decay product of a new particle, the pion.C.F. Powell was awarded the Nobel Prize in Physics for this discovery in 1950. By that time there was a strong tradition of collaboration between Italian and British physicists, which became very important after World War II.}

On October 12, 1938,  Bruno Rossi and his wife Nora Lombroso left Italy after the enactment of fascist racist laws. They spent some time in Copenhagen with Niels Bohr.Bohr had always enabled refugee physicists to come to his institute since 1933,  and was constantly  searching  for positions for scientists fleeing the Nazis.\footnote{For a detailed reconstruction of Rossi's forced emigration from Italy see  \cite{Bonolis:2011fk}.\label{Bonolis:2011fk}).}  Despite the painful situation, in Bohr's Institute Rossi fully experienced   the international character of physics, and the deep concern of many physicists for the plight of the refugees from Nazi Germany and Fascist Italy (`The human interests, the lively intellectual climate, the sane vision of political events that were the essence of the `Spirit of Copenhagen'  went a long way toward clearing our minds, and strengthening our confidence  in the future')  [\dots]'.\footnote{\cite{Rossi:1990aa} (note \ref{Rossi:1990aa}), 40--41.} 
While he was in Copenhagen, Bohr called a meeting which was attended by many cosmic-ray physicists: `I could swear, recalled Rossi, that one of Bohr's motives was to give me the opportunity of meeting people who might help me find a job. In any case, this is exactly what happened, because shortly thereafter Patrick Blackett wrote, inviting me to Manchester.'

 Bruno and Nora Rossi moved to Great Britain the following December. In Manchester, where he remained until the early Summer of 1939, Rossi  was offered a  fellowship from the {\it Society for the Protection of Science and Learning} by Blackett, who had just succeeded W.L. Bragg as the Langworthy Professor of Physics.  Rossi's arrival in Manchester, where he found A.C.B. Lovell, G.D. Rochester, J.C. Wilson, and  the Hungarian physicist Lajos J\'anossy, was instrumental in triggering a brand-new programme of cosmic ray research, especially focused on the newly discovered mesotron.

During that period, Rossiformed a lifelong friendship with Blackett, and the two began to collaborate on research in particle physics. Both had a deep grasp of the essential theoretical concepts and elegance of experimental technique, which they conveyed in an article on the problem of the instability of the mesotron appearing on the December 3 issue of {\it Nature}  \cite{Rossi:1938rz}. This short article,  which saw Rossi again catching up with the forefront of research after a difficult period of forced `rest',\footnote{After moving to the University of Padua on the chair of Experimental Physics, Rossi became deeply involved in planning and supervising the construction and equipping of his new Physics Institute,  which he had to abandon a short time after its inauguration.} 
was his first explicit contribution to the problem of the mesotron's decay.  This topic would  later be at the core of his research programme once he moved to the U.S.A. 

In June 1939  Rossi and his wife Nora left for the United States, where they found Enrico Fermi.  Fermi had left Italy with his wife Laura Capon in December 1938, after having been awarded the Nobel Prize in Stockholm. Rossi had been invited by Compton to participate to an international congress on cosmic rays to be held in Chicago. 
This was the first major international conference on cosmic rays. It took
place from June 27--30, 1939, and some sixty active researchers from most of the
leading groups participated.
In total, about three hundred people attended the conference. The electromagnetic cascade theory had left completely open the problem of  the hard  component of cosmic rays. The enigma was solved by 
Anderson and Seth Neddermeyer using a Blackett-Occhialini cloud chamber, notwithstanding Millikan's well-known distrust of Geiger-M\"uller counters. In an article received on 30 March 1937, they suggested that   either positive and negative electrons `possess some property' according to which the standard Bethe-Heitler theory failed, or `there exist particles 
of unit charge but with a mass (which may not have a unique value) larger 
than that of a normal free electron and much smaller than that of a proton'.\footnote{\cite{Neddermeyer:1937fk}, 886. In a note added in proof they mentioned the `Excellent experimental evidence' for an analogous particle `obeying the Bethe-Heitler theory' reported by Street and Stevenson at the Meeting of the American Physical Society.} 
Soon after,  Jabez C. Street and Edward C. Stevenson, who used a three-counter telescope to select the penetrating particles triggering the cloud chamber and a lead filter for removing shower particles, had published the first photograph of a  mesotron \cite{Street:1937qf,Street:1937fk}.  This   again confirms how the Blackett-Occhialini hybrid combination and the know-how about $n$-fold coincidences was becoming an established practise:  `[\dots] for many people it  was the most impressive piece of evidence for the new particle'.\footnote{\cite{Galison:1983fk} (note \ref{Galison:1983fk}), 301--302.  In 1935 Street and Stevenson had used a cloud chamber triggered by two counters to show that `at least 90\% of the coincident counts for such an arrangement are directly due to the passage of single electrons through the apparatus, thus employing visual means to validate all results obtained up to that moment with coincidence circuits \cite[p. 643]{Stevenson:1935fk}}. They actually specifically challenged Anderson, Millikan, Neddermeyer, and Pickering's paper criticizing Blackett and Occhialini and especially Rossi's results built on coincidences due to the passage of a single particles through counters (\cite{Anderson:1934lr}).
From measurement of its ionization and momentum, Street and Stevenson deduced a value of about 130 electron masses, to be compared to the current accepted value 
of $105.65836715\pm0.0000038)$ MeV/$c^2$ \cite{Beringer:2012fk}.

The marriage between the visual and electronic tradition  had provided a `convincing' experimental proof   for distinguishing  the soft from the hard penetrating component of cosmic rays. The discovery of this new particle 
 was actually both an arrival and a departure for new physics.  From an 
analogy with the well-established  $\beta$-decay 
process of radioactive substances, it was assumed that cosmic-ray mesotrons  should be unstable, and disintegrate spontaneously each into an electron and a neutrino  \cite{Bhabha:1938uq}. For the first time physicists were facing such a phenomenon. And in fact, an entire day of the symposium was devoted to the problem of the radioactive instability of mesotrons. 
However, the participants at the Chicago meeting had agreed that there was no conclusive experimental evidence for such a decay, and Rossi recognized
 that to obtain such confirmation their absorption with increasing altitude in the
 atmosphere should be investigated.\footnote{Immediately after his arrival in the USA, Rossi had become  more and more involved in early research regarding cosmic-rays as a source of  `elementary particles'. By that time he was considered a leading figure in the field, and was asked to prepare  reviews on cosmic-ray problems of the moment  \cite{Rossi:1939zl,Rossi:1939fk,Rossi:1940pd,Rossi:1941la}.}

 Impressed with his analysis,  Compton invited Rossi and his wife to spend a few days with him and his wife Betty  in their summer cottage on Otsego Lake in Upper Michigan. 
Rossi explained his ideas more fully, and Compton became convinced that he should immediately organize an expedition to Mount Evans, in the Colorado Rocky Mountains.  As one of the tallest mountains in the United States, with its peak over 4,000 meters above sea level, Mount Evans would be an ideal site for such an experiment.  There was a road to its summit, where a small cabin had
 been built for earlier scientific research. Rossi was `taken aback,' since it was already the
 middle of July, and snow might begin to fall in the Colorado mountains by the end
of August. However, he could not but accept the challenge offered by Compton's proposal. By that time Rossi had accepted  a research associateship from
 the University of Chicago  paid for by the Committee
 in Aid of Displaced Foreign Scholars. In accepting this position Rossi had
made a final decision to remain in the United States.

Rossi built the three Geiger-M\"uller 
 counters and coincidence circuit he would need, and with the help of two of Compton's physicist friends, Norman Hilberry from New York University and Barton Hoag from the University of Chicago, loaded everything into an old
 bus that Compton borrowed from the Zoology Department. They left Chicago for Colorado on August 26, 1939, only a few days before Germany's invasion of Poland would set off World War II.

Now, for the first time, physicists were dealing with the phenomenon of spontaneous instability of an elementary particle. Establishing the reality of such a  process ---and the accurate determination of the mesotron's mean life---  became one of the outstanding problems of cosmic ray research, which Rossi had tackled since he had left Italy with his first article on this subject written with Blackett. 

After this first mountain experiment, between 1939 and 1941 the problem was solved by Rossi and his collaborators in three successive steps.  They performed a series of experiments which exemplify his style, and which remain  classics in the history of the logic tradition.  The experiments gave definitive proof of mesotron decay, demonstrated the relativistic dilation of the lifetime of mesotrons in flight, and culminated in the first precise measurement of the mean life of mesotrons at rest.\footnote{\cite{Rossi:1939lr,Rossi:1940gf,Rossi:1941wd,Rossi:1942ai,Rossi:1942dp}.}
In the last experiment, performed with Norris Nereson, they observed several hundred decays.  This made it possible  to plot   the first decay curve of an elementary particle ever measured, showing an exponential decay with a lifetime of about 2 $\mu$s. It  provided much firmer evidence of the reality of the process, since the only prior observation of mesotrons decaying at the end of their range  had been a couple of  cloud-chamber tracks photographed in 1940 by  Williams and  Roberts\footnote{ \cite{Williams:1940rt}. Rossi's new electronic circuit, the first of the time-measuring devices later known as  {\it time-to-amplitude} converters (TAC), was instrumental in providing the experiment with the necessary statistical accuracy.  The experimentrequired the measurement of time intervals between the discharges of the G-M counters ranging from a fraction of one to several $\mu$s.}

In the autumn of 1940, Rossi left the University of Chicago. Thanks to Hans Bethe, his old acquaintance since the early 1930s, Rossi was appointed to fill a vacant associate professorship at Cornell University in Ithaca, New York. There he began a brand new  career in the United States, becoming internationally renowned, attracting researchers, and training a new brilliant generation of physicists from all over the world.

\section{Conclusion}

 Notwithstanding the international tensions in the aftermath of World War I,  scientists gradually   tried to find ways to resume  relations, recovering the international practise interrupted by the war and by the ostracism against German and Austrian scientists.
In the case of physics, the emergence of an international community around  the fields of atomic physics and quantum mechanics certainly helped to re-establish official relationships.  In part this can be attributed to the preeminent contributions German physicists made to these fields, These relationships were also helped by the involvement of the new generation  of physicists born around the beginning of the 20th century, who generally began their university studies after World War I.
The beginnings of their scientific lives were strongly interrelated with an explosive development  in physics which saw a growing institutionalisation of the traditional internationalism of the discipline.  This internationalism was helped by the creation of IUPAP, and by its  action in promoting international cooperation and free circulation of scientists, information, and ideas.

It also appears  that this trend was influenced by the informal and unofficial contacts which became the most reasonable and natural way of rebuilding international science which had, in repudiation of the universalist ideals which had characterized  the end of the 19th century, been deliberately restricted after World War I. International scientific congresses, which are the primary and most explicit intersection between the national and international dimensions of scientists' activities, contributed to the establishment of personal ties and a general exchange of ideas. Conferences clearly had a role also in helping senior physicists to talk about their younger collaborators, and in deciding about future sojourns in other European centres. 
All this coincided with the reorganization of the Rockefeller Foundation, which, from the beginning of the 1930s, extended its longstanding attachment to science, increasingly committing itself to scientific research as a means of enabling human progress. The guiding concept of  `the advancement of knowledge' characterizing the Foundation also became the base for the extensive circulation of scientists, especially of young scientists.  This helped to promote a new era in international relationships, which promised the creation of a more established international  research network.  

The glimpse of the backstage of physics research during the `happy 1930s' provided by Bruno Rossi's and Enrico Fermi's personal interaction with several prominent physicists from Europe and United States ---and the influences of these contacts--- tells us a lot about the  life of the international community, and  reflects the personal and social aspects of science. In a period when quantum electrodynamics, nuclear physics, and cosmic ray studies   were still  strongly interlaced,  and  particle physics was beginning to take its first steps, exploring the scientific discussions and institutional movements of physicists can shed light on  a flourishing period of modern physics. 

Enrico Fermi played a central role in the rebirth of Italian physics, both at a national and international level, during the 1920sand1930s. Fermi was a major contributor to both theory and experiment. During the end of the 1910s and beginning of the 1920s, he was  one of the few Italian physicists aware of the importance discoveries being made at  an international level, and was anxious that his work not be restricted to Italy. On the other hand, his scientific personality and style of tackling problems starting from the phenomenology fascinated many young theoreticians,  and contributed to the development of a new attitude towards experimental results. 

At the same time, Fermi's influence helped to focus international attention on the new Italian scientific community linked to the developments in modern physics. He immediately understood the relevance of Rossi's research, which for some time certainly was at the forefront of physics in Italy, and asked him to give the introductory talk at the session dedicated to cosmic rays in the Conference of nuclear physics organised in Rome in 1931. Having already established some relationships with members of the international scientific community during his stay in Berlin, Rossi now saw his work fully recognised at a higher level, having the occasion of being trusted by Arthur Compton, and of influencing his view on the problem of cosmic rays which led to the world wide campaign studying the latitude effect. On the same occasion he could reinforce  his familiarity with Heisenberg, directly contributing to his reflections on quantum electrodynamics and the high energy processes, which led to Heisenberg's first article on cosmic rays. Rossi probably benefited from Heisenberg's appreciation, becoming more self-confident about his results, which at the moment were contested by Millikan and his collaborators. Convinced of the soundness of Rossi's experiments, Heisenberg  favoured the publication of the results on the secondary shower experiments which otherwise would have been delayed, or even impeded. Heisenberg's bonds with the Italian physicists continued to be  very strong after World War II, when European physicists began to reorganise their activities.

Through Heisenberg, information about Rossi's experiments was certainly disseminated to other members of the scientific community, notably Bohr. The interest of other  theoreticians in the  implications of the study of cosmic rays also had a fundamental role in creating a brand new relationship between theory and experiment.  This new relationship had not been favoured by scientists of the previous generation, including Marie Curie and Ernest Rutherford, who had a well known suspicious attitude towards `theorizing.' 
On the other hand, Bruno Rossi himself had a good theoretical education, which facilitated his interactions with theoreticians like Heisenberg, Bethe, and Bhabha.  For his own part, Bothe's extraordinary experience and skill as an experimenter made him immediately aware of Ross's capabilities, even if the latter was very young when he arrived in Berlin in the summer 1930. This sojourn, during which Bothe fully shared his knowledge with Rossi and discussed the problem of cosmic rays, was an unforgettable experience for the Italian physicist. On the other side, the  elegance and rigour of Rossi's work, which later aroused the interest of  physicists like  Blackett, Bothe, Leprince-Ringuet, and Auger, as well as  U. S. physicists like Johnson, Street, and Stevenson, directly or indirectly contributed to the spread of electronic  methods in the study of cosmic rays and nuclear physics. 

Incidentally, both Fermi and Rossi had a strong impact on Italian physics, and both contributed to the development of a tradition whose main characteristics can still be traced to their influence today.\footnote{The well-known experiment of Marcello Conversi, Ettore Pancini, Oreste Piccioni \cite{Conversi:1947uq}, carried out in Rome, and establishing  that the cosmic ray mesotron did not behave in agreement with   the theory that identified it with the strongly interacting meson hypothesized by Hideki Yukawa, but, rather, as a weakly interacting particle, is part and parcel of Rossi's legacy. In Luis W. Alvarez' opinion, it was the beginning of `modern particle physics' (L. W. Alvarez, Nobel Lecture, December 11, 1968, available at http://nobelprize.org/nobel\_prizes/physics/laureates/1968/alvarez-lecture.html).  For a general discussion on the observation of the spontaneous decay of mesotrons see \cite{Monaldi:2005qy}.}

From Rossi's experience, it is clear  that the personal contacts he established during conferences and visits abroad not only strengthened pre-existing ties, but had a fundamental role in providing a most favourable occasion for planning his ---and his collaborators'--- sojourns in other European laboratories. His  stay in Paris not only introduced coincidence methods in France, but opened a new research perspective for Leprince-Ringuet and Auger whose importance cannot be underestimated. The agreement with Blackett, which led to the latter's collaboration with Rossi's brilliant student Giuseppe Occhialini, was instrumental for introducing  GM counters and the electronic coincidence technique to the Cavendish Laboratory. The successive development of the new experimental technique of the cloud chamber triggered by a coincidence circuit represented a milestone in physics detecting technology , and contributed greatly to the theoretical understanding of cosmic ray phenomenology and to the early steps of particle physics. 

The Blackett-Occhialini technique not only created a new tradition, but helped to clarify how previous studies performed with only the use of counters and electronic devices had a firm experimental base. The electronic and the visual tradition continued to develop and to intermingle depending on personal preferences or experimental needs. Rossi, for example, stuck to the logic tradition for his entire scientific life, up to the pioneering experiments of the early 1960s, which confirmed the existence of the solar wind and led to the discovery of the first extra-solar source of X-rays.\footnote{These experiments are discussed in the article mentioned in note \ref{astroparticlejournal}.}

As a pioneer of cosmic rays, Rossi made a great contribution to the creation of a new scientific community which set the basis for the development of modern particle physicsRossi remained a leading figure in the field until the early 1950s, when accelerators became the main source of high energy particles. The first period of his scientific life in Italy, during which he became a well-known and appreciated member of the international scientific community, came abruptly to an end in 1938, when he was dismissed from his Chair of Experimental Physics in Padua following Mussolini's racist laws. The ties established from 1930 until 1938 proved to be of vital importance when he was obliged to leave his country, as widely testified by the correspondence of the period.\footnote{\cite{Bonolis:2011fk} (note \ref{Bonolis:2011fk}).} At that time, a new wave of forced migration in Europe nearly destroyed the young generation of physicists in Italy, and reshaped the landscape of world physics. The solidarity of many members of the international scientific community gave Rossi the courage to begin a new life in the United States, where he could continue  research and teaching. This solidarity, which owed much to the commitment of some members of the scientific community like Niels Bohr,  is evidence of a changed attitude in  mutual relations within the physicists' community. 

Only about ten years had elapsed since when German and Austrian scientists were still excluded from international meetings.  Rossi himself described his  personal feelings about  the relationships between Italy and the international community:\footnote{Footnote \ref{Brown-Hoddeson:1983uq}, 286} 
 `I really did not feel isolated at all in Italy. For one thing, I was working in Florence, and Florence is very close to Rome, and in Rome there was Fermi and his group, with whom I had fairly frequent contact. Then I was traveling abroad. There were meetings in Switzerland very often, and I spent one summer in Germany. I felt a part of the European team, not of the italian team [\dots]'  We can imagine that very similar feelings were shared by many other members of the scientific community.   
 
 After the war Rossi became an influential figure on the world stage, and he was able to make important  contributions to the great developments of post-war physics.  This was thanks in part to the strong personal ties he continued to cultivate with his friends on the other side of the Ocean. He was very active in inviting young researchers from all over the world to work in his laboratory at MIT, thus providing a most important opportunity of high-level training for the post-war generation. The presence of many European physicists in the United States before the war certainly contributed to the strengthening of world-wide ties.

 Extensive studies in different national contexts might provide new keys for investigating  the dynamics of scientific research. One of the most natural results  would be, for example, a better understanding of the mechanisms underlying the rebirth of the European scientific community after the war, which clearly benefited from relationships similar to those outlined here.

At that time, European scientists became aware of the continuously increasing gap between the means available in Europe to the various fields of science and those available in the United States. It was evident that the gap was even greater with respect to the applied sciences and technology. It also became evident to a certain number of scientists that this situation could be changed only by a common effort made by many European countries.
Some European physicists like Edoardo Amaldi, a member of the first nucleus of Fermi's group in Rome, had during the war developed a strong awareness about the importance of resisting the temptation to find a position in the United States.  Amaldi decided to remain in Italy and tackle a difficult and dramatic situation. This feeling, together with his  longstanding acquaintance with prominent physicists all over the world, became the basis of his action when he became a leader of Italian physics and a promoter of the reconstruction of European science. In particular, Amaldi was one of the founding fathers of CERN.\footnote{The  extensive correspondence with physicists of the whole world preserved in Edoardo Amaldi's Archive in the Physics Department of Sapienza University in Rome, is in itself a  gold mine for investigating on similar related issues.}

 The relationships established before the war clearly represented the premise and the basis for the rapid reorganization and the flourishing of post-war physics in Europe.  European physicists became aware that the different tasks of reconstruction and of competition with the powerful  community of U.S. physicists could be tackled mainly by organizing in international groups.  This  happened with the famous  G-Stack experiment of 1954,\cite{Olivotto:2009fk}  which saw several European groups collaborating in the launch of a giant stack of emulsions, the final of a series which became the training ground for later similar enterprises. Such activities opened the way to  more extensive international cooperation within the newly established laboratory of CERN at Geneva, created in the early 1950s to foster pure science and international collaboration between people of different countries, traditions, and mentalities.

\section{Acknowledgments}

I am very thankful to Helmut Rechenberg for the stimulating and fruitful exchange of ideas on these issues during the last years. I would also like to thank Michael Eckert, Francesco Guerra, Dieter Hoffmann, Matteo Leone, Pietro Nastasi, and Nadia Robotti  for various interesting discussions on several related topics. I am  particularly grateful to the MIT Institute Archives \& Special Collections in Cambridge M.A., to Wilhelm F\"ussl of the Deutsches Museum Archiv in Munich, and to the Archives of the Max-Planck-Gesellschaft in Berlin for their assistance.  Special thanks go to Renaud Huynh and Natalie Pigeard of the historical Archives of the Institut Curie, to Olivier Azzola of the Archives of the \'Ecole Polyt\'echnique, and to Florence Greffe of the Archive of the Acad\'emie des Sciences in Paris. Last but not least I also thank Giovanni Battimelli and Giovanni Paoloni for many useful conversations  on various problems regarding  historical archives and the history of 20th century physics. Many thanks to Sergio De Benedetti's family for providing his unpublished autobiographical notes. I am also quite grateful  to  two anonymous referees for their relevant comments and helpful criticism. Last but not least, I owe a special debt to an anonymous editor for a careful and thoughtful editorial work on the paper. The final version of this work was performed at Max Planck Institute for the History of Science in Berlin.

\bibliographystyle{plain}

\bibliography{BiblioRossiAnnals} 

\begin{thebibliography}{100}

\bibitem{Alvarez:1933yq}
L.~Alvarez and A.~H. Compton.
\newblock {A Positively Charged Component of Cosmic Rays}.
\newblock {\em Physical Review}, 43(10):835--836, May 1933.

\bibitem{Anderson:1934lr}
C.~D. Anderson, R.~A. Millikan, S.~Neddermeyer, and W.~Pickering.
\newblock {The Mechanism of Cosmic-Ray Counter Action}.
\newblock {\em Physical Review}, 45(6):352--363, 1934.

\bibitem{Anderson:1932fk}
Carl~D. Anderson.
\newblock {The Apparent Existence of Easily Deflectable Positives}.
\newblock {\em Science}, 76(1967):238--239, 1932.

\bibitem{Auger:1936qf}
P.~Auger.
\newblock Analyse des gerbes de rayons cosmiques par l'utilisation de leur
  divergence.
\newblock {\em Comptes Rendus de l'Acad\'emie des Sciences}, 203:246--248,
  1936.

\bibitem{Auger:1936bh}
P.~Auger.
\newblock \'etude sur les gerbes cosmiques en haute altitude.
\newblock {\em Comptes Rendus de l'Acad\'emie des Sciences}, 203:1082--1084,
  1936.

\bibitem{Auger:1938fr}
P.~Auger.
\newblock Les grandes gerbes cosmiques de l'atmosph\`ere.
\newblock {\em Comptes Rendus de l'Acad\'emie des Sciences}, 207:228--230,
  1938.

\bibitem{Auger:1935fk}
P.~Auger and P.~Ehrenfest.
\newblock Clich\'es de rayons cosmiques obtenus avec une chambre de
  wilson-blackett dans des conditions sp\'eciales.
\newblock {\em Journal de Physique Radium}, 6(6):255--256, June 1935.

\bibitem{Auger:1937fp}
P.~Auger, P.~Ehrenfest, A.~Freon, and A.~Fournier.
\newblock {Sur la distribution angulaire des rayons corpusculaires cosmiques
  durs}.
\newblock {\em Comptes Rendus de l'Acad\'emie des Sciences}, 204:257--259,
  January 25 1937.

\bibitem{Auger:1937cr}
P.~Auger, P.~Ehrenfest, A.~Freon, and T.~Grivet.
\newblock M\'ecanisme de production des gerbes cosmiques.
\newblock {\em Comptes Rendus de l'Acad\'emie des Sciences}, 204:1797--1799,
  1937.

\bibitem{Auger:1933ys}
P.~Auger and L.~Leprince-Ringuet.
\newblock \'etude de la variation du rayonnement cosmique entre les latitudes
  45$\,^{\circ}$ nord et 38$\,^{\circ}$ sud.
\newblock {\em Comptes Rendus de l'Acad\'emie des Sciences}, 197:1242--1244,
  1933.

\bibitem{Auger:1935kx}
P.~Auger, L.~Leprince-Ringuet, and P.~Ehrenfest.
\newblock Absorption de la fraction molle du rayonnement corpusculaire
  cosmique.
\newblock {\em Comptes Rendus de l'Acad\'emie des Sciences}, 200:1747--1749,
  May 1935.

\bibitem{Auger:1936uq}
P.~Auger, L.~Leprince-Ringuet, and P.~Ehrenfest.
\newblock Analyse du rayonnement cosmique \`a l'altitude de 3 500 m\'etres.
\newblock {\em Journal de Physique Radium}, 7(2):58--64, February 1936.

\bibitem{Auger:1938rt}
P.~Auger, R.~Maze, and T.~Grivet-Meyer.
\newblock Grandes gerbes cosmiques atmosph\`eriques contenant des corpuscules
  ultrap\`enetrants.
\newblock {\em Comptes Rendus de l'Acad\'emie des Sciences}, 206:1721--1723,
  1938.

\bibitem{Auger:1935uq}
P.~Auger and A.~Rosenberg.
\newblock Sur les effets secondaires des rayons cosmiques.
\newblock {\em Comptes Rendus de l'Acad\'emie des Sciences}, 200:447--449,
  February 1935.

\bibitem{Auger:1936dq}
P.~Auger and A.~Rosenberg.
\newblock Sur les propri\'et\'es des corpuscules cosmiques du groupe
  p\'en\'etrant.
\newblock {\em Comptes Rendus de l'Acad\'emie des Sciences}, 202:1923--1925,
  1936.

\bibitem{Auger:1929fr}
P.~Auger and D.~V. Skobeltzyn.
\newblock Sur la nature des rayons ultrap\'en\'etrants (rayons cosmiques).
\newblock {\em Comptes Rendus de l'Acad\'emie des Sciences}, 189:55--57, July 1
  1929.

\bibitem{Auger:1934kx}
P.~V. Auger and L.~Leprince-Ringuet.
\newblock {\'Etude par la m\'ethode des coincidences de la variation du
  rayonnement cosmique suivant la latitude}.
\newblock {\em J. Phys. Radium}, 5(5):193--198, 1934.

\bibitem{Beringer:2012fk}
J.~Beringer and et~al. (Particle Data~Group).
\newblock {Review of Particle Physics}.
\newblock {\em Physical Review}, D86(1):010001, 2012.

\bibitem{Bethe:1930aa}
H.~Bethe.
\newblock {Zur Theorie des Durchgangs schneller Korpuskularstrahlen durch
  Materie}.
\newblock {\em Annalen der Physik}, 397(3):325--400, 1930.

\bibitem{Bethe:1932aa}
H.~Bethe and E.~Fermi.
\newblock \"uber die wechselwirkung von zwei elektronen.
\newblock {\em Zeitschrift f{\"u}r Physik}, 77:296--306, 1932.

\bibitem{Bethe:1934fj}
H.~Bethe and W.~Heitler.
\newblock {On the Stopping of Fast Particles and on the Creation of Positive
  Electrons}.
\newblock {\em Proceedings of the Royal Society of London A}, 146(856):83--112,
  Aug 1 1934.

\bibitem{Bethe:2002fk}
H.~A. Bethe and H.~Bethe.
\newblock Enrico fermi in rome, 1931--32.
\newblock {\em Physics Today}, 55(6):28--29, June 2002.

\bibitem{Bhabha:1933fj}
H.~J. Bhabha.
\newblock {Zur Absorption der H\"ohenstrahlung}.
\newblock {\em Zeitschrift f\"ur Physik}, 86(1):120--130, 1933.

\bibitem{Bhabha:1938uq}
H.~J. Bhabha.
\newblock {Nuclear Forces, Heavy Electrons and the $\beta$-Decay}.
\newblock {\em Nature}, 141(7278):118, 1938.

\bibitem{Bhabha:1936uq}
H.~J. Bhabha and W.~Heitler.
\newblock {Passage of Fast Electrons through Matter}.
\newblock {\em Nature}, 138(3488):401, 5 Sep 1936.

\bibitem{Bhabha:1937yq}
H.~J. Bhabha and W.~Heitler.
\newblock {The Passage of Fast Electrons and the Theory of Cosmic Showers}.
\newblock {\em Proceedings of the Royal Society of London A},
  159(898):432--458, 1937.

\bibitem{Blackett:1969fk}
P.~M.~S. Blackett.
\newblock {The Old Days of the Cavendish}.
\newblock {\em Rivista del Nuovo Cimento}, I(Special Issue):xxxii--xxxix, 1969.

\bibitem{Blackett:1933rt}
P.~M.~S. Blackett, J.~Chadwick, and G.~P.~S. Occhialini.
\newblock {New Evidence for the Positive Electron}.
\newblock {\em Nature}, 131(3309):473--473, April 1 1933.

\bibitem{Blackett:1932mz}
P.~M.~S. Blackett and G.~P.~S. Occhialini.
\newblock {Photography of Penetrating Corpuscular Radiation}.
\newblock {\em Nature}, 130(3279):363--363, Sep 3 1932.

\bibitem{Blackett:1933fj}
P.~M.~S. Blackett and G.~P.~S. Occhialini.
\newblock {Some Photographs of the Tracks of Penetrating Radiation}.
\newblock {\em Proceedings of the Royal Society of London A},
  139(839):699--726, March 3 1933.

\bibitem{Bonolis:2011fk}
L.~Bonolis.
\newblock {Bruno Rossi and the racial laws of fascist Italy}.
\newblock {\em Physics in Perspective}, 13(1):58--90, 2011.

\bibitem{Bonolis:2011ak}
L.~Bonolis.
\newblock Walther bothe and bruno rossi: The birth and development of
  coincidence methods in cosmic-ray physics.
\newblock {\em American Journal of Physics}, 79(11):1133--1150, 2011.

\bibitem{Bothe:1930lr}
W.~Bothe.
\newblock {Zur Vereinfachung von Koinzidenzz\"ahlungen}.
\newblock {\em Zeitschrift f{\"u}r Physik}, 59(1):1--5, Jan 1930.

\bibitem{Bothe:1932fk}
W.~Bothe.
\newblock {Bemerkungen \"uber die Ultra-Korpuskularstrahlung}.
\newblock In {\em {Convegno di Fisica nucleare}}, pages 153--154. Reale
  Accademia d'Italia, 1932.

\bibitem{Bothe:1928vn}
W.~Bothe and W.~Kolh{\"o}rster.
\newblock {Eine neue Methode f\"ur Absorptionsmessungen an sekund\"aren
  $\beta$-Strahlen}.
\newblock {\em Die Naturwissenschaften}, 16(49):1045, 1928.

\bibitem{Bothe:1929aa}
W.~Bothe and W.~Kolh\"orster.
\newblock {Das Wesen der H\"ohenstrahlung}.
\newblock {\em Zeitschrift f{\"u}r Physik}, 56(11-12):751--777, Nov 1929.

\bibitem{Bothe:1929fk}
W.~Bothe and W.~Kolh\"orster.
\newblock {The nature of the Penetrating Radiation}.
\newblock {\em Nature}, 123(3104):638--638, April 27 1929.

\bibitem{Bothe:1930uq}
W.~Bothe and W.~Kolh\"orster.
\newblock {Vergleichende H\"ohenstrahlungsmessungen auf n\"ordlichen Meeren}.
\newblock {\em Sitzber. Preuss. Akad.}, 26:450--456, 1930.

\bibitem{Brown-Hoddeson:1983uq}
L.~M. Brown and L.~Hoddeson, editors.
\newblock {\em {The Birth of Particle Physics}}.
\newblock Cambridge University Press, New York, 1983.

\bibitem{Brown:1991fk}
L.~M. Brown and H~Rechenberg.
\newblock Quantum field theories, nuclear forces, and the cosmic rays
  (1934--1938).
\newblock {\em American Journal of Physics}, 59(595--605), 1991.

\bibitem{Bustamante:1994aa}
M.~C. Bustamante.
\newblock {Bruno Rossi} au d\'ebut des ann\`ees trente: une \'etape d\'ecisive
  dans la physique des rayons cosmiques.
\newblock {\em Archives Internationales d'Histoire des Sciences}, 44:92--115,
  1994.

\bibitem{Bustamante:2007kx}
M.~C. Bustamante.
\newblock {Giuseppe Occhialini and the history of cosmic-ray physics in the
  1930s: From Florence to Cambridge}.
\newblock In P.~Redondi, G.~Sironi, P.~Tucci, and G.~Vegni, editors, {\em {The
  scientific legacy of Beppo Occhialini}}, pages 35--49, Bologna, 2007.
  Societ\`{a} Italiana di Fisica.

\bibitem{Bustamante:2010zr}
M.~C. Bustamante.
\newblock {Cosmic rays in France during the thirties: Pierre Auger and Louis
  Leprince-Ringuet}.
\newblock In A.~Troper, A.~A.~P. Videira, and C.~Leite~Vieira, editors, {\em
  {Os 60 anos do CBPF e a G\^enese do CNPq}}, pages 95--115, Rio de Janeiro,
  2010. CBPF.

\bibitem{Carlson:1937kx}
J.~F. Carlson and J.~R. Oppenheimer.
\newblock {On Multiplicative Showers}.
\newblock {\em Physical Review}, 51(4):220--231, Feb 1937.

\bibitem{Carlson:2011uq}
P.~Carlson and A.~De~Angelis.
\newblock Nationalism and internationalism in science: The case of the
  discovery of cosmic rays.
\newblock {\em European Physical Journal H}, 35(4):309--329, 2011.

\bibitem{Chadwick:1932fj}
J.~Chadwick.
\newblock {Possible Existence of a Neutron}.
\newblock {\em Nature}, 129(3252):312, Feb. 27 1932.

\bibitem{Chadwick:1932kx}
J.~Chadwick.
\newblock {The Existence of a Neutron}.
\newblock {\em Proceedings of the Royal Society of London}, 136(830):692--708,
  1932.

\bibitem{Clark:2000fk}
G.~W. Clark.
\newblock {Bruno Benedetto Rossi, 13 April 1925. 21 November 1993}.
\newblock {\em Proceedings of the American Philosophical Society},
  144(3):329--341, Sep 2000.

\bibitem{Clay:1927fk}
J.~Clay.
\newblock {Penetrating Radiation I}.
\newblock {\em Proceedings of the Royal Academy of Sciences Amsterdam},
  30:1115--1127, 1927.

\bibitem{Clay:1928fk}
J.~Clay.
\newblock {Penetrating Radiation II}.
\newblock {\em Proceedings of the Royal Academy of Sciences Amsterdam},
  31:1091--1097, 1928.

\bibitem{Compton:1932kx}
Arthur~H. Compton.
\newblock Progress of cosmic-ray survey.
\newblock {\em Physical Review}, 41(5):681--682, Sep 1932.

\bibitem{Compton:1933vn}
Arthur~H. Compton.
\newblock {A Geographic Study of Cosmic Rays}.
\newblock {\em Physical Review}, 43(6):387--403, Mar 1933.

\bibitem{Conversi:1947uq}
M.~Conversi, E.~Pancini, and O.~Piccioni.
\newblock {On the Disintegration of Negative Mesons}.
\newblock {\em Physical Review}, 71(3):209--210, Feb 1947.

\bibitem{Curie:1933ys}
I.~Curie and F.~Joliot.
\newblock Contribution \`a l'\'etude des \`electrons positifs.
\newblock {\em Comptes Rendus de l'Acad\'emie des Sciences}, 196:1105--1107,
  Apr 1933.

\bibitem{Curie:1933vn}
I.~Curie and F.~Joliot.
\newblock \'electrons de mat{\'e}rialisation et de transmutation.
\newblock {\em J. Phys. Radium}, 4(8):494--500, 1933.

\bibitem{Curie:1933yt}
I.~Curie and F.~Joliot.
\newblock \'electrons positifs de transmutation.
\newblock {\em Comptes Rendus de l'Acad\'emie des Sciences}, 196:1885--1887,
  1933.

\bibitem{Curie:1933kx}
I.~Curie and F.~Joliot.
\newblock Recherches sur le rayonnement ultrap\'en\'etrant \`a la station
  scientifique du jungfraujoch.
\newblock {\em J. Phys. Radium}, 4(8):492--493, 1933.

\bibitem{Curie:1933zr}
I.~Curie and F.~Joliot.
\newblock Sur l'origine des \'electrons positifs.
\newblock {\em Comptes Rendus de l'Acad\'emie des Sciences}, 196:1581--1583,
  May 1933.

\bibitem{De-Benedetti:1935ve}
S.~De~Benedetti.
\newblock Production de positrons dans diff\`erents \'el\'ements.
\newblock {\em Comptes Rendus de l'Acad\'emie des Sciences}, 200:1389--1391,
  1935.

\bibitem{De-Benedetti:1936vn}
S.~De~Benedetti.
\newblock Recherches sur l'\'emission des positons.
\newblock {\em Journal de Physique Radium}, 7(5):205--210, 1936.

\bibitem{De-Benedetti:1936ys}
S.~De~Benedetti.
\newblock Sur l'\'emission de positons par une source de thb+c.
\newblock {\em Comptes Rendus de l'Acad\'emie des Sciences}, 202:50--52, 1936.

\bibitem{De-Benedetti:1941bh}
Sergio De~Benedetti.
\newblock Simultaneous emission of particles and pair production.
\newblock {\em Physical Review}, 59(5):463--463, 1941.

\bibitem{De-Maria:1985vn}
M.~De~Maria, M.~G. Ianniello, and A.~Russo.
\newblock {The Discovery of Cosmic Rays: Rivalries and Controversies Between
  Europe and the United States}.
\newblock {\em Rivista di Storia della Scienza}, 2:237--286, 1985.

\bibitem{De-Maria:1989aa}
M.~De~Maria and A.~Russo.
\newblock Cosmic ray romancing: the discovery of the latitude effect and the
  {Compton--Millikan} controversy.
\newblock {\em Historical Studies in the Physical and Biological Sciences},
  19:211--266, 1989.

\bibitem{Fermi:1934uq}
E.~{Fermi}.
\newblock {Radioattivit{\`a} indotta da bombardamento di neutroni}.
\newblock {\em La Ricerca Scientifica}, 5(1):283, 1934.

\bibitem{Fermi:1934qy}
E.~Fermi.
\newblock Versuch einer theorie der $\beta$-strahlen. i.
\newblock {\em Zeitschrift f\"ur Physik}, 88(3):161--177, 1934.

\bibitem{Fermi:1933kx}
E.~Fermi and F.~Rasetti.
\newblock Uno spettrografo per raggi ``gamma'' a cristallo di bismuto.
\newblock {\em La Ricerca Scientifica}, 4((2)):299--302, 1933.

\bibitem{Fermi:1933fk}
E.~Fermi and G.~E. Uhlenbeck.
\newblock On the recombination of electrons and positrons.
\newblock {\em Physical Review}, 44(6):510--511, 09 1933.

\bibitem{Fermi:1932lr}
Enrico Fermi.
\newblock {Quantum Theory of Radiation}.
\newblock {\em Reviews of Modern Physics}, 4(1):87--132, 1932.

\bibitem{Fick:2009fk}
D.~Fick and H.~Kant.
\newblock Walther bothe's contributions to the understanding of the
  wave-particle duality of light.
\newblock {\em Studies In History and Philosophy of Science Part B: Studies In
  History and Philosophy of Modern Physics}, 40(4):395--405, 2009.

\bibitem{Galdabini:1989lr}
S.~Galdabini and G.~Giuliani.
\newblock {Physics in Italy between 1900 and 1940: The universities,
  physicists, funds, and research}.
\newblock {\em Historical Studies in Physical Sciences}, 19:115--136, 1989.

\bibitem{Galison:1983fk}
P.~Galison.
\newblock The discovery of the muon and the failed revolution against quantum
  electrodynamics.
\newblock {\em Centaurus}, 26:262--316, 1983.

\bibitem{Galison:1987kx}
P.~Galison.
\newblock {\em {How Experiments End}}.
\newblock The University of Chicago Press, Chicago, 1987.

\bibitem{Galison:1997rm}
P.~Galison.
\newblock {\em {Image and Logic. A Material Culture of Microphysics}}.
\newblock University of Chicago Press, Chicago and London, 1997.

\bibitem{Geiger:1928kx}
H.~Geiger and W.~M\"uller.
\newblock {Elektronenz\"ahlrohr zur Messung schw\"achster Aktivit\"aten}.
\newblock {\em Die Naturwissenschaften}, 16(31):617--618, Aug 1928.

\bibitem{Gembillo:1993lr}
G.~Gembillo.
\newblock Un carteggio inedito tra {Werner K. Heisenberg} e {Bruno Rossi}.
\newblock {\em Scienza e Storia}, 9:113--122, 1993.

\bibitem{Goodstein:2001fk}
J.~R. Goodstein.
\newblock {A conversation with Franco Rasetti}.
\newblock {\em Physics in Perspective}, 3:271--313, 2001.

\bibitem{Goudsmit:1932fk}
S.A. Goudsmit.
\newblock {Present difficulties in the theory of hyperfine structure}.
\newblock In {\em {Convegno di Fisica nucleare}}, pages 33--49, Rome, 1932.
  Reale Accademia d'Italia.

\bibitem{Greinacher:1924lr}
H.~Greinacher.
\newblock {\"Uber die akustische Beobachtung und galvanometrische Registrierung
  von Elementarstrahlen und Einzelionen}.
\newblock {\em Zeitschrift f{\"u}r Physik}, 23(1):361--378, 1924.

\bibitem{Greinacher:1926lr}
H.~Greinacher.
\newblock {Eine neue Methode zur Messung der Elementarstrahlen}.
\newblock {\em Zeitschrift f{\"u}r Physik}, 36(5):364--373, 1926.

\bibitem{Guerra:2011uq}
F.~Guerra, M.~Leone, and N.~Robotti.
\newblock {The Discovery of Artificial Radioactivity}.
\newblock {\em Physics in Perspective}, 14:33--58, 2012.

\bibitem{Guerra:2009fk}
F.~Guerra and N.~Robotti.
\newblock {Enrico Fermi's Discovery of Neutron-Induced Artificial
  Radioactivity: The Influence of His Theory of Beta Decay}.
\newblock {\em Physics in Perspective}, 11:379--404, 2009.

\bibitem{Heisenberg:1932ys}
W.~{Heisenberg}.
\newblock {Theoretische {\"U}berlegungen zur H{\"o}henstrahlung}.
\newblock {\em Annalen der Physik}, 405:430--452, 1932.

\bibitem{Heisenberg:1993qy}
W.~Heisenberg.
\newblock {\em {Gesammelte Werke/Collected Works. Series A/Part II}}.
\newblock Springer, 1993.

\bibitem{Hoffmann:1927fj}
G.~Hoffmann and E.~Steinke.
\newblock {Die Maximalh\"arte der Hessschen Ultra-$\gamma$-Strahlung}.
\newblock {\em Die Naturwissenschaften}, 15(51):995--995, Dec. 1927.

\bibitem{Johnson:1933fj}
T.~H. Johnson.
\newblock {The Azimuthal Asymmetry of the Cosmic Radiation}.
\newblock {\em Physical Review}, 43(10):834--835, May 1933.

\bibitem{Johnson:1932qv}
T.~H. Johnson, Fleischer, and J.~C. Street.
\newblock {Minutes of the Washington Meeting April 28-30, 1932}.
\newblock {\em Physical Review.}, 40(6):1024--1057, Jun 1932.

\bibitem{Johnson:1932uq}
Thomas~H. Johnson and J.~C. Street.
\newblock {The Production of Multiple Secondaries in Lead by Cosmic Radiation}.
\newblock {\em Physical Review}, 40(4):638--639, May 1932.

\bibitem{Joliot:1933fk}
F.~Joliot.
\newblock Preuve exp\'erimentale de l'annihilation des \'electrons positifs.
\newblock {\em Comptes Rendus de l'Acad\'emie des Sciences}, 197:1622--1625,
  1933.

\bibitem{Joliot:1934qf}
F.~Joliot.
\newblock \'etude des rayons de recul radioactifs par la m\'ethode des
  d\'etentes de wilson.
\newblock {\em Journal de Physique Radium}, 5(5):219--224, May 1934.

\bibitem{Joliot:1934uq}
F.~Joliot.
\newblock Sur la d\'emat\'erialisation de paires d'\'electrons.
\newblock {\em Comptes Rendus de l'Acad\'emie des Sciences}, 198:81--83, 1934.

\bibitem{Joliot:1934fk}
F.~Joliot and I.~Curie.
\newblock {Artificial Production of a New Kind of Radio-Element}.
\newblock {\em Nature}, 133(3354):201--202, February 10 1934.

\bibitem{Kampert:2012fk}
K.-H. Kampert and A.A. Watson.
\newblock {Extensive Air Showers and Ultra High-Energy Cosmic Rays: A
  Historical Review}.
\newblock {\em European Physics Journal H}, 37(3):359--412, 2012.

\bibitem{Klein:1929fj}
O.~Klein and T.~Nishina.
\newblock {\"Uber die Streuung von Strahlung durch freie Elektronen nach der
  neuen relativistischen Quantendynamik von Dirac}.
\newblock {\em Zeitschrift f{\"u}r Physik}, 52(11):853--868, 1929.

\bibitem{Kragh:1992uq}
H.~Kragh.
\newblock {Relativistic Collisions: The Work of Christian M\o ller in the Early
  1930s}.
\newblock {\em Archives for History of Exact Sciences}, 43(4):299--328, 1992.

\bibitem{Lemaitre:1933qy}
G.~Lema\^itre and M.~S. Vallarta.
\newblock {On Compton's Latitude Effect of Cosmic Radiation}.
\newblock {\em Physical Review}, 43(2):87--91, 1933.

\bibitem{Leone:2011kx}
M.~Leone.
\newblock Particles that take photographs of themselves: The emergence of the
  triggered cloud chamber technique in early 1930s cosmic-ray physics.
\newblock {\em American Journal of Physics}, 79:454--460, 2011.

\bibitem{Leone:2005aa}
M.~Leone, A.~Mastroianni, and N.~Robotti.
\newblock {Bruno Rossi and the Introduction of the Geiger--M{\"u}ller Counter
  in Italian Physics: 1929--1934}.
\newblock {\em Physis}, XLII(2):453--480, 2005.

\bibitem{Leone:2008fk}
M.~Leone and N.~Robotti.
\newblock {P. M. S. Blackett, G. Occhialini and the invention of the
  counter-controlled cloud chamber (1931--32)}.
\newblock {\em European Journal of Physics}, 29(2):177--189, 2008.

\bibitem{Leone:2010qf}
M.~Leone and N.~Robotti.
\newblock {Fre\'ed\'eric Joliot, Ir\`ene Curie and the early history of the
  positron (1932--33)}.
\newblock {\em Eur. J. Phys.}, 31:975--987, 2010.

\bibitem{Leprince-Ringuet:1931fk}
L.~Leprince-Ringuet.
\newblock L'amplificateur \`a lampes et la d\'etection des rayonnements
  corpusculaires isol\'es.
\newblock {\em Annales des Postes T\'el\'egraphes et T\'elephones}, pages
  480--492, Jun 1931.

\bibitem{Leprince-Ringuet:1931lr}
L.~Leprince-Ringuet.
\newblock {Relation entre le parcours d'un proton rapide dans l'air et
  l'ionisation qu'il produit. Application \`a l'\'etude de la
  d\'esint\'egration artificielle des \'el\'ements}.
\newblock {\em Comptes Rendus de l'Acad\'emie des Sciences}, 192:1543--1545,
  Jun 15 1931.

\bibitem{Leprince-Ringuet:1935zr}
L.~Leprince-Ringuet.
\newblock Sur le signe et la nature des particules ultra-p\'en\'etrantes du
  rayonnement cosmique.
\newblock {\em Comptes Rendus de l'Acad\'emie des Sciences}, 201:1184--1186,
  1935.

\bibitem{Leprince-Ringuet:1935vn}
L.~Leprince-Ringuet.
\newblock Sur les changements brusques de vitesse et de direction pr\'esent\'es
  par les rajectoires d'\'electrons de grande \'energie.
\newblock {\em Comptes Rendus de l'Acad\'emie des Sciences}, 200:1524--1526,
  April 1935.

\bibitem{Leprince-Ringuet:1936kx}
L.~Leprince-Ringuet.
\newblock Etude de la partie ultra p\'en\'etrante corpusculaire du rayonnement
  cosmique dans le champ magn\'etique de l'\'electro-aimant de bellevue.
\newblock {\em Journal de Physique Radium}, 7(2):67--70, 1936.

\bibitem{Leprince-Ringuet:1991nx}
L.~Leprince-Ringuet.
\newblock {\em Noces de diamant avec l'atome}.
\newblock Flammarion, Paris, 1991.

\bibitem{Millikan:1928vn}
R.~A. Millikan and G.~H. Cameron.
\newblock {Evidence for the Continuous Creation of the Common Elements out of
  Positive and Negative Electrons}.
\newblock {\em Proceedings of the National Academy of Sciences},
  14(6):445--450, 1928.

\bibitem{Millikan:1928uq}
R.~A. Millikan and G.~H. Cameron.
\newblock {High Altitude Tests on the Geographical, Directional, and Spectral
  Distribution of Cosmic Rays}.
\newblock {\em Physical Review}, 31(2):163--173, Feb 1928.

\bibitem{Millikan:1928aa}
R.~A. Millikan and G.~H. Cameron.
\newblock {The Origin of the Cosmic Rays}.
\newblock {\em Physical Review}, 32(4):533--557, Oct 1928.

\bibitem{Millikan:1933uq}
R.A. Millikan.
\newblock {New Techniques in the Cosmic-Ray Field and Some of the Results
  Obtained With Them}.
\newblock {\em Physical Review}, 43(8):661--669, Apr 1933.

\bibitem{Monaldi:2005qy}
D.~Monaldi.
\newblock {Life of $\mu$: The Observation of the Spontaneous Decay of Mesotrons
  and Its Consequences, 1938-1947}.
\newblock {\em Annals of Science}, 62:419--455, 2005.

\bibitem{Mott-Smith:1931fk}
L.~M. Mott-Smith.
\newblock {An Attempt to Measure the Energy of the Cosmic Electrons by Magnetic
  Deflection}.
\newblock {\em Physical Review}, 37(8):1001--1003, Apr 1931.

\bibitem{Mott-Smith:1932uq}
L.~M. Mott-Smith.
\newblock {On an Attempt to Deflect Magnetically the Cosmic-Ray Corpuscles}.
\newblock {\em Physical Review}, 39(3):403--414, Feb 1932.

\bibitem{Mott-Smith:1931kx}
L.~M. Mott-Smith and G.~L. Locher.
\newblock {A New Experiment Bearing on Cosmic-Ray Phenomena}.
\newblock {\em Physical Review}, 38(8):1399--1408, Oct 1931.

\bibitem{Neddermeyer:1937fk}
S.~H. Neddermeyer and C.~D. Anderson.
\newblock {Note on the Nature of Cosmic-Ray Particles}.
\newblock {\em Physical Review}, 51(10):884--886, May 1937.

\bibitem{Olivotto:2009fk}
C.~Olivotto.
\newblock The g--stack collaboration (1954): an experiment of transition.
\newblock {\em Historical Studies in the Natural Sciences}, 39(1):63--103,
  2009.

\bibitem{Orlando:1998jt}
L.~Orlando.
\newblock {Physics in the 1930s : Jewish physicists' contribution to the
  realization of the new tasks of physics in Italy}.
\newblock {\em Historical Studies in the Physical and Biological Sciences},
  29(1):141--181, 1998.

\bibitem{Ortner:1929uq}
G.~Ortner and G.~Stetter.
\newblock {{\"U}ber den elektrischen Nachweis einzelner Korpuskularstrahlen}.
\newblock {\em Zeitschrift f{\"u}r Physik}, 54(7):449--476, 1929.

\bibitem{Pigeard:2012fk}
N.~Pigeard~Micault.
\newblock {The Curie's Lab and its Women (1906--1934)}.
\newblock {\em Ann. Sci.}, pages 1--30, 2012.

\bibitem{Pontecorvo:1938uq}
B.~Pontecorvo.
\newblock {Isomeric Forms of Radio Rhodium}.
\newblock {\em Nature}, 141(3574):785--786, 1938.

\bibitem{Pontecorvo:1938kx}
B.~Pontecorvo.
\newblock {Nuclear Isomerism and Internal Conversion}.
\newblock {\em Physical Review}, 54(7):542--542, 1938.

\bibitem{Pontecorvo:1939fk}
B.~Pontecorvo and A.~Lazard.
\newblock Isom\'erie nucl\'eaire produite par les rayons x du spectre continu.
\newblock {\em Comptes Rendus de l'Acad\'emie des Sciences}, 208:99--101, 1939.

\bibitem{Puyo:1976uq}
Jean Puyo, editor.
\newblock {\em {Le bonheur de chercher. Jean Puyo interroge Louis
  Leprince-Ringuet}}.
\newblock Le Centurion, 1976.

\bibitem{Rechenberg:2010kx}
H~Rechenberg.
\newblock {\em {Werner Heisenberg -- Die Sprache der Atome}}.
\newblock Springer, 2010.

\bibitem{nature:1931fk}
Unknown reviewer.
\newblock The volta conference at rome.
\newblock {\em Nature}, 128(3238):861, 1931.

\bibitem{Roque:1992fk}
X.~Roqu\'e.
\newblock {M\o ller Scattering: a Neglected Application of Early Quantum
  Electrodynamics}.
\newblock {\em Archives for History of Exact Sciences}, 44(3):197--264, 1992.

\bibitem{Roque:1997qy}
X.~Roqu\'e.
\newblock {The Manufacture of the Positron}.
\newblock {\em Studies in History and Philosophy of Modern Physics},
  28(1):73--129, 1997.

\bibitem{Rossi:1930mz}
B.~Rossi.
\newblock {Method of registering multiple simultaneous impulses of several
  {Geiger} Counters}.
\newblock {\em Nature}, 125(3156):636, 26 April 1930.

\bibitem{Rossi:1930ve}
B.~Rossi.
\newblock {On the magnetic Deflection of Cosmic Rays}.
\newblock {\em Physical Review}, 36:606, 1930.

\bibitem{Rossi:1930qf}
B.~Rossi.
\newblock {\"Uber den {Ursprung} der durchdringenden {Korpuskularstrahlung} der
  {Atmosph{\"a}re}}.
\newblock {\em Die Naturwissenschaften}, 18:1096--1097, 1930.

\bibitem{Rossi:1931lq}
B.~Rossi.
\newblock {Magnetic Experiments on the Cosmic Rays}.
\newblock {\em Nature}, 128(3225):300--301, 1931.

\bibitem{Rossi:1931dq}
B.~Rossi.
\newblock {Measurements on the Absorption of the penetrating corpuscular Rays
  coming from inclined Directions}.
\newblock {\em Nature}, 128(3227):408, 1931.

\bibitem{Rossi:1931pd}
B.~Rossi.
\newblock Ricerche sull'azione del campo magnetico terrestre sopra i corpuscoli
  della radiazione penetrante.
\newblock {\em Atti della Reale Accademia Nazionale dei Lincei. Rendiconti},
  XIII:47--52, 1931.

\bibitem{Rossi:1931ul}
B.~Rossi.
\newblock {\"U}ber den {Ursprung} der durchdringenden {Korpuskularstrahlung}
  der {Atmosph{\"a}re}.
\newblock {\em Zeitschrift f{\"u}r Physik}, 68(1--2):64--84, Jan. 1931.

\bibitem{Rossi:1932cr}
B.~Rossi.
\newblock Absorptionsmessungen der durchdringenden {Korpuskularstrahlung} in
  einem {Meter Blei}.
\newblock {\em Die Naturwissenschaften}, 20(4):65, Jan. 22 1932.

\bibitem{Rossi:1931rr}
B.~Rossi.
\newblock Il problema della radiazione penetrante.
\newblock In {\em {Convegno di Fisica nucleare}}, pages 51--64. Reale Accademia
  d'Italia, 1932.

\bibitem{Rossi:1932fk}
B.~Rossi.
\newblock Nachweis einer {Sekund\"arstrahlung} der durchdringenden
  {Korpuskularstrahlung}.
\newblock {\em Physikalische Zeitschrift}, 33(16):304--305, Apr 1932.

\bibitem{Rossi:1932kx}
B.~Rossi.
\newblock Ricerche sulla radiazione secondaria della radiazione corpuscolare
  penetrante.
\newblock {\em La Ricerca Scientifica}, 3(2):234--264, 1932.

\bibitem{Rossi:1932uq}
B.~Rossi.
\newblock Sugli effetti secondari della radiazione corpuscolare penetrante.
\newblock {\em Atti della Reale Accademia Nazionale dei Lincei. Rendiconti},
  15:734--741, 1932.

\bibitem{Rossi:1933fk}
B.~{Rossi}.
\newblock {I risultati della missione scientifica in {E}ritrea per lo studio
  dei raggi cosmici}.
\newblock {\em La Ricerca Scientifica}, IV(2):365--368, 1933.

\bibitem{Rossi:1933qy}
B.~Rossi.
\newblock Interaction between cosmic rays and matter.
\newblock {\em Nature}, 132(3326):173--174, 1933.

\bibitem{Rossi:1933vn}
B.~Rossi.
\newblock \"uber die {Eigenschaften} der durchdringenden {Korpuskularstrahlung}
  in {Meeresniveau}.
\newblock {\em Zeitschrift f{\"u}r Physik}, 82(3--4):151--178, 1933.

\bibitem{Rossi:1933uq}
B.~{Rossi}.
\newblock {\"Uber die Wirkungen der Ultrastrahlung auf die Materie}.
\newblock {\em Helvetica Physica Acta}, 6:440--445, 1933.

\bibitem{Rossi:1934aj}
B.~Rossi.
\newblock {I risultati della {M}issione scientifica in {E}ritrea per lo studio
  della radiazione penetrante (Raggi cosmici). Misure sulla distribuzione
  angolare di intensit\`a della radiazione penetrante all'Asmara}.
\newblock {\em La Ricerca Scientifica. {Supplemento}}, 1(9--10):579--589, May
  1934.

\bibitem{Rossi:1939fk}
B.~Rossi.
\newblock {The Disintegration of Mesotrons}.
\newblock {\em Reviews of Modern Physics}, 11(3-4):296--303, Jul 1939.

\bibitem{Rossi:1940pd}
B.~Rossi.
\newblock {System of Units for Nuclear and Cosmic-Ray Phenomena}.
\newblock {\em Physical Review}, 57(7):660, Apr 1940.

\bibitem{Rossi:1966aa}
B.~Rossi.
\newblock {\em Cosmic Rays}.
\newblock McGraw--Hill, Inc., London, 1966.

\bibitem{Rossi:1981aa}
B.~Rossi.
\newblock {Early Days in Cosmic Rays}.
\newblock {\em Physics Today}, 34:35--43, 1981.

\bibitem{Rossi:1990aa}
B.~Rossi.
\newblock {\em {Moments in the Life of a Scientist}}.
\newblock Cambridge University Press, New York, 1990.

\bibitem{Rossi:1938rz}
B.~{Rossi} and P.~M.~S. Blackett.
\newblock {Further Evidence for the Radioactive Decay of Mesotrons}.
\newblock {\em Nature}, 142(3505):992--993, Dec 3 1938.

\bibitem{Rossi:1939zl}
B.~{Rossi} and P.~M.~S. Blackett.
\newblock {Some Recent Experiments on Cosmic Rays}.
\newblock {\em Reviews of Modern Physics}, 11(3--4):277--281, Jul 1939.

\bibitem{Rossi:1933rt}
B.~Rossi and E.~Fermi.
\newblock Azione del campo magnetico terrestre sulla radiazione penetrante.
\newblock {\em Atti della Reale Accademia Nazionale dei Lincei. Rendiconti},
  17:346--350, 1933.

\bibitem{Rossi:1941la}
B.~Rossi and K.~Greisen.
\newblock {Cosmic-Ray Theory}.
\newblock {\em Reviews of Modern Physics}, 13(4):240--309, Oct 1941.

\bibitem{Rossi:1942ai}
B.~Rossi, K.~Greisen, J.~C. Stearns, D.~K. Froman, and P.~G. Koontz.
\newblock {Further Measurements of the Mesotron Lifetime}.
\newblock {\em Physical Review}, 61(11-12):675--679, Jun 1942.

\bibitem{Rossi:1941wd}
B.~Rossi and D.~B. Hall.
\newblock {Variation of the Rate of Decay of Mesotrons with Momentum}.
\newblock {\em Physical Review}, 59(3):223--228, Feb 1941.

\bibitem{Rossi:1939lr}
B.~Rossi, N.~Hilberry, and J.~B. Hoag.
\newblock {The Disintegration of Mesotrons}.
\newblock {\em Physical Review}, 56(8):837--838, Oct 1939.

\bibitem{Rossi:1940gf}
B.~Rossi, N.~Hilberry, and J.~B. Hoag.
\newblock {The Variation of the Hard Component of Cosmic Rays with Height and
  the Disintegration of Mesotrons}.
\newblock {\em Physical Review}, 57(6):461--469, Mar 1940.

\bibitem{Rossi:1942dp}
B.~Rossi and N.~Nereson.
\newblock {Experimental Determination of the Disintegration Curve of
  Mesotrons}.
\newblock {\em Physical Review}, 62(9-10):417--422, Nov 1942.

\bibitem{Russo:2000ab}
A.~Russo.
\newblock {\em Le reti dei fisici. Forme dell'esperimento e modalit\`a della
  scoperta nella fisica del Novecento}.
\newblock La Goliardica Pavese, Pavia, 2000.

\bibitem{Segre:1993qy}
E.~Segr\`e.
\newblock {\em {A Mind Always in Motion. The Autobiography of Emilio Segr\`e}}.
\newblock University of California Press, Berkeley, 1993.

\bibitem{Skobeltzyn:1927fk}
D.~V. Skobeltzyn.
\newblock {Die Intensit\"atsverteilung in dem Spektrum der $\gamma$-Strahlen
  von Ra C}.
\newblock {\em Zeitschrift f{\"u}r Physik}, 43(5):354--378, 1927.

\bibitem{Skobeltzyn:1929uq}
D.~V. Skobeltzyn.
\newblock {\"Uber eine neue Art sehr schneller $\beta$-Strahlen}.
\newblock {\em Zeitschrift f{\"u}r Physik}, 54(9):686--702, 1929.

\bibitem{Skobeltzyn:1985vn}
D.~V. Skobeltzyn.
\newblock {The early stage of cosmic ray particle research}.
\newblock In Y.~Sekido and H.~Elliot, editors, {\em {Early History of Cosmic
  Ray Studies: Personal Reminiscences with old Photographs}}, pages 47--52,
  Dordrecht, 1985. D. Reidel.

\bibitem{Skobeltzyn:1930fj}
D.~V. {Skobelzyn}.
\newblock {Die Richtungsverteilung der von gestreuten {$\gamma$}-Strahlen
  erzeugten R{\"u}cksto{\ss}strahlen}.
\newblock {\em Zeitschrift f\"ur Physik}, 65(11-12):773--798, Nov 1930.

\bibitem{Stevenson:1935fk}
E.~C. Stevenson and J.~C. Street.
\newblock Minutes of the new york meeting, february 22-23, 1935.
\newblock {\em Physical Review}, 47(8):637--646, 1935.

\bibitem{Street:1937fk}
J.~C. Street and E.~C. Stevenson.
\newblock {New Evidence for the Existence of a Particle of Mass Intermediate
  Between the Proton and Electron}.
\newblock {\em Physical Review}, 52(9):1003--1004, 1937.

\bibitem{Street:1937qf}
J.~C. Street and E.~C. Stevenson.
\newblock {Penetrating Corpuscular Component of the Cosmic Radiation. Minutes
  of the Washington Meeting, April 29, 30 and May 1, 1937}.
\newblock {\em Physical Review}, 51(11):1005, Jun 1937.

\bibitem{Trenn:1986lr}
T.~J. Trenn.
\newblock {The Geiger-M\"uller Counter of 1928}.
\newblock {\em Annals of Science}, 43(2):111--135, 1986.

\bibitem{Ward:1929lr}
F.~A.~B. Ward, C.~E. Wynn-Williams, and H.~M. Cave.
\newblock {The Rate of Emission of Alpha Particles from Radium}.
\newblock {\em Proceedings of the Royal Society (London)}, A125(799):713--730,
  Nov 1 1929.

\bibitem{Wick:1934fk}
G.C. Wick.
\newblock {Sugli elementi radioattivi di F. Joliot e I. Curie}.
\newblock {\em Atti R. Acc. Naz. Lincei. Rend.}, 19:319--324, 1934.

\bibitem{Williams:1940rt}
E.~J. Williams and G.~E. Roberts.
\newblock {Evidence for Transformation of Mesotrons into Electrons}.
\newblock {\em Nature}, 145(3664):102--103, 1940.

\end{thebibliography}

\end{document}